%% file: SciPost-version.tex
\documentclass[submission, Phys]{SciPost}

\usepackage{mathtools,amsthm,amsmath,amsfonts,amssymb}
\DeclareMathOperator{\Tr}{Tr}
\usepackage{scalerel}
\usepackage{graphicx}
\usepackage{subcaption}
\usepackage{mathrsfs}
\usepackage{bbold}
\usepackage[]{units}
\usepackage{bm}
\usepackage{physics}
\usepackage{makecell}
\usepackage{enumitem}
\usepackage{upgreek}
\usepackage{blindtext}
\usepackage{verbatim} % for math
\usepackage[normalem]{ulem}
\usepackage[noend]{algpseudocode}
\usepackage[dvipsnames]{xcolor}
\usepackage{orcidlink}
\usepackage{tikz}
\usepackage{tikz-3dplot}
    \usetikzlibrary{shapes.geometric,arrows.meta,decorations.pathmorphing,shadows.blur,calc}
    \usetikzlibrary{math,graphs, graphs.standard,shadings}
\usepackage{bbm}
\usepackage{array}
\usepackage{multirow}
\usepackage{tabularx}
\usepackage{float}
\usepackage{dcolumn}% Align table columns on decimal point
\usepackage{slashed}
\usepackage{resizegather}
\usepackage[export]{adjustbox}
\usepackage{caption}

\pgfdeclareradialshading[tikz@ball]{ball}{
	\pgfqpoint{-20bp}{20bp}}{%
	color(0bp)=(tikz@ball!0!white);
	color(17bp)=(tikz@ball!0!white);
	color(21bp)=(tikz@ball!70!black);
	color(25bp)=(black!70);
	color(30bp)=(black!70)}
\makeatother

\newtheorem*{theorem}{Theorem}

\DeclareMathOperator{\arcosh}{arcosh}

\DeclareMathOperator{\sign}{sign}
\DeclareMathOperator{\mpsi}{\hat{\chi}}

\begin{document}
	
	% TODO: write your article's title here.
	% The article title is centered, Large boldface, and should fit in two lines
	\begin{center}{
    \Large \textbf{
		Solving L\'evy Sachdev-Ye-Kitaev Model
	}}
    \end{center}
	%\sout{Large$-N$ Theory of}
	% TODO: write the author list here. Use initials + surname format.
	% Separate subsequent authors by a comma, omit comma at the end of the list.
	% Mark the corresponding author with a superscript *.
	
	\begin{center}
		Budhaditya Bhattacharjee\textsuperscript{1,2*},
		William. E. Salazar\textsuperscript{3,4*},
		Alexei Andreanov\textsuperscript{2,5,6} and
		Dario Rosa\textsuperscript{4}
	\end{center}
	
	% TODO: write all affiliations here.
	% Format: institute, city, country
	\begin{center}
		{\bf 1} Department of Physics and Materials Science, University of Luxembourg, Ave. de la F\"aiencerie, Luxembourg 1511
		\\
		{\bf 2} Center for Theoretical Physics of Complex Systems, Institute for Basic Science(IBS), Daejeon 34126, Republic of Korea
		\\
		{\bf 3} Center for Quantum Technology, National University of Singapore, 3 Science Drive 2, Block S15, Singapore 117543
		\\
		{\bf 4} ICTP South American Institute for Fundamental Research, Instituto de F\'{i}sica Te\'{o}rica,\\ UNESP - Univ. Estadual Paulista, Rua Dr. Bento Teobaldo Ferraz 271, 01140-070, S\~{a}o Paulo, SP, Brazil 
		\\ 
		{\bf 5} Center for Trapped Ions Quantum Science, Institute for Basic Science, Daejeon 34126, Republic of Korea
		\\
		{\bf 6} Basic Science Program, Korea University of Science and Technology (UST), Daejeon 34113, Republic of Korea
		\\

		% TODO: provide email address of corresponding author
		{*budhaditya.bhattacharjee@uni.lu}
		\\
		{*william\_esteban@u.nus.edu}
	\end{center}
	
	\begin{center}
		\today
	\end{center}
	
	% For convenience during refereeing: line numbers
	%\linenumbers
	
	\section*{Abstract}
	\textbf{
		% TODO: write your abstract here.
		We present an exact solution in the large-\(N\) limit of the L\'{e}vy Sachdev-Ye-Kitaev (LSYK) model introduced in Ref.~\cite{bhattacharjee2025levy}, wherein the couplings are drawn from a  L\'{e}vy Stable distribution parameterized by a tail exponent \(\mu \in [0, 2]\). 
		Starting from the Hamiltonian and its associated partition function, we highlight the key differences from the standard Gaussian SYK model and derive the large-\(N\) Schwinger-Dyson equations via a bosonic oscillator representation of the action. 
		These equations are solved both numerically and analytically in the large-\(q\) and infrared limits. 
		We subsequently analyze the chaotic properties of the model by computing the Krylov exponent from the large-\(q\) Green's function and extracting the Lyapunov exponent from the \(4\)-point function. 
		The parameter \(\mu\) continuously interpolates between a free theory at \(\mu = 0\) and the conventional, maximally chaotic Gaussian SYK model at \(\mu = 2\), with non-maximal chaos persisting throughout the intermediate regime \(0 < \mu < 2\). 
		Thermodynamic quantities, including the entropy, free energy, average energy, and specific heat capacity, are computed and compared with their Gaussian SYK counterparts. 
		The interpretations of the thermodynamics are discussed with respect to the holographic dual and non-Fermi liquid theory.
		Finally, we discuss an alternative representation of the LSYK model based on a distinct decomposition of the L\'{e}vy Stable distribution, which establishes a non-trivial connection to Gaussian SYK, and provide supporting analytical and numerical results in the appendices.
	}

	% TODO: include a table of contents (optional)
	% Guideline: if your paper is longer that 6 pages, include a TOC
	% To remove the TOC, simply cut the following block
	
	\newpage
	\vspace{10pt}
	\noindent\rule{\textwidth}{1pt}
	\tableofcontents\thispagestyle{fancy}
	\noindent\rule{\textwidth}{1pt}
	\vspace{10pt}

	\section{Introduction}
	\label{intro}
	
	The study of quantum chaos forms a central research direction in quantum mechanics, quantum field theory and quantum gravity.
	In quantum mechanics, the central dogma of quantum chaos is the \emph{Bohigas-Giannoi-Schmidt (BGS)} conjecture~\cite{bohigas1984characterization}, connecting quantum chaos to random matrix theory~\cite{haake1991quantum, mehta2004random}.
	The basic idea of this conjecture is that the spectral correlations in quantum systems that are ``chaotic'' are faithfully represented by the spectral correlations of random matrices.
	This has been successfully tested in a large number of systems, most notably in billiard systems demonstrating semi-classical chaos~\cite{guhr1988random}.
	Treating the BGS conjecture as a \emph{de-facto definition} of quantum chaos, it has been explored extensively in quantum \emph{many-body systems}.
	Several distinct approaches (apart from random matrix universality) are used to study many-body chaos, such as eigenstate thermalization~\cite{srednicki1994chaos,deutsch2018eigenstate}, free probability~\cite{fava2025designs}, entanglement~\cite{wang2004entanglement}, out-of-time-ordered correlators~\cite{xu2024scrambling} and Krylov complexity~\cite{parker2018a}, among many others. 
	
	A quantum mechanical many-body model that has proven to be particularly useful in the study of many-body chaos is the \emph{Sachdev-Ye-Kitaev (SYK)} model~\cite{sachdev1993spin,kitaevLectures}.
	This is a model described by a quenched, disordered Hamiltonian with all-to-all interacting fermions.
	The most extensively studied instance of the model consists of Gaussian disorder and Majorana fermions~\cite{polchinski2016the,maldacena2016remarks,cotler2016black,garcia2016spectral}.
	The unique feature of this model is that it is \emph{solvable} in the $N\rightarrow\infty$ limit, where $N$ is the number of Majorana fermions.
	This is part of the reason why the model is of significant interest also from the perspective of holography~\cite{sekino2008fast,shenker2014black,mertens2023solvable}.
	This model has been discussed in several comprehensive reviews~\cite{sarosi2018ads2,rosenhaus2019an,chowdhury2022sachdev,jha2025introduction}.
	From the perspective of many-body quantum chaos, the model and its' modifications have been extensively studied~\cite{tezuka2023binary,garcia-garcia2018chaotic,garcia2021sparse,andreanov2025from}.
	For a system with $N$ Majorana fermions, the SYK Hamiltonian is written as 
	\begin{align}
		\hat{H} = \frac{i^{\lfloor q/2 \rfloor}}{q!} \sum_{i_1,i_2,\dots,i_q}^{N} J_{i_1,i_2,\dots,i_q} \hat{\chi}_{i_1}\hat{\chi}_{i_2}\cdots\hat{\chi}_{i_q}\,.
	\end{align}
	where $q$ is an even integer denoting the number of Majorana fermions that interact in a single term of the Hamiltonian.
	The random interactions are denoted by $J_{i_1,i_2,\dots,i_q}$.
	
	A particularly interesting modification of the SYK model with a crossover from chaotic to integrable signatures is one with \emph{sparse} couplings~\cite{garcia2021sparse,tezuka2023binary}.
	In this variant of the model, the Gaussian interaction tensor $J_{i_1 i_2 \cdots i_q}$ comes equipped with an additional sparsity factor $p \in [0,1]$, where $p=0$ represents no interaction and $p = 1$ indicates dense connectivity.
	In practice, every random variable $J_{i_1 i_2 \cdots i_q}$ is kept with probability $p$ and set to zero with probability $1 - p$, thus reducing the connectivity of the model.
	The introduction of the sparsity parameter effectively eliminates a finite fraction of interactions between the Majorana fermions, and this is found to induce non-chaotic behaviour. This comes at the cost of loosing the inherent \emph{solvability} of the model in the large-\(N\) limit.

	An open problem was to find a model that had all the salient solvability features of the SYK model, and still featured a crossover from chaotic to integrable quantum many-body behaviour in a controlled manner by effectively emulating sparse SYK.

	Such model was proposed in Ref.~\cite{bhattacharjee2025levy}.
	To briefly recap, the main modification is the replacement of the Gaussian disorder $J_{i_1 i_2 \cdots i_q}$ by a \emph{L\'evy-Stable} distribution.
	L\'evy distributions are known to posses fat tails (usually controlled by a parameter $\mu \in (0,2]$, where $\mu=2$ is Gaussian) and therefore display large deviations in their sample sets.
	Treating the random realisations of the interaction in a hierarchical manner~\cite{monthus2016localization}, it is possible to construct a scaling theory for eigenvalue correlations~\cite{kutlin2024anatomy,bhattacharjee2025levy} which demonstrates that there is an $N-$ dependence on values of $\mu$ that cause deviation from chaotic behaviour.
	This is exactly the scenario also obtained in sparse SYK with respect to the sparsity parameter.
	Normalising the Hamiltonian by the largest of $J_{i_1 i_2 \cdots i_q}$ in a given realisation, an approximate sparse network is obtained since the largest values are \emph{parameterically larger} than the typical values.
	However, the main difference from the sparse SYK is that all-to-all models with L\'evy Stable disorder are \emph{solvable}, as evidences (for example) in L\'evy Spin glass problems~\cite{janzen2008replica,janzen2010the,janzen2010thermodynamics} suggest.
	In this article, we solve the L\'evy Sachdev-Ye-Kitaev model~\cite{bhattacharjee2025levy} in the large$-N$ limit, for any value of the parameter $\mu$.
	A brief summary of our findings is given below.

	\subsection*{Summary} 

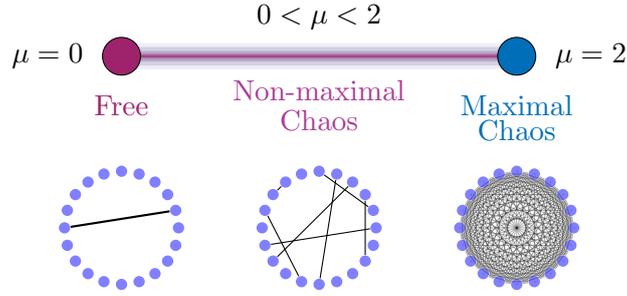
\begin{figure}[ht]
		\centering
		\input{Figures/rg-flow.tikz}
		\caption{
			%        Chaos Profile of L\'evy Sachdev-Ye-Kitaev model.
			Phase diagram of the L\'evy Sachdev-Ye-Kitaev model.
			Top: for \(\mu=0\) the model is free.
			Chaoticity increases with \(\mu\) reaching maximum for \(\mu=2\), the standard Sachdev-Ye-Kitaev model.
			Bottom: sparsity pattern of the model, as explained in the main text.
		}
		\label{fig:lsyk-chaos-profile}
	\end{figure}

	We present the solution of the L\'evy Sachdev-Ye-Kitaev model in the large$-N$ limit and discuss several of its aspects.
	In Section~\ref{sec:ham-parition-function}, we introduce the Hamiltonian and the salient features of the disorder distribution.
	The partition function is derived for the Hamiltonian and the main difference from the Gaussian SYK case is highlighted.
	In Section~\ref{sec:bosonic-oscillator}, a method involving bosonic oscillators is introduced which casts the partition function in an appropriate form for deriving the closed form Schwinger-Dyson equations.
	In Section~\ref{sec:sd-eqn}, the Schwinger-Dyson equations in the large-\(N\) limit are derived from the action derived in Section.~\ref{sec:bosonic-oscillator}.
	The solution of the equations is found numerically, as well as analytically in the large$-q$ and conformal (deep IR) limit.
	In Section~\ref{sec:chaos-exp}, the Krylov exponent is evaluated from the large$-q$ Green's function and the Lyapunov exponent is extracted from the $4-$point function calculation.
	These indicators can be used to clarify the chaoticity of the model : the results are schematically represented in the Figure.~\ref{fig:lsyk-chaos-profile}.
	This schematic shows that there is a family of SYK models parameterised by a continuous parameter \(\mu\in [0,2]\)~\footnote{Which parameterises the tail behaviour of the L\'evy Stable distribution}.
	The model is free for $\mu = 0$ and maximally chaotic for $\mu = 2$~\cite{maldacena2016a,parker2018a}.
	The emergent sparsity structure (where only the parametrically dominant ``bonds'' are retained) is also presented in the schematic.
	The network in the schematic denotes $2-$body interactions, but the picture is easily generalised to $q-$ body interactions.
	In the intermediate regime $0 < \mu < 2$, the model is still chaotic, but in a non-maximal sense.

	In Section~\ref{sec:thermo}, thermodynamic quantities such as entropy, free energy, average energy and specific heat capacity of the system are studied numerically and in the large$-q$ limit. Their behaviour is also compared and contrasted with that of Gaussian SYK.
	We conclude by reviewing the main results and mentioning future directions in Section.~\ref{sec:conclusions}.
	In the Appendix.~\ref{app:alternate-desc}, we provide an alternative description of the L\'evy SYK based on a different representation of the L\'evy Stable distribution.
	This connects the model in a non-trivial way to Gaussian SYK.
	In Appendix.~\ref{app:sch-prob}, we briefly discuss the Schr\"odinger problem that arises in the evaluation of the Lyapunov exponent.
	Finally, Appendix.~\ref{app:num-res} presents some supporting numerical results. 
	%\emph{Introduction:} 

	\section{Hamiltonian and Partition Function}
	\label{sec:ham-parition-function}

	The L\'evy Sachdev-Ye-Kitaev model is described by all-to-all interacting Majorana fermions, connected via $q-$body interactions whose strengths are sourced from the L\'evy distribution:
	\begin{align}
		\hat{H} = \frac{i^{\lfloor q/2 \rfloor}}{q!}\sum_{I}J_I\Psi_I\,,
		\label{eq:lsyk_hamiltonian}
	\end{align}
	where $J_{I} = J_{i_1,i_2,\dots,i_q}$ is the interaction strength and $\Psi_{I} = \mpsi_{i_1}\mpsi_{i_2}\dots\mpsi_{i_q}$ is the $q-$body Majorana fermion term and each index $i_k$ ranges from $1$ to $N$.
	The interaction $J_I$ is sampled from the L\'evy Stable distribution with stability index $\mu$, which is described by the following probability density function
	\begin{align}
		\mathrm{d}P_{\mu}[X] =\frac{\mathrm{d}X}{2\pi} 
		\int \mathrm{d}k \, \exp\left(ikX-|\sigma k|^{\mu}\right)\,.
		\label{eq:def_stable_dist_1}
	\end{align}
	This is a special case of the L\'evy Stable distribution, which is generally defined using the following additional parameters:
	$\eta \in [-1,1]$, known as the "skewness", and $\delta \in R$, known as the shift~\cite{nolan2020univariate}. 
	The full PDF is often denoted as $\mathbf{L}_\mu (\eta,\sigma,\delta)$.
	We set these parameters to $0$ for simplicity.
	As is clear, a closed-form expression for the PDF does not exist for general $\mu$.
	The key feature of this distribution is that for $\mu < 2$, the variance diverges, and for $\mu < 1$, the mean also diverges.
	The tail of this distribution is captured via the power law behaviour $\mathrm{P}_{\mu}(X \gg 1) \sim \vert X \vert^{-1 - \mu}$.
	For a smooth flow of the scale parameter $\sigma$ into the Gaussian SYK variance~\cite{maldacena2016remarks} and to ensure the extensivity of the Hamiltonian~\eqref{eq:lsyk_hamiltonian}, we choose the factor as 
	\begin{align}
		\sigma = J\left(\frac{2 q}{N}\binom{N}{q}\right)^{\frac{-1}{\mu}}\,.
	\end{align}
	It is convenient to write $\sigma^\mu = \frac{J^\mu}{\lambda \mathcal{N}}$. Here we denote $\mathcal{N} = \binom{N}{q}$ and $\lambda = \frac{2 q}{N}$. 
	%For LSYK, we denote the scale parameter $\sigma$ in \eqref{eq:def_stable_dist_1} as $\mathcal{J}_L$.
	From the Hamiltonian Eqn.~\eqref{eq:lsyk_hamiltonian}, the partition function $Z(\beta)$, with the inverse temperature $\beta$, is evaluated.
	Let us write the (disorder averaged) path integral as~\footnote{In this work we focus on the annealed average. 
		The quenched average, and its comparison to the annealed one is deferred to future work~\cite{anscheutz25stronglt}.}
	\begin{align}
		\langle Z(\beta) \rangle = \langle \Tr\exp\{-\beta \hat{H}\}\rangle_{J_I}\,.
		\label{eq:partition-function-beta}
	\end{align}
	It is known from the study of L\'evy spin glass~\cite{neri2010the} that $Z(\beta)$ diverges for real $\beta$.
	The way around this problem, as discussed in Ref.~\cite{janzen2008replica}, involves Wick rotation to imaginary temperature $\beta = -i k$ and performing the averaging in this frame.
	In terms of the Grassmann fields, the imaginary temperature partition function is written as 
	\begin{align}
		{Z}(k,\{J_I\}) = \int \mathcal{D}\psi \exp\left\{\int_{0}^{k}\mathrm{d}t\left(-\sum_{i = 1}^{N}\frac{1}{2}\psi_i \partial_t \psi_i + i H\right)\right\}\,.
		\label{eq:parition-function-k}
	\end{align}
	Using \eqref{eq:lsyk_hamiltonian}, we write it as
	\begin{align}
		&{Z}(k,\{J_I\}) = \int \mathcal{D}\psi \exp\left\{\int_{0}^{k}\mathrm{d}t\left(-\sum_{i = 1}^{N}\frac{1}{2}\psi_i \partial_t \psi_i + i^{1 + \lfloor q/2 \rfloor}\sum_{I}J_I \Psi_I\right)\right\}\,.
	\end{align}
	To proceed, we need to perform the ensemble average over the random Levy variables, \textit{i.e.} we need to evaluate 
	\begin{align}
		\langle {Z}(k) \rangle \equiv \int \prod_I \mathrm{d}J_I P(J_I) Z(k, \left\{ J_I\right\})\,.
	\end{align}
	To this end, we use the characteristic function of L\'evy variables, given by $\int \mathrm{d}X P_\mu (X)e^{i t X} = \exp\{-\vert \sigma t \vert^\mu\}$, which allows us to find that the averaged partition function is given by
	\begin{align}
		\langle{Z}(k)\rangle = \int\mathcal{D}\psi \exp\left\{-\int_{0}^{k}\mathrm{d}t \sum_{i = 1}^{N}\frac{1}{2}\psi_i \partial_t \psi_i - \sum_I\left(\sigma^2 V(G_I)\right)^{\frac{\mu}{2}}\right\} \equiv \int \mathcal{D}\psi\exp\{-S_\mu(\{\psi\})\}\,,
		\label{eq:partition-function-k-avg}
	\end{align}
	where we define the functional $V(G_I) \equiv \Big\vert\int_0^{k}\mathrm{d}t \Psi_I (t)\Big\vert^2$.
	Note that the presence of the overall exponent $\mu/2$ marks the difference from the usual, Gaussian SYK.
	Setting $\mu = 2$, we recover the Gaussian SYK action~\cite{maldacena2016remarks}.
	Having evaluated the partition function, the Wick rotation is inverted by replacing $\beta = i k$, which gives us $S_\mu(\{\psi\})$ in terms of $\beta$.

	\section{Bosonic Oscillators}
	\label{sec:bosonic-oscillator}

	Apart from the case $\mu = 2$, when the model boils down to the standard SYK model, the action $S_\mu (\{\psi\})$ is not in a suitable form from which the SD equations can be derived.
	To obtain such a form, we recall the following result for bosonic modes.
	Consider the bosonic creation-annihilation operators $a^\dagger, a$ endowed with the vacuum $\ket{0}$: $a\ket{0} = 0$.
	We can construct a unitary displacement operator~\footnote{It shifts $a$ and $a^\dagger$ by the complex number $z$ and $z^{*}$ respectively.} (given a $c-$number $z$) $D(z) = \exp\{z a^\dagger - z^{*}a\}$ which satisfies
	\begin{align}
		\bra{0}D^\dagger (z) A(a^\dagger, a) D(z) \ket{0} = A(z^{*},z)
	\end{align}
	for any operator $A(a^\dagger,a)$.
	For our analysis, it is convenient to consider
	\begin{align}
		\bra{0}A(a^\dagger,a)\ket{0} = \lim_{\beta_c \rightarrow \infty}\frac{\Tr\left(e^{-\beta_c a^\dagger a}A(a^\dagger,a)\right)}{\Tr\left(e^{-\beta_c a^\dagger a}\right)}
	\end{align}
	where we have introduced another ``temperature'' $\beta_c$. Considering $A(a^\dagger,a) = F(a^\dagger)F(a)$, we can write 
	\begin{align}
		\bra{0}D^\dagger (z) F(a^\dagger)F(a) D(z) \ket{0}= \lim_{\beta_c \rightarrow \infty} \frac{\mathrm{Tr}\left(e^{-\beta_c (a^\dagger - z^{*}) (a - z)}F(a^\dagger)F(a)\right)}{\mathrm{Tr}\left(e^{-\beta_c a^\dagger a}\right)} 
	\end{align}
	This expression can now be cast in the path integral form based on bosonic coherent states $\ket{\phi} = e^{\phi a^\dagger}\ket{0}$.
	The full expression for the same is given by
	\begin{align}
		A(z^{*}, z) = F(z^*)F(z) = \lim_{\beta_c \rightarrow \infty}\int \mathcal{D}\phi\,\exp\{-\mathcal{S}(\phi,z,z^{*})\}
	\end{align}
	where the action $\mathcal{S}$ is given by
	\begin{align}
		&\mathcal{S}(\phi,z,z^{*}) = \int_0^{\beta_c} \mathrm{d}\tau \left(\bar{\phi}\partial_\tau \phi + \vert \phi(\tau) - z \vert^2\right)\notag\\
		&- \int_{0}^{\beta_c}\mathrm{d}\tau\left[\delta(\tau-\beta_c)f(\bar{\phi}(\tau)) + \delta(\tau)f(\phi(\tau))\right]\,. 
	\end{align}
	Here $F \equiv e^f$ defines the function $f$ in $\mathcal{S}$.
	
	\begin{figure}%[ht]
		\centering
		\includegraphics[width=0.95\columnwidth]{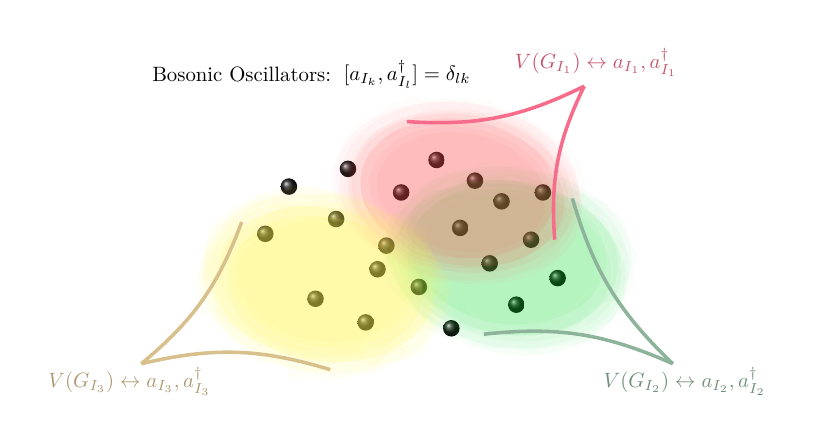}
		\caption{
			Schematic representation of the Bosonic oscillator approach.
			Each colored bubble represents a collective mode $V(G_I) = \int \psi_{i_1}\cdots \psi_{i_q} \mathrm{d}\tau$ which are linked to Bosonic creation-annihilation pair $a_I, a^\dagger_I$.
			The hard spheres represent the Majorana fermions.
		}
		\label{fig:schem-osc}
	\end{figure}
	
	This approach can now be applied to the partition function~\eqref{eq:partition-function-k-avg} to
	replace the non-linear exponent \(\exp\{V^{\mu/2}(G_I)\}\) in~\eqref{eq:partition-function-k-avg} by an integral over bosonic modes, that is quadratic in \(V(G_I)\).
	This approach is similar to the Hubbard-Stratonovich method.
	The potential term in the action can be written as
	\begin{align}
		e^{\sum_I \frac{1}{\lambda\mathcal{N}}(J^2 V(G_I))^{\frac{\mu}{2}}} = \prod_I \bra{0}_I D^\dagger (z_I)F(a^\dagger_I)F(a_I)D(z_I)\ket{0}_I \notag
	\end{align}
	where $F(a) = \exp\{\frac{1}{2 \lambda \mathcal{N}}a^{\frac{\mu}{2}}\}$, and similarly for $F(a^\dagger)$.
	For each $I$, an oscillator $a_I$ is introduced and the parameter $z_I = J^2 V(G_I)$ is chosen. 
	This is presented schematically in Fig.~\ref{fig:schem-osc}.
	The full vacuum is the direct product of the individual vacuum states $\ket{\Omega} = \ket{0}\otimes\ket{0}\otimes\cdots\otimes\ket{0}$.
	All-in-all, the full partition is written as
	\begin{align}
		\langle{Z}(\beta)\rangle = \int \mathcal{D}\psi \exp\left\{\left(-\int_{0}^{\beta}\mathrm{d}\tau \sum_{i = 1}^{N}\frac{1}{2}\psi_i \partial_\tau \psi_i\right)\right\}\lim_{\beta_c \rightarrow\infty}\int\mathcal{D}\phi \exp\left\{-S_{\mathrm{b},\beta_c}\left(\{\bar{\phi},\phi\}\right)\right\}
	\end{align}
	where the bulk action $S_\mathrm{b}$ is given as
	\begin{align}
		S_{\mathrm{b},\beta_c}&(\{\bar{\phi}_I,\phi_I\}) = \sum_I \int_0^{\beta_c} \mathrm{d}\tau' \left(\bar{\phi}_I\partial_\tau\phi_I + \vert \phi_I - J^2 V(G_I) \vert^2\right)\notag\\
		&- \sum_I \frac{1}{2 \lambda \mathcal{N}}\int_0^{\beta_c} \mathrm{d}\tau' \left(\delta(\tau'-\beta_c)\bar{\phi}^\frac{\mu}{2}_{I} + \delta(\tau')\phi^{\frac{\mu}{2}}_{I}\right)\,.
		\label{eq:saddle-phi-full}
	\end{align}
	The full LSYK action is recovered only in the limit $\beta_c \rightarrow \infty$.
	We note that, this approach is not specific to the LSYK model, but rather can be applied to any model.
	However, to advance further and compute anything useful, the bosonic modes $\phi_I$ have to be integrated over:
	the feasibility of this integration depends strongly on the details of the model.
	For LSYK, it is not possible to directly integrate over the fields $\phi_I$.
	Therefore, we will make a choice of saddles $\{\phi^*_I\}$ of $S_{\mathrm{b},\beta_c}$ to obtain $\langle Z(\beta)\rangle$ in an appropriate form.
	We emphasize that while the large-\(N\) saddle point equations we derive in the following sections are exact, their solutions' manifold is complex and requires careful analysis to characterize completely.
	We expect that there can potentially be a large number of distinct saddles, corresponding to solutions with distinct features.
	In the following sections we focus on one such saddle and discuss the physical motivation of selecting that particular saddle.

	\subsection{Static saddles}
	
	We begin by imposing some constraints on the bosonic oscillator modes $\phi_{I}$, with regard to their distribution over the indices $I$, and evaluate the action.
	There are several possible choices: Motivated by the all-to-all interaction of the Hamiltonian, we focus on $\phi_I$ that posses an exact or approximate symmetry in the indices $I$.
	The first natural choice is to assume that $\phi_I = \phi \;\;\forall\;I$, i.e. complete site independence.
	Given this assumptions, the bulk action~\eqref{eq:saddle-phi-full} is given by
	\begin{align}
		S_\mathrm{b} = &\mathcal{N}\int_0^{\beta_c}\mathrm{d}\tau' \left\{\bar{\phi} \dot{\phi} + \vert \phi \vert^2 - \frac{1}{2 \lambda \mathcal{N}}(\delta(\tau'-\beta_c)\bar{\phi}^{\frac{\mu}{2}} + \delta(\tau')\phi^\frac{\mu}{2})\right\}\notag\\
		&+ \sum_I \int_0^{\beta_c}\mathrm{d}\tau' \left(J^{4} V(G_I)^2 - J^2 V(G_I)(\phi + \bar{\phi})\right)
		\label{eq:sb-condensate1}
	\end{align}
	There is a constant piece, which is independent of $\tau'$ leading to the divergence $\lim_{\beta_c \rightarrow\infty}\int_0^{\beta_c}\mathrm{d}\tau' \equiv \mathrm{Vol}(\mathrm{R}^{+})$.
	With this, the bulk action can be written as
	\begin{align}
		&S_\mathrm{b} = \mathcal{N}\int_0^{\beta_c}\mathrm{d}\tau' \left\{\bar{\phi} \dot{\phi} + \vert \phi \vert^2 - \frac{1}{2 \lambda\mathcal{N}}(\delta(\tau'-\beta_c)\bar{\phi}^{\frac{\mu}{2}} + \delta(\tau')\phi^\frac{\mu}{2})\right\}\notag\\
		&+ \sum_I \left(\mathrm{Vol}(\mathrm{R}^{+})J^{4} V(G_I)^2  - J^2 V(G_I)\int_0^{\beta_c}\mathrm{d}\tau' (\phi + \bar{\phi})\right)\,.
		\label{eq:action-frozen-condensate}
	\end{align}
	The $\phi-$independent term in Eqn.~\eqref{eq:action-frozen-condensate} can be ignored for the time being since it does not appear in the saddle point equations.
	Additionally, note that 
	\begin{align}
		\sum_I V(G_I)
		&= \frac{1}{q!}\int_{0}^{\beta}\mathrm{d}\tau\mathrm{d}\tau' \left(\sum_{i}\psi_{i}(\tau) \psi_{i}(\tau')\right)^q \\
		\sum_{I}V(G_I)^2 &= \frac{1}{q!}\int_0^{\beta}\mathrm{d}\tau\mathrm{d}\tau'\mathrm{d}\tau''\mathrm{d}\tau'''\notag\left(\sum_i \psi_i(\tau)\psi_i(\tau')\psi_i(\tau'')\psi_i(\tau''')\right)^q
	\end{align}
	In particular, we can integrate over the terms that depend on $G_I$, by performing the integral $\int\mathcal{D}\psi$ over the Majorana fermions.
	The result of the integral on $\sum_{I} V(G_I)$ is defined as $\mathcal{N}V[G]$, by introducing the collective functional $V[G]$.
	The resulting equation for $\phi$, $\bar{\phi}$ is then given by
	\begin{align}
		\dot{\phi}(\tau') + \phi(\tau') - \frac{\mu}{4 \lambda \mathcal{N}}\delta(\tau'-\beta_c)\bar{\phi}^{\frac{\mu}{2}-1}(\tau') - J^2 V[G] &= 0
		\label{eq:saddle-phi} \\
		\dot{\bar{\phi}}(\tau') - \bar{\phi}(\tau') + \frac{\mu}{4 \lambda \mathcal{N}}\delta(\tau')\phi^{\frac{\mu}{2}-1}(\tau') + J^2 V[G] &= 0
		\label{eq:saddle-phi-bar}
	\end{align}
	Apart from the boundary $\delta$-function, these equations describes an asymptotically decaying mode $\phi(\tau') \sim e^{-\tau'}\phi(0)$ and a growing mode $\bar{\phi}(\tau') \sim e^{\tau'}\bar{\phi}(0)$.
	The general solution for these are written as
	\begin{align}
		\phi(\tau) = v + c e^{-\tau} \;\;,\;\; \bar{\phi}(\tau) = v + c' e^{\tau}\,.
		\label{eq:phi-phibar-soln}
	\end{align}
	Here we have used the shorthand $v = J^2 V[G]$, while $c, c'$ are constants.
	To fix these constants, we integrate Eqns.~\eqref{eq:saddle-phi} and~\eqref{eq:saddle-phi-bar} across the boundary points \(\tau' = \beta_c\) and \(0\) respectively and find:
	\begin{align}
		c(1 - e^{-\beta_c}) = \frac{\mu}{4\lambda \mathcal{N}}\bar{\phi}^{\frac{\mu}{2}-1}(\beta_c)\;\;,\;\; c'(1- e^{\beta_c}) = -\frac{\mu}{4\lambda\mathcal{N}}\phi^{\frac{\mu}{2}-1}(0)\,.
		\label{eq:c-cprime-soln}
	\end{align}
	The above is rewritten as non-linear self-consistent conditions:
	\begin{align}
		c &\sim \frac{\mu}{4\lambda\mathcal{N}}\left(v + \frac{\mu}{4\lambda\mathcal{N}}(v+c)^{\frac{\mu}{2}-1}\right)^{\frac{\mu}{2}-1}\,,
		\label{eq:c-rel} \\
		c'e^{\beta_c} &\sim \frac{\mu}{4\lambda\mathcal{N}}\left(v + \frac{\mu}{4\lambda\mathcal{N}}(v+c'e^{\beta_c})^{\frac{\mu}{2}-1}\right)^{\frac{\mu}{2}-1}\,.
		\label{eq:cprime-rel}
	\end{align}
	For this the periodic boundary conditions on the thermal circle $\tau \in [0,\beta_c]$ are applied since \(\phi\) are bosons.
	We also use the limit of large $\beta_c$.
	The $\delta$ function discontinuity acts as a boundary divergence for $\phi,\bar{\phi}\rightarrow 0$ and $\mu < 2$.
	Using this solution of $\phi,\bar{\phi}$, we can evaluate the action Eqn.~\eqref{eq:action-frozen-condensate}.
	This gives us
	\begin{align}
		\frac{S_\mathrm{b}}{\mathcal{N}} = -v \int_0^{\beta_c} \mathrm{d}\tau' \phi(\tau') + \left(\frac{\mu - 2}{4\lambda\mathcal{N}}\right)\bar{\phi}^{\frac{\mu}{2}}(\beta_c) - \frac{1}{2\lambda\mathcal{N}}\phi^{\frac{\mu}{2}}(0) + \Gamma\,.
	\end{align}
	Here we denote the divergent term appearing with $\text{Vol}(\mathrm{R}^+)$ with $\Gamma$. Using Eqn.~\eqref{eq:phi-phibar-soln} and Eqn.~\eqref{eq:c-cprime-soln}, we can further write this as
	\begin{align}
		\frac{S_\mathrm{b}}{\mathcal{N}} = -\frac{\mu v}{4\lambda\mathcal{N}}\bar{\phi}^{\frac{\mu}{2}-1}(\beta_c) + \left(\frac{\mu - 2}{4\lambda\mathcal{N}}\right)\bar{\phi}^{\frac{\mu}{2}}(\beta_c) - \frac{1}{2\lambda\mathcal{N}}\phi^{\frac{\mu}{2}}(0) + \Gamma - v^2\beta_c\,.
	\end{align}
	This expresses the action entirely in terms of the boundary contributions.
	We have to evaluate the boundary fields to obtain the expression for $S_\mathrm{b}$ in terms of $v$.
	This results in the following expression
	\begin{align}
		\frac{S_\mathrm{b}}{\mathcal{N}} &= \frac{\mu-2}{4\lambda\mathcal{N}}\left(v +\frac{\mu}{4\lambda\mathcal{N}}\left(v + \frac{\mu}{4\lambda\mathcal{N}}(v+c'e^{\beta_c})^{\frac{\mu}{2}-1}\right)^{\frac{\mu}{2}-1}\right)^{\frac{\mu}{2}} + \Gamma - v^2\beta_c \notag\\
		&-\frac{1}{2\lambda\mathcal{N}}\left(v + \frac{\mu}{4\lambda\mathcal{N}}\left(v + \frac{\mu}{4\lambda\mathcal{N}}(v+c)^{\frac{\mu}{2}-1}\right)^{\frac{\mu}{2}-1}\right)^{\frac{\mu}{2}}\notag\\
		&-\frac{\mu v}{4\lambda\mathcal{N}}\left(v +\frac{\mu}{4\lambda\mathcal{N}}\left(v + \frac{\mu}{4\lambda\mathcal{N}}(v+c'e^{\beta_c})^{\frac{\mu}{2}-1}\right)^{\frac{\mu}{2}-1}\right)^{\frac{\mu}{2}-1}\,.
		\label{eq:action-frozen} 
	\end{align}
	This expression depends on the coefficients $c,c'$ which are defined recursively through Eqns.~\eqref{eq:c-rel}-\eqref{eq:cprime-rel}, and can be constructed to higher orders in powers of $\frac{\mu}{4\lambda\mathcal{N}}$.
	The term $\Gamma - v^2\beta_c$ is proportional to $\mathrm{Vol}(\mathbb{R}^{+})$ in the limit $\beta_c\rightarrow \infty$. 
	Therefore it is divergent, which is a reflection of the fact that our choice of spatially frozen bosonic modes is too n\"aive, particularly in ignoring fluctuations of $\phi$ in the ``spatial'' index $I$ , which we expect to be relevant in LSYK with expected emergence of sparsity.
	It is straightforward to verify that in the limit $\mu = 2$, we obtain $S_\beta \sim -\frac{v}{\lambda}$ (aside from the divergent $\Gamma - v^2\beta_c$ term), as expected for Gaussian SYK.

	\subsection{Fluctuating saddles}
	
	Therefore ``spatial'' dependence on \(I\) need to be present and we introduce fluctuations to the bosonic modes:
	\begin{align}
		\phi_{I}(\tau) = \phi(\tau) + \xi_I.
	\end{align}
	We assume that the fluctuations $\xi_I$ are independent of $\tau$: keeping them \(\tau\)-dependent is equivalent to keeping all the \(\phi_I\) time dependent, contrary to our expectation that permutation symmetry, i.e. equivalence of sites, is restored in the saddle point solution of the averaged partition function.
	However, since all $I$ are statistically equivalent (on average) in Eqn.~\eqref{eq:lsyk_hamiltonian} (since the model is $0$-dimensional) there cannot exist a global fluctuation, otherwise this symmetry is violated.
	This is enforced by a $\delta$-function constraint:
	\begin{align}
		\sum_I \xi_I = 0.
	\end{align}
	The constraint on $\xi_I$ is incorporated in the partition function by an integral representation of the Dirac \(\delta\)-function
	\begin{align}
		\delta \left(\sum_I \xi_I\right) = \int \mathcal{D} K e^{-\int K\sum_I \xi_I}
	\end{align}
	where we have introduced a \emph{Lagrange multiplier} $K$.
	The full partition function is given by
	\begin{align}
		\langle{Z}(\beta)\rangle = \int &\mathcal{D}\psi \exp\left\{\left(-\int_{0}^{\beta}\mathrm{d}\tau \sum_{i = 1}^{N}\frac{1}{2}\psi_i \partial_\tau \psi_i\right)\right\}\notag\\
		&\times\lim_{\beta_c \rightarrow\infty}\int\mathcal{D}\phi \mathcal{D}K \mathcal{D}\xi \exp\left\{-\tilde{S}_{\mathrm{b},\beta_c}\left(\{\bar{\phi},\phi,\bar{\xi},\xi\}\right)\right\}
	\end{align}
	where the new action is $\tilde{S}_{\mathrm{b},\beta_c}(\{\bar{\phi},\phi,\bar{\xi},\xi\})$ is given by
	\begin{align}
		\tilde{S}_{\mathrm{b},\beta_c} & (\{\bar{\phi},\phi,\bar{\xi},\xi\}) = \sum_I \int_0^{\beta_c} \mathrm{d}\tau' \left((\bar{\phi} + \bar{\xi}_I)\partial_\tau(\phi + \xi_I) + \vert \phi + \xi_I - J^2 V(G_I) \vert^2\right)\notag\\ 
		&- \sum_I \frac{1}{2 \lambda\mathcal{N}}\int_0^{\beta_c} \mathrm{d}\tau' \left(\delta(\tau'-\beta_c)(\bar{\phi} + \bar{\xi}_I)^\frac{\mu}{2} + \delta(\tau')(\phi + \xi_I)^{\frac{\mu}{2}}\right)\notag\\
		&+ \sum_I\int_0^{\beta_c}\mathrm{d}\tau' (K \xi_I + \bar{K}\bar{\xi}_I)
	\end{align}
	The action consists of two parts: the index-independent dynamical field $\phi(\tau),\bar{\phi}(\tau)$ and the index-dependent frozen fluctuations $\xi_I,\bar{\xi}_I$.
	The dynamical field part is given by
	\begin{align}
		\tilde{S}^{(1)}_{\mathrm{b},\beta_c}(\phi,\bar{\phi}) = \sum_I\int_0^{\beta_c}\mathrm{d}\tau' \left(\bar{\phi}\dot{\phi} + \vert \phi - J^2 V(G_I) \vert^2 - \frac{1}{2\lambda\mathcal{N}}\delta(\tau' - \beta_c)\bar{\phi}^{\frac{\mu}{2}} - \frac{1}{2\lambda\mathcal{N}}\delta(\tau')\phi^{\frac{\mu}{2}} \right)\,.
	\end{align}
	Introducing the fluctuation $\phi_I \rightarrow \phi + \xi_I$ and performing a Taylor expansion.
	This expansion is again justified by the expected restoration of the permutation symmetry in the index \(I\) after averaging over disorder.
	Since the simple choice of \(\phi_I=\phi\) does not produce the correct saddle, we resort to the next best choice: $\xi_I \ll \phi$ and $\partial_\tau\xi_I = 0$.
	This ensures that \emph{dynamically} all the collective modes evolve identically.
	The smallness of $\xi_I$ also ensures that no collective mode significantly dominates over the others in the sum $\sum_I V^{\mu/2}(G_I)$.
	\begin{align}
		(\phi + \xi_I)^\frac{\mu}{2} \sim \phi^\frac{\mu}{2}\left(1 + \frac{\mu \xi_I}{2 \phi} + \binom{\mu/2}{2}\frac{\xi^2_I}{\phi^2} + \cdots\right)\,
	\end{align}
	gives the following frozen fluctuation action in addition to the fluctuation-free action (same as Eq.~\eqref{eq:sb-condensate1})
	\begin{align}
		\tilde{S}^{(2)}_{\mathrm{b},\beta_c}(\{\bar{\xi},\xi\}) &= \sum_{I}\int_0^{\beta_c} \mathrm{d}\tau' \left(\vert\xi_I\vert^2 - J^2 V(G_I)(\xi_I + \bar{\xi}_I)\right) + \sum_I\int_0^{\beta_c}\mathrm{d}\tau' (K \xi_I + \bar{K}\bar{\xi}_I) \notag\\
		&-C\sum_I \int_0^{\beta_c}\mathrm{d}\tau'\left(\delta(\tau')\phi^{\frac{\mu}{2}-2}(\tau')\xi^2_I + \delta(\tau'-\beta_c)\bar{\phi}^{\frac{\mu}{2}-2}(\tau')\bar{\xi}^2_I\right)\,.
		\label{eq:sfluc1}
	\end{align}
	Here $K,\bar{K}$ are Lagrange multipliers, and also integrated over.
	The constant is $C = \frac{1}{2\lambda\mathcal{N}}\binom{\mu/2}{2}$.
	To deal with the frozen fluctuation modes, we can choose the saddle of~\eqref{eq:sfluc1}.
	The saddle equations are given by
	\begin{align}
		\xi_I - 2 C\delta(\tau' - \beta_c)\bar{\phi}^{\frac{\mu}{2}-2}(\tau')\bar{\xi}_I + \bar{K} - J^2 V(G_I) &= 0\\
		\bar{\xi}_I - 2 C\delta(\tau')\phi^{\frac{\mu}{2}-2}(\tau')\xi_I + K - J^2 V(G_I) &= 0
	\end{align}
	The solution to this pair of linear equations is
	\begin{align}
		\xi_I &= J^2 V(G_I) - K + 2 C\delta(\tau'-\beta_c)\bar{\phi}^{\frac{\mu}{2}-2}(\tau')\left(J^2 V(G_I) - \bar{K}\right) \\
		\bar{\xi}_I &= J^2 V(G_I) - \bar{K} + 2 C\delta(\tau')\phi^{\frac{\mu}{2}-2}(\tau')\left(J^2 V(G_I) - K\right)
	\end{align}
	A key step in finding this relatively simple relation is to note that any term of the form $\delta(\tau')\delta(\tau'-\beta_c)$ vanishes.
	This solution is not fully time independent, instead has bumps at $\tau' = 0, \beta_c$.
	Since we eventually consider the limit $\beta_c \rightarrow \infty$, the fluctuations are frozen for all $0 < \tau' < \infty$.
	In the limit $\beta_c \rightarrow \infty$, this term~\eqref{eq:sfluc1} becomes (by solving the saddle equations for $\xi_I,\bar{\xi}_I$ and $K,\bar{K}$)
	\begin{align}
		\lim_{\beta_c \rightarrow \infty} \frac{\tilde{S}^{(2)}_{\mathrm{b},\beta_c}}{\mathcal{N}} = -\mathrm{Vol}(\mathbb{R}^{+}) J^{4}\left(\sum_I V(G_I)^4 - V[G]^2\right)\,.
		\label{eq:eq-fluc-action}
	\end{align}
	The frozen condensate action $\tilde{S}^{(1)}$ only depends on $\phi$, and thus is identical to the previous result~\eqref{eq:action-frozen}.
	The action~\eqref{eq:eq-fluc-action} exactly cancels out the divergent term, and retains the regular piece.
	The full bulk on-shell action $S_{\mathrm{b},\mathrm{on-shell}} = \tilde{S}^{(1)}(\phi,\bar{\phi}) + \tilde{S}^{(2)}(\xi,\bar{\xi})$ is then given by 
	\begin{align}
		S_{\mathrm{b},\text{on-shell}} &= \frac{\mu-2}{4\lambda}\left(v +\frac{\mu}{4\lambda\mathcal{N}}\left(v + \frac{\mu}{4\lambda\mathcal{N}}(v+c'e^{\beta_c})^{\frac{\mu}{2}-1}\right)^{\frac{\mu}{2}-1}\right)^{\frac{\mu}{2}} \notag\\
		&-\frac{1}{2\lambda}\left(v + \frac{\mu}{4\lambda\mathcal{N}}\left(v + \frac{\mu}{4\lambda\mathcal{N}}(v+c)^{\frac{\mu}{2}-1}\right)^{\frac{\mu}{2}-1}\right)^{\frac{\mu}{2}}\notag\\
		&-\frac{\mu v}{4\lambda}\left(v +\frac{\mu}{4\lambda\mathcal{N}}\left(v + \frac{\mu}{4\lambda\mathcal{N}}(v+c'e^{\beta_c})^{\frac{\mu}{2}-1}\right)^{\frac{\mu}{2}-1}\right)^{\frac{\mu}{2}-1}\,.
		\label{eq:final-action}
	\end{align}
	This concludes the derivation of the bulk on-shell action in terms of the Majorana functional $v = J^2 V[G]$.
	Using this action, we derive the Schwinger-Dyson equations in the limit \(N\rightarrow\infty\) in the next section.

	\subsubsection*{Comments on the choice of saddles}
	
	So far in the analysis, we have not discussed the features of the saddle manifold of the bosonic oscillator action.
	A detailed analysis is beyond the scope of this work. However, we can make some comments here.
	The motivation of choosing saddles that are invariant in the index $I$ is two-fold.
	First, invariance in the index \(I\) respects the statistical independence of the (L)SYK Hamiltonian (on average) with respect the indexing rule the Majorana fermions that we choose.
	This is due to the fact that the model is all-to-all/\(0\)-dimensional.
	Second, in the Gaussian SYK there is an emergent $O(N)$ symmetry in the large$-N$ limit.
	Since we would like to smoothly connect our solution for arbitrary $\mu$ to the Gaussian SYK at $\mu = 2$, the simplest choice is to use a saddle that leads to an $O(N)$ invariant action for all $\mu$. 
	
	The choice of static saddle, while correct in the sense that it is a valid saddle of Eqn.~\eqref{eq:saddle-phi-full}, leads to a divergent contribution.
	The contribution at the next order is what we look for (which we expect to be finite), and that is obtained by imposing the frozen fluctuations $\xi_I$ (which do not depend on $\tau$). 
	The saddle will be stable if the Hessian of small fluctuations near the saddle is positive definite.
	Let us now consider a fluctuation around the saddle $\phi_I = \phi + \eta_I$. 
	The action Eqn.~\eqref{eq:saddle-phi-full} evaluated to second order in $\eta_I$ (since the first order vanishes) is given by
	\begin{align}
		\delta^2 S_{\mathrm{b},\beta_c} = \sum_I \int_0^{\beta_c} \mathrm{d}\tau' \bar{\eta}_I\left(\partial_{\tau'} + 1\right)\eta_I-  \frac{\mu (\mu-2)}{8 \lambda \mathcal{N}}\left((\bar{\phi}(\beta_c))^{\frac{\mu}{2}-2}\bar{\eta}^2_I + (\phi(0))^{\frac{\mu}{2}-2}\eta^2_I\right)\,.
		\label{eq:saddle-phi-fluc}
	\end{align}
	In the large$-N$ limit, $\bar{\phi}(\beta_c)\approx \phi(0) \approx v + \frac{\mu}{4\lambda\mathcal{N}}v^{\frac{\mu}{2}-1} \sim v$.
	This can be inserted in Eqn.~\eqref{eq:saddle-phi-fluc} to give
	\begin{align}
		\delta^2 S_{\mathrm{b},\beta_c} = \sum_I \int_0^{\beta_c} \mathrm{d}\tau' \bar{\eta}_I\left(\partial_{\tau'} + 1\right)\eta_I
		- \frac{\mu (\mu-2) v^{\frac{\mu}{2}-2}}{8 \lambda \mathcal{N}}\left(\bar{\eta}^2_I + \eta^2_I\right)\,.\label{eq:saddle-phi-fluc-2}
	\end{align}
	
	Note that the first term is the operator $\partial_\tau + 1$, which has eigenvalues with real part $+1$.
	This is the term that solely contributes for $\mu = 2$.
	The second term is proportional to $-\frac{\mu (\mu-2) v^{\frac{\mu}{2}-2}}{8 \lambda \mathcal{N}}$, which is positive for $\mu < 2$.
	Therefore, the Hessian of $S_{\mathrm{b},\beta_c}$ is positive definite (trivially in the large$-N$ limit where the boundary term can be ignored) and the saddle is stable.
	This analysis serves as a heuristic justification for the choice of the saddle of the oscillator action. 
	
	%In a related direction, it is useful to consider the difference between an $O(N)-$invariant potential in the action and the actual potential. This is evaluated as
	%\begin{align}
	%    \sum_I V(G_I)^\alpha - \left(\sum_I V(G_I)\right)^\alpha \sim \mathcal{N}V^\alpha - \mathcal{N}^\alpha V^\alpha \sim \mathcal{O}(\mathcal{N}-\mathcal{N}^\alpha)V^\alpha\,
	%\end{align}
	%where the (typical) magnitude of the scalar $V(G_I) \sim V$. Since at the saddle $\sum_I V(G_I) \sim \mathcal{O}(\mathcal{N})$, each individual $V(G_I)$ is $\mathcal{O}(1)$. Here $\alpha = \mu/2 < 1$ is the exponent of interest. For $\alpha$ close to $1$, the 

	\section{The Schwinger-Dyson Equations}
	\label{sec:sd-eqn}
	
	The next step in the analysis is to derive the Schwinger-Dyson equations for the Green's function and self-energy.
	Let us recall from Gaussian SYK~\cite{maldacena2016remarks} that the procedure involves the partition function which can be written as
	\begin{align}
		Z(\beta) = \int \mathcal{D}\psi \exp\left\{-\frac{1}{2}\sum_i\int_0^\beta \mathrm{d}\tau\,\psi_i\partial_\tau \psi_i + R\left(\sum_I V(G_I)\right)\right\}
	\end{align}
	where $R$ is the appropriate functional form of the bulk action.
	Then we introduce the following resolution of identity
	\begin{align}
		1 = \int \mathrm{d}G \mathrm{d}\Sigma \exp\left\{-\frac{N}{2}\int \mathrm{d}\tau \mathrm{d}\tau' \left(G  - \frac{1}{N}\sum_i\psi_i(\tau)\psi_i(\tau')\right)\right\}\,.
	\end{align}
	This integral forces $G (\tau,\tau') = \frac{1}{N}\sum_{i}\psi_i (\tau)\psi_i (\tau')$, allowing us to recast the function $R$ entirely in terms of $G$ and therefore integrate over $\psi_i$ among the terms outside $R$.
	We can split the action as 
	\begin{align}
		Z(\beta) = &\int \mathrm{d}G\mathrm{d}\Sigma \exp\left\{R -\frac{N}{2}\int \mathrm{d}\tau \mathrm{d}\tau' \Sigma G \right\}\notag\\
		&\times\int\mathcal{D}\psi \exp\left\{-\frac{1}{2}\sum_i \int \mathrm{d}\tau\mathrm{d}\tau' \psi_i(\tau) \delta(\tau - \tau')\partial_\tau \psi(\tau) - \psi_i(\tau)\psi(\tau')\right\} \notag\\
		&= \int \mathrm{d}G\mathrm{d}\Sigma \exp\left\{R - \frac{N}{2}\int \mathrm{d}\tau \mathrm{d}\tau' \Sigma G\right\}[\det(\partial_\tau - \Sigma)]^{N/2},
	\end{align}
	and write the effective action as
	\begin{align}
		I_{\text{eff}} = -\frac{1}{2}\log[\det(\partial_\tau - \Sigma)] - \frac{R}{N} + \frac{1}{2}\int \mathrm{d}\tau\mathrm{d}\tau' \Sigma G\,,
		\label{eq:eff-action-largeN}
	\end{align}
	producing the following Schwinger-Dyson equations
	\begin{align}
		G = \frac{1}{\partial_\tau - \Sigma} \;\;,\;\; \Sigma = \frac{2}{N}\frac{\delta R}{\delta G}.
	\end{align}
	For the L\'evy SYK, the functional $R$ is written as 
	\begin{align}
		R = -S_{\mathrm{b},\mathrm{on-shell}} \sim \frac{v^{\mu/2}}{\lambda}\,,
	\end{align}
	where we have used the limit $\mathcal{N} \rightarrow \infty$ in Eqn.~\eqref{eq:final-action} to only keep the leading order terms.
	Recalling that $v = J^2 \mathcal{N}^{-1} \sum_I V(G_I) \sim J^2 \int_0^{\beta}\mathrm{d}\tau\mathrm{d}\tau' G^{q}(\tau,\tau')$, we can write the functional derivative as 
	\begin{align}
		\frac{\delta R}{\delta G} = \frac{\mu}{2}J^{\mu}v^{\frac{\mu}{2}-1}G^{q-1}\,.
	\end{align}
	Therefore, the Schwinger-Dyson equations can be expressed as
	\begin{align}
		G(\tau,\tau') &= \frac{1}{\delta(\tau-\tau')\partial_{\tau} -\Sigma(\tau,\tau')}\,,
		\label{eq:sd-eqn-1}\\
		\Sigma(\tau,\tau') &= \frac{\mu}{2}J^{2}\left(J^{2}\int_0^{\beta}\mathrm{d}\tau\mathrm{d}\tau' G^{q}(\tau,\tau')\right)^{\frac{\mu}{2}-1}G^{q-1}(\tau,\tau')\,.
		\label{eq:sd-eqn-2}
	\end{align}
	Finally, using translation invariance: $G(\tau,\tau') \rightarrow G(\tau - \tau')$, and renaming $\tau - \tau'$ as $\tau$, we obtain the equations
	\begin{align}
		G(\tau) &= \frac{1}{\partial_{\tau} -\Sigma(\tau)}
		\label{eq:sd-eqn-3}\\
		\Sigma(\tau) &= \frac{\mu}{2}J^{2}\left(\beta J^{2}\int_0^{\beta}\mathrm{d}\tau G^{q}(\tau)\right)^{\frac{\mu}{2}-1}G^{q-1}(\tau)\,.
		\label{eq:sd-eqn-4}
	\end{align}
	The main difference between the usual Gaussian SYK and the L\'evy SYK is the presence of the overall factor in the expression for self-energy.
	This factor is non-trivial: it is independent of $\tau$, but it does depend on $\beta$.
	For any $\mu < 2$, it becomes divergent for $\beta\rightarrow0$.
	Let us denote this factor by $A_\beta$.
	\begin{align}
		A_\beta \equiv J^2\int_0^\beta\mathrm{d}\tau \mathrm{d}\tau' G^q(\tau,\tau')\,.
	\end{align}
	This expression simplifies for the solutions $G_*, \Sigma_*$ of the Schwinger-Dyson equations.
	\begin{align}
		A^{\frac{\mu}{2}}_\beta &= \frac{2\beta}{\mu} \lim_{\tau \rightarrow 0_+}\partial_\tau G_{*}(\tau)\,.
		\label{eq:abeta-defn}
	\end{align}
	This gives an expression for the integral factor in terms of the solution of the Schwinger-Dyson equation.
	Using the Galitski-Migdal-Koltun sum rule, this factor can be related to the average energy:
	\begin{align}
		A^{\frac{\mu}{2}}_\beta = \frac{2\beta q}{N \mu}\langle H \rangle\,.
		\label{eq:a-beta-H}
	\end{align}
	It is straightforward to see that these results are consistent in the appropriate limits.
	In particular, we find that for $\mu = 2$, the Eqns.~(\ref{eq:sd-eqn-1}-\ref{eq:sd-eqn-2}) reduce exactly to the Gaussian SYK result~\cite{maldacena2016remarks}.
	For $\mu = 0$, the theory becomes free, as is also expected from the action itself, which becomes a constant. 
	
	%\section{Solution of the Schwinger-Dyson Equations}
	
	\begin{figure}[t]
		\begin{subfigure}[t]{0.49\textwidth}
			\centering
			\includegraphics[height=0.62\linewidth,width=\linewidth]{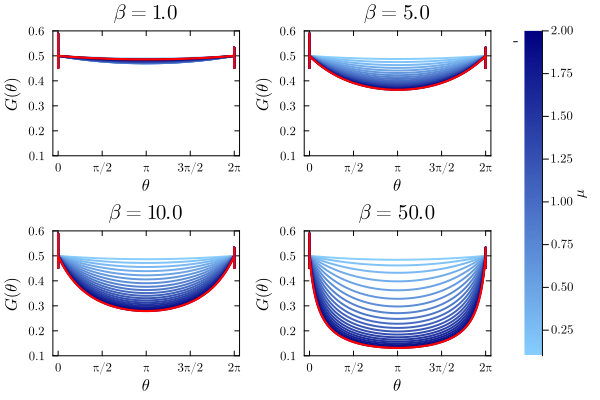}
			\caption{}
		\end{subfigure}%
		\hfill
		\begin{subfigure}[t]{0.49\textwidth}
			\centering
			\includegraphics[height=0.62\linewidth,width=\linewidth]{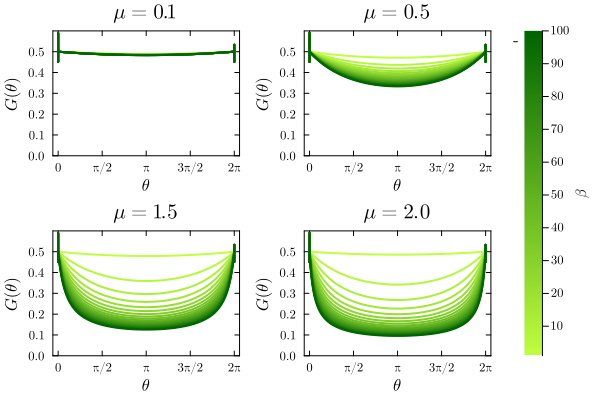}
			\caption{}
		\end{subfigure}
		\caption{
			Numerical solution of the Schwinger-Dyson equations~(\ref{eq:sd-eqn-1}-\ref{eq:sd-eqn-2}): the Green function \(G(\theta=\frac{2\pi \tau}{\beta})\) as a function of \(\mu\) (a) and inverse temperature \(\beta\) (b), \(J=1\).
			Left, (a): \(G(\theta)\) tends towards the free theory solution (independent of \(\beta\)) for any \(\beta\) as \(\mu\) decreases. 
			The red curve is the Gaussian SYK result.
			Right, (b): \(G(\theta)\) changes strongly with \(\beta\) for \(\mu\) close to \(2\).
			Away from \(\mu=2\) the \(\beta\)-dependence of \(G\) is weaker. 
		}
		\label{fig:sd-soln-1}
	\end{figure}
	
	The Schwinger-Dyson (SD) equations Eqns.~(\ref{eq:sd-eqn-3}-\ref{eq:sd-eqn-4}) can be solved numerically.
	This is achieved by transforming Eqn.~\eqref{eq:sd-eqn-3} to the frequency domain through discrete Fourier transform at Matsubara frequency $\omega_n = \frac{2\pi}{\beta}(n + \frac{1}{2})$.
	An inverse transform gives us $G(\tau)$ through which we can evaluate $\Sigma(\tau)$ using Eqn.~\eqref{eq:sd-eqn-4}.
	This iteration is performed recursively until the solution converges.
	We used an initial choice $G_{\mathrm{in}}(\tau) = \frac{1}{2}\mathrm{sgn}(\tau)$, corresponding to the free theory, and computed $G(\tau),\Sigma(\tau)$ via the above recursive algorithm.
	The numerical results for the solution of the Schwinger-Dyson equations are presented in Fig.~\ref{fig:sd-soln-1}.
	The results in Fig.~\ref{fig:sd-soln-1} indicate the the solution approaches a free theory as \(\mu\) decreases.
	Similarly, the inverse temperature \(\beta\), which controls the effective interaction strength of the theory, also takes the model closer to a free theory as it decreases.

	\subsection{Large$-q$ solution}
	
	Inspired by the Gaussian SYK, where analytical results are derived in the large-\(q\) limit, we present the large-\(q\) solution the SD equations Eqn.~(\ref{eq:sd-eqn-3}-\ref{eq:sd-eqn-4}).
	We  start with the ansatz~\cite{maldacena2016remarks}, motivated by the Gaussian SYK:
	\begin{align}
		G(\tau) = \frac{1}{2}\text{sgn}(\tau)e^{g(\tau)/(q-1)}
	\end{align}
	Inserting this ansatz into the equation~\eqref{eq:sd-eqn-3}, the following Liouville equation is obtained
	\begin{align}
		\partial^2_\tau \left(\frac{1}{2 q}\mathrm{sgn}(\tau)g(\tau)\right) = \frac{\mu J^{2}}{2^{q}}\mathrm{sgn}(\tau)e^{g(\tau)}\left(\frac{\beta J^{2}}{2^q}\int_0^{\beta} \mathrm{d}\tau e^{g(\tau)q/(q-1)}\right)^{\frac{\mu}{2}-1}
	\end{align}
	The integral on the RHS is \(\tau\)-independent, and  can be expressed via the \(\beta\)-dependent constant \(A_\beta\)~\eqref{eq:abeta-defn}.
	Therefore the overall equations are of the Liouville type, and can be written as
	\begin{align}
		\partial^2_\tau \left(\frac{1}{2 q}\mathrm{sgn}(\tau)g(\tau)\right) = \frac{\mu J^{2}A^{\frac{\mu}{2}-1}_\beta}{2^{q}}\mathrm{sgn}(\tau)e^{g(\tau)}\,,
		\label{eq:lveqn-1v}
	\end{align}
	and are solved by the usual general form of the solution
	\begin{align}
		e^{g(\tau)} = \frac{c^2}{K^2 [\sin(c(\vert\tau\vert + \tau_0))]^2}\,.
		\label{eq:lveqn-v1}
	\end{align}
	
	Using the boundary conditions on the thermal circle $G_*(0) = G_*(\beta)$ and introducing the parameter $\nu$ defined via $c = \pi\nu/\beta$ and $\tau_0 = \beta(1-\nu)/2\nu$, we obtain the following constraint equations for $\nu$ and $K$ in terms of $\beta$ and $J$
	\begin{align}
		\mu q  \left(\frac{\beta^2  J^{2}}{2^{q}}\right)^{\frac{\mu}{2}} &= \left(\frac{\pi\nu}{\cos(\pi\nu/2)}\right)^2\left(\frac{\pi\nu}{\sin(\pi\nu)}\right)^{\frac{\mu}{2}-1}\,,\notag\\
		K^{\mu} &= \frac{J^{\mu}(\pi\nu)^{\frac{\mu}{2}-1}}{2^{\frac{q\mu}{2}}}(2\tan(\pi\nu/2))^{\frac{\mu}{2}-1}\,.
		\label{eq:eff-int-mu-s}
	\end{align}
	For $\mu = 2$, it reduces to the condition $\beta K = \frac{\pi\nu}{\cos(\pi\nu/2)}$, fixing $K_{\mu = 2} = \frac{\sqrt{q} J}{2^{\frac{q-1}{2}}}$.
	In that limit, $\nu = 1$ corresponds to $\beta K \rightarrow \infty$ and $\nu = 0$ corresponds to $\beta K = 0$.
	For  \(\mu < 2\), the LHS corresponds to the effective interaction strength raised to the power $\mu$, and the RHS corresponds to the generalization of $\pi\nu/\cos(\pi\nu/2)$ raised to power $\mu$.
	It is interesting to note that for any $0 < \mu \leq 2$, the limit $\nu \rightarrow 0$ (on the LHS) goes as $2^{1 -\frac{\mu}{2}}\pi^2 \nu^2$ and therefore converges to $0$.
	Hence $\nu = 0$ corresponds to the free theory for any $\mu \in (0,2]$.
	Similarly as $\nu \rightarrow 1$, the LHS behaves as $\pi^{\frac{\mu}{2}-1}(1-\nu)^{-2}(4\nu + \mu(1-\nu^2))$.
	This is divergent and corresponds to a theory with divergent effective interaction strength as $\nu \rightarrow 1$ for $\mu \in (0,2]$.
	The special case $\mu = 0$ forces $\nu = 0$ in Eqn.~\eqref{eq:eff-int-mu-s}, leading to $g(\tau) = 0$ and thus $G(\tau) = \frac{1}{2}\mathrm{sgn}(\tau)$ i.e. the free theory.
	
	\begin{figure}%[t]
		\centering
		\includegraphics[width=0.75\columnwidth]{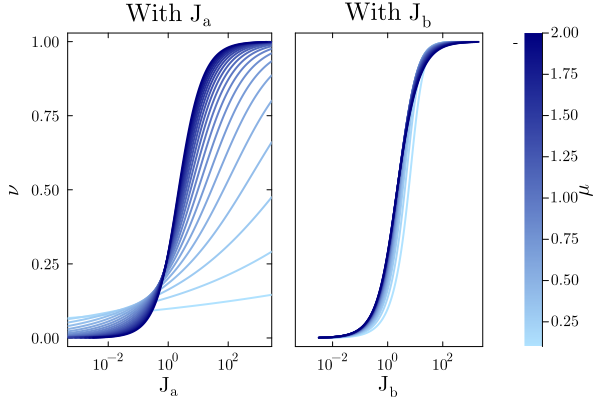}
		\caption{
			Solution \(\nu\)~\eqref{eq:eff-int-mu-s} as a function of \(\mathrm{J}_a\) and \(\mathrm{J}_b\).
			The non-trivial dependence of \(\nu\) on \(\mu\) present on the plot vs. \(\mathrm{J}_a\) weakens, e.g. different curves almost collapse, on the plot vs. \(\mathrm{J}_a\).
		}
		\label{fig:eff-int-mu-soln}
	\end{figure}
	
	In order to draw a close analogy to the Gaussian SYK, we define the effective interaction strength $\mathcal{J}_\mu = q^{1/\mu}J/2^{\frac{q-1}{2}}$ allowing us to write the Eqn.~\eqref{eq:eff-int-mu-s} as
	\begin{align}
		\sqrt{\frac{\mu}{2^{\frac{\mu}{2}}}} \left(\beta\mathcal{J}_{\mu}\right)^{\frac{\mu}{2}}=  \left(\frac{\pi\nu}{\cos(\pi\nu/2)}\right)\left(\frac{\pi\nu}{\sin(\pi\nu)}\right)^{\frac{\mu-2}{4}}\,.
		\label{eq:eff-int-mu-J}
	\end{align}
	This expression highlights the similarities and differences with the Gaussian SYK.
	We present the solution of Eqn.~\eqref{eq:eff-int-mu-J} in Fig.~\ref{fig:eff-int-mu-soln} by as a function of $\mathrm{J}_a = \beta\mathcal{J}_\mu$ and $\mathrm{J}_{b} = (\beta\mathcal{J}_\mu)^{\mu/2} = \mathrm{J}^{\mu/2}_a$ for different $\mu$. Using this, we find the following expression for $e^{g(\tau)}$
	\begin{align}
		e^{g(\tau)} = \left(\frac{\cos(\frac{\pi\nu}{2})}{\cos \left(\pi \nu \left(\frac{1}{2} - \frac{\vert\tau\vert}{\beta}\right)\right)}\right)^2\,,
		\label{eq:green-f-euctime}
	\end{align}
	where the parameter \(\nu\) is given by the solution of Eqn.~\eqref{eq:eff-int-mu-J}.
	
\begin{figure}[t]
		\centering
		\includegraphics[width=0.85\columnwidth]{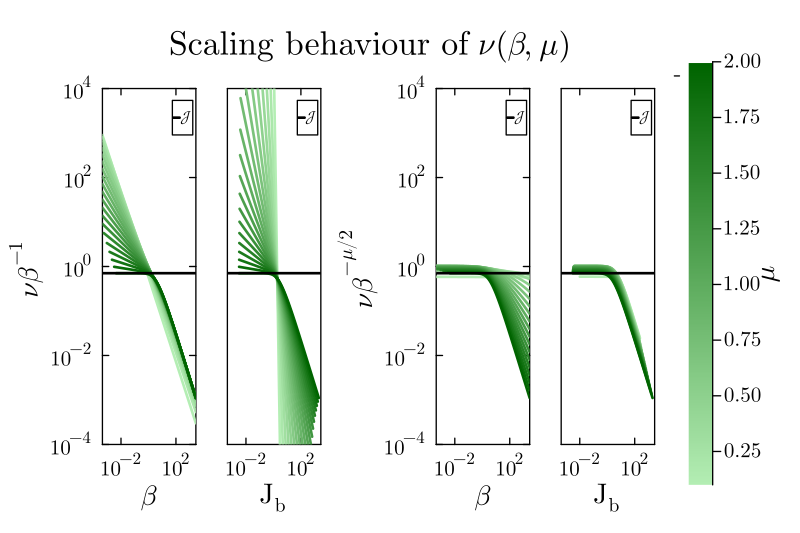}
		\caption{
			The scaling of $\nu$~\eqref{eq:eff-int-mu-s} with \(\beta\) (left) and \(\beta^{\mu/2}\) (right) vs \(\beta\) and \(\mathrm{J}_b\sim \beta^{\mu/2}\).
			A nearly perfect collapse is observed for the plot of \(\nu/\beta^{\nu/2}\) vs \(\mathrm{J}_b\).
			The solid black line $\mathcal{J}$ is the limit $\lim_{\beta\rightarrow0}\nu/\beta$ for the Gaussian SYK.
		}
		\label{fig:nu0-vs-beta-and-betamu}
	\end{figure}

	For the Greens' function~\eqref{eq:green-f-euctime}, the analytical continuation to real time is achieved by replacing $\vert\tau\vert$ by $\frac{\beta}{2} + i t$.
	This gives us
	\begin{align}
		e^{g(t)} = \frac{\cos^2(\frac{\pi\nu}{2})}{\cosh^2 \left(\frac{\pi \nu t}{\beta}\right)}\,.
	\end{align}
	In the large-\(q\) limit, we can approximate $e^{g(t)/(q-1)}$ by $1 + \frac{g(t)}{q}+\cdots$.
	This allows us to express the large-$q$ Green's function as
	\begin{align}
		G_R(t) = \frac{1}{2}\theta(t) + \frac{1}{q}\ln\left(\mathrm{sech}\left(\frac{\pi \nu t}{\beta}\right)\right) + \cdots\,.
		\label{eq:greenf-largeq-exp}
	\end{align}
	This result differs from the corresponding Gaussian SYK Green's function due to implicit dependence of \(\nu\) on \(\mu\).
	Therefore the object of interest is the scaling of $\nu$ with $\mu$ for a fixed $\beta$.
	To analyze this, we first consider the related problem of the scaling of $\nu$ with $\beta$ for fixed $\mu$ shown in Fig.~\ref{fig:nu0-vs-beta-and-betamu}.
	In particular this tells us about the expected behaviour of the theory in the infinite temperature limit.
	
	These results indicate that $\nu/\beta$ diverges as \(\beta\to 0\) for $\mu < 2$ (as a function of both $\beta$ and $\mathrm{J_b}$) while $\nu/\beta^{\mu/2}$ converges to a constant in the same limit as a function of both $\beta$ and $\mathrm{J_b}$.
	To see this behaviour from the constraint equation~\eqref{eq:eff-int-mu-J}, we consider the derivative $\partial\nu/\partial\beta$.
	%This gives us
	%\begin{align}
	%    \beta\frac{\partial\nu}{\partial\beta} = \frac{2 \nu \mu }{(2-\mu )\pi   \nu  \cot (\pi  \nu )+\mu +2 \pi  \nu  \tan \left(\frac{\pi  \nu }{2}\right)+2} \equiv P_{\mu}(\nu)\,.
	%    \label{eq:nu-der}
	%\end{align}
	It is straightforward to see that $\partial\nu/\partial\beta$ is a positive bounded quantity for all $\nu \in[0,1]$ and all $0 < \mu < 2$. 
	Thus $\nu$ increases monotonically with $\beta$in agreement with numerical results in Fig.~\ref{fig:eff-int-mu-soln}.
	Additionally, the saturation of \(\nu/\beta^{\mu/2}\) for small \(\beta\) can also be seen by evaluating $\partial (\nu \beta^{-\eta})/\partial \beta$ in the limit $\beta \rightarrow 0$ for some real $\eta$.
	In the small $\mu$ limit, we find that 
	\begin{align}
		\beta\frac{\partial (\nu \beta^{-\eta})}{\partial\beta} \xrightarrow[\beta\rightarrow0]{} \left(\frac{\mu}{2}-\eta\right)\nu^{1-\frac{2 \eta}{\mu}} + \cdots\,.
		\label{eq:beta-scaling}
	\end{align}
	This converges for $\eta \geq \frac{\mu}{2}$ (since $\nu\rightarrow0$ as $\beta\rightarrow0$), and vanishes particularly fast for $\eta = \mu/2$ due to the leading order term being $0$.
	This explains the scaling behaviour observed in Fig.~\ref{fig:nu0-vs-beta-and-betamu}.

	\subsection{The Infrared Regime}
	
\begin{figure}%[ht] 
		\centering
		\includegraphics[width=0.95\linewidth]{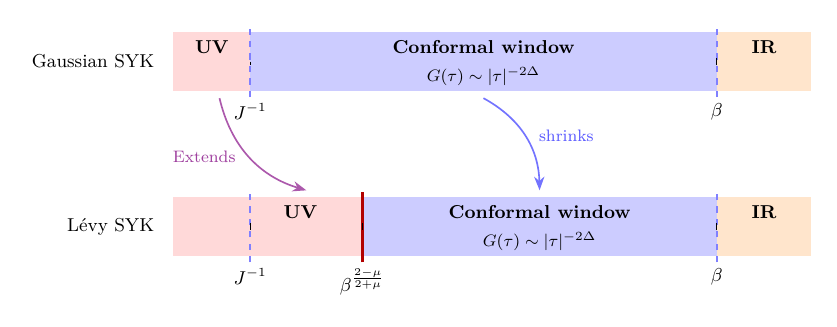}
		\caption{
			Schematic representation of the shrinking of the conformal window in the L\'evy SYK model due to L\'evy disorder.
		}
		\label{fig:cnflm-window}
	\end{figure}

	The SYK model is known to feature emergent symmetries in the infrared regime, which is achieved at strong interaction strength or large time-scales~\cite{maldacena2016remarks}.
	In this section, we investigate the properties of LSYK in the same regime and discuss the related symmetries and spontaneous symmetry breaking.
	Let us recall that the large$-N$ effective action for LSYK is given by
	\begin{align}
		\frac{S_{\mathrm{b},\text{on-shell}}}{N } = -\frac{1}{2}\log[\det(\partial_\tau - \Sigma)] - \frac{1}{2 q}\left(J^2 \int_0^\beta \mathrm{d}\tau\mathrm{d}\tau' G^q (\tau,\tau')\right)^{\frac{\mu}{2}} + \frac{1}{2}\int \mathrm{d}\tau\mathrm{d}\tau' \Sigma G\,.
		\label{eq:eff-action-2}
	\end{align}
	From this, we obtain the Schwinger-Dyson equations~(\ref{eq:sd-eqn-1}-\ref{eq:sd-eqn-2}).
	In the limit $\tau \gg J^{-\mu}$ (strong coupling limit), the derivative term $\partial_\tau$ can be ignored and  the equations become invariant under reparameterization $\tau \rightarrow f(\tau)$ with a conformal dimension $\Delta = 1/q$. 
	The L\'evy contribution $A^{\frac{\mu}{2}-1}_\beta$ is also manifestly invariant under the same reparameterization.
	This result is identical to that of Gaussian SYK. 
	The conformal ans\"atze~\cite{maldacena2016remarks} for both finite-temperature, first expression, and infinite-temperature (\(\beta\to 0\)) , second expression read:
	\begin{align}
		G_{c,\beta}(\tau) = b\left(\frac{\pi}{\beta\sin\left(\frac{\pi\tau}{\beta}\right)}\right)^{2\Delta}\mathrm{sgn}(\tau)\;\;\;,\;\;\; G_{c}(\tau) = \frac{b}{\vert\tau\vert^{2\Delta}}\mathrm{sgn}(\tau)\,.
		\label{eq:c-anz}
	\end{align}
	This ansatz spontaneously breaks the reparameterisation symmetry down to $SL(2,R)$.
	The finite and infinite temperatures solutions can be connected by the reparameterisation $\tau \rightarrow f(\tau) = \tan\left(\frac{\pi \tau}{\beta}\right)$.
	The conformal self-energy similarly has the form
	\begin{align}
		\Sigma_{c,\beta}(\tau) = \frac{\mu}{2}\mathrm{sgn}(\tau)J^{2} A_\beta^{\frac{\mu}{2}-1}\frac{b^{q-1}\pi^{2\Delta(q-1)}}{\left(\beta\sin\left(\frac{\pi\tau}{\beta}\right)\right)^{2\Delta(q-1)}}\;\;,\;\; \Sigma_c(\tau) = \frac{\mu}{2}J^2 A^{\frac{\mu}{2}-1}_0 \frac{b^{q-1}}{\vert\tau\vert^{2\Delta(q-1)}}\mathrm{sgn}(\tau)\,.
		\label{eq:sigma-c}
	\end{align}
	To fix the coefficient $b$, we evaluate the relation $\int \mathrm{d}\tau' \Sigma(\tau_1,\tau')G(\tau',\tau_2) = -\delta(\tau_1 - \tau_2)$, which follows from Eqn.~\eqref{eq:sd-eqn-1} by ignoring the $\partial_\tau$ term.
	In Fourier space this becomes simply $\tilde{G}(\omega)\tilde{\Sigma}(\omega) = -1$.
	We compute $\tilde{G}(\omega), \tilde{\Sigma}(\omega)$ for the Matsubara frequencies on the thermal circle, \(\omega_n = \frac{2\pi}{\beta}(n + 1/2)\):
	\begin{eqnarray}
		\tilde{G}(\omega_n) &=& b\int_{0}^{\beta}\mathrm{d}\tau\,\left(\frac{\pi}{\beta \sin\left(\frac{\pi\tau}{\beta}\right)}\right)^{2\Delta} e^{\iota\omega_n \tau} \notag\\
		&\underset{\omega_n >0}{=}& b\frac{4^{\Delta } e^{i \pi  \Delta } \pi ^{2 \Delta } \beta^{1- 2 \Delta} \left(-1+e^{i \left(2 \pi  \Delta +\beta  \omega _n\right)}\right) \Gamma \left(\Delta +\frac{\beta  \omega _n}{2 \pi }\right)}{\left(-1+e^{4 i \pi  \Delta }\right) \Gamma (2 \Delta ) \Gamma \left(-\Delta +\frac{\beta  \omega _n}{2 \pi }+1\right)}\notag\\
		&=& i\pi b \csc(\pi\Delta)\left(\frac{2\pi}{\beta}\right)^{2\Delta-1}\frac{\Gamma \left(\Delta +\frac{\beta  \omega _n}{2 \pi }\right)}{\Gamma(2\Delta)\Gamma \left(1 -\Delta +\frac{\beta  \omega _n}{2 \pi }\right)}\,.
		\label{eq:g-tilde}
	\end{eqnarray}
	Using Eqn.~\eqref{eq:g-tilde}, we can also determine $\tilde{\Sigma}(\omega)$ by replacing $\Delta$ by $\Delta(q-1)$ and the appropriate additional factors~\footnote{Note that the integral of $\Sigma$ is divergent and so the result has to be obtained via an analytic continuation}.
	Using the conformal constraint $\tilde{G}(\omega)\tilde{\Sigma}(\omega) = -1$, we obtain the following relation for $b$:
	\begin{align}
		b^{q} = \frac{2}{\pi \mu J^2}\left(\frac{1}{2} - \frac{1}{q}\right)\tan\left(\frac{\pi}{q}\right)A^{1 - \frac{\mu}{2}}_\beta\,.
		\label{eq:b^q-eqn}
	\end{align}
	The primary difference from the usual Gaussian SYK case is the factor of $A^{1-\frac{\mu}{2}}_\beta$, which vanishes for $\mu = 2$.
	At first glance, this makes the coefficient $b$ dependent on $\beta$: However, we can use the conformal solution to evaluate $A_\beta$ by analytic continuation, producing the following result:
	\begin{align}
		A_\beta = J^2 \int_0^\beta \mathrm{d}\tau \mathrm{d}\tau' G_c^{q}(\tau,\tau') = J^2 b^q \pi \int_0^\pi \frac{\mathrm{d}x}{\sin^2(x)}\,.
	\end{align}
	Naturally, this expression is divergent owing to the integration from $0$ to $\beta$: the conformal solution is invalid at the endpoints.
	Denoting the divergent integral by $\kappa$, we can write $A_\beta = J^2 b^q \kappa$, and rewrite~\eqref{eq:b^q-eqn} in a \(\beta\)-independent form:
	\begin{align}
		b^{\frac{q\mu}{2}} = \frac{2}{\mu \pi J^{\mu}}\left(\frac{1}{2} - \frac{1}{q}\right)\tan\left(\frac{\pi}{q}\right)\kappa^{1 - \frac{\mu}{2}}\,,
	\end{align}
	This result reduces to the Gaussian SYK relation for $\mu = 2$.
	We note that the divergence $\kappa$ can be controlled under appropriate regularization: We return to the Schwinger-Dyson equations Eqn.~(\ref{eq:sd-eqn-1}-\ref{eq:sd-eqn-2}), where we focus on the coefficient of $G^{q-1}$ in the equation for $\Sigma$. 
	The conformal solution is valid in the regime where $\vert\Sigma(i\omega_n)\vert \gg \vert\omega_n\vert$, which gives us the following lower bound on the timescale $\vert\tau - \tau'\vert$ for the IR regime
	\begin{align}
		\vert \tau - \tau' \vert \gg \frac{1}{J \sqrt{\frac{\mu}{2}}A^{\frac{\mu-2}{4}}_\beta} \equiv \xi\,.
	\end{align}
	Using this limit, we can write a self-consistency equation for $A_\beta$
	\begin{align}
		A_\beta = b^q J^2 \beta \int_0^\beta \mathrm{d}\tau \left(\frac{\pi}{\beta \sin\left(\frac{\pi \tau}{\beta}\right)}\right)^2 \xrightarrow{\xi \ll \vert\tau\vert \sim \epsilon \ll 1} b^q J^2 \beta \int_{\xi}^{\beta} \frac{\mathrm{d}\epsilon}{\epsilon^2} \sim \frac{b^q J^2 \beta}{\xi}\,, 
	\end{align}
	where we have kept the UV cutoff term $\xi$.
	Inserting the expression for $b^q$ from Eqn.~\eqref{eq:b^q-eqn} (and writing $b^q$ as $b^q = \frac{c_q}{J^2}A^{1-\frac{\mu}{2}}_\beta$, introducing the factor $c_q$ for simplicity) we have the following temperature dependent scaling of $A_\beta$ and the corresponding $\beta-$ dependence of the UV cutoff valid for large $\beta$ and $\mu > 0$:
	\begin{align}
		A_\beta \sim \beta^{\frac{4}{\mu + 2}}\left(c_q J \sqrt{\frac{\mu}{2}}\right)^{\frac{4}{\mu + 2}} \;\; \implies \;\; \xi \sim \beta^{\frac{2-\mu}{2+\mu}}\,.
	\end{align}
	The case of $\mu = 0$ is trivial. 
	This cutoff implies a shrinking conformal window (for $\mu < 2$) for L\'evy SYK, as presented schematically in Fig.~\ref{fig:cnflm-window}.
	With this $\beta-$dependence of the UV cutoff, we can consider fluctuations around the conformal saddle and construct the effective theory of reparmeterisations. 
	
	The theory of reparametrisation is one of the remarkable features of the Sachdev-Ye-Kitaev model~\cite{maldacena2016remarks} and is the primary reason for the near-conformal limit of the SYK model being dual to a Jackiw–Teitelboim gravity in $1+1$ dimensions.
	The core idea is to study the effect of reparameterization $\tau \rightarrow f(\tau)$ on the full effective action $S_{\mathrm{eff}} = N I_{\mathrm{eff}}$ close to the conformal saddles $G_c, \Sigma_c$.
	
	The last two terms in Eqn.~\eqref{eq:eff-action-2} are reparameterization invariant.
	The only non-trivial contribution coming from reparameterization is from the $\log\det(\partial_\tau - \Sigma) = \Tr \log (\partial_\tau - \Sigma)$ term.
	Let us denote the reparameterised $G,\Sigma$ fields as
	\begin{align}
		G_{f} = [f'(\tau_1)f'(\tau_2)]^{\Delta}G_{c}(f(\tau_1),f(\tau_2))\;\;\;,\;\;\;\Sigma_{f} = [f'(\tau_1)f'(\tau_2)]^{\Delta(q-1)}\Sigma_{c}(f(\tau_1),f(\tau_2))\,.
	\end{align}
	Here $G_c, \Sigma_c$ are the finite-temperature conformal solutions~(\ref{eq:c-anz}-\ref{eq:sigma-c}) with $SL(2,R)$ invariance.
	The action $S_\mathrm{eff}$ is evaluated at $\Sigma_f, G_f$. 
	To extract the reparameterization-dependent part, consider the kinetic part of the action.
	\begin{align}
		S_{f} = -\frac{N}{2}\Tr\log(\partial_\tau - \Sigma_f) = -\frac{N}{2}\Tr \log(-\Sigma_f) -\frac{N}{2}\Tr\log(1 - \Sigma^{-1}_f\partial_\tau)\,.
	\end{align}
    The first term is also reparameterization invariant and the non-trivial contribution comes from the second term.
	\begin{align}
		\Tr \log(1 - \Sigma^{-1}_f\partial_\tau) 
		= -\underset{=0}{\underbrace{\Tr(\Sigma^{-1}_f \partial_\tau)}} + \frac{1}{2}\Tr(\Sigma^{-1}_f \partial_\tau \Sigma^{-1}_f \partial_\tau) + \cdots\,.
		\label{eq:tr-term}
	\end{align}
	We have to evaluate the reparameterization $\tau \rightarrow f(\tau)$ of $\Sigma^{-1}$, for which we use the condition $\int\mathrm{d}\tau_3 \Sigma^{-1}(\tau_1,\tau_3) \Sigma(\tau_3,\tau_2) \sim \delta(\tau_1 - \tau_2)$. 
    Using the conformal solution, we find that $\Sigma^{-1}_c \propto \mathrm{sgn}(\tau)\vert\tau\vert^{-2\Delta}$ where $\Delta = 1/q$.
    Thus, we now have to evaluate the derivative $\partial_{\tau_1}\Sigma_f^{-1}(\tau_1,\tau_2)$.
    To capture the UV effects, we define~\footnote{The translation invariance $G(\tau_1,\tau_2) = G(\tau_1 - \tau_2)$ is broken by the reparameterization.} $\tau = \frac{\tau_1 + \tau_2}{2}$ and $\epsilon = \tau_1 - \tau_2$ and perform the expansion $\epsilon$:
	\begin{align}
		\partial_{\tau_{1,2}}\Sigma^{-1}_f(\tau_1,\tau_2) &= \frac{2}{\mu J^2 B^{\frac{\mu}{2}-1}_\beta b^{q-1}}\left(\frac{1}{2}\frac{\partial}{\partial\tau} \pm \frac{\partial}{\partial \epsilon}\right)\frac{\left(f'(\tau + \epsilon/2)f'(\tau - \epsilon/2)\right)^{\Delta}}{\left(\beta \sin\left[\frac{\pi}{\beta}(f(\tau + \epsilon/2) - f(\tau - \epsilon/2))\right]\right)^{2\Delta}}\notag\\
		&=\frac{2}{\mu J^2 B^{\frac{\mu}{2}-1}_\beta b^{q-1}}\left(\frac{1}{2}\frac{\partial}{\partial\tau} \pm \frac{\partial}{\partial \epsilon}\right) \frac{1}{\epsilon^{2\Delta}}\left(1 + \frac{\Delta\epsilon^2}{6}\{f,\tau\} + \cdots\right).
	\end{align}
	This can then be expanded to leading powers of $\epsilon$ to give the final result
	\begin{align}
		\partial_{\tau_1}\Sigma^{-1}_f(\tau_1,\tau_2) \partial_{\tau_2}\Sigma^{-1}_f(\tau_2,\tau_1)\sim \frac{8\gamma^2 b^{2-2q}}{\mu^2 J^4 B^{\mu-2}_\beta } \frac{1}{\epsilon^{2(2\Delta+1)}}\left(1 - \frac{(1-\Delta)^2 \epsilon^2}{6^2}\{f,\tau\} + \cdots\right).
	\end{align}
	Integrating this over $\mathrm{d}\tau\mathrm{d}\epsilon$ and using the same cutoff $\xi$, we note that the terms scale as
	\begin{align}
		&\int\mathrm{d}\tau_1 \mathrm{d}\tau_2\partial_{\tau_1}\Sigma^{-1}_f(\tau_1,\tau_2) \partial_{\tau_2}\Sigma^{-1}_f(\tau_2,\tau_1)\notag\\
        &\sim (\#_1)\beta^{\frac{2\mu}{\mu + 2} + o(1/q)} + (\#_2)\beta^{o(1/q)} + \underset{\sim \beta^{\frac{2-\mu}{2+\mu}}\int \{f,\tau\}\mathrm{d}\tau}{\underbrace{(\#_3)\beta^{-\frac{2\mu}{\mu + 2} + o(1/q)}}}  + \cdots\,,
	\end{align}
	where we have corrections of order $1/q \,(= \Delta)$ in the exponent.
    The coefficients, which are $\beta-$independent are simply denoted by $(\#_{i})$.
    We also indicate the term that arises from the the Schwarzian and use the thermodynamic reparameterization $f(\tau) = \tan(\pi\tau/\beta)$ to evaluate its' $\beta-$contribution.
    The $1/q$ corrections will disappear in the large$-q$ limit, as will be demonstrated by the scaling of the large$-q$ thermodynamics.
	
	In particular, we note that the reparametrisation action has the identical analytical form as the Gaussian case.
    The difference is in the overall factor (that we write as $\alpha_S$) in front of the Schwarzian, which depends on the temperature $\beta$ and gives the result
	\begin{align}
		S_{f} \sim -\frac{N \alpha_S (\mu,q,\beta)}{\mathcal{J}^\mu}\int\mathrm{d}\tau \{f(\tau),\tau\}\,.\label{eq:schwarzian-lsykl}
	\end{align}
	The behaviour of the factor $\alpha_S$ can be determined numerically from the saddle solutions $G_*, \Sigma_*$. 
	In a related direction, we can consider the $\beta-$contributions arising from the other terms in the action.
    The second and third terms $\int \Sigma G$ and $(J^2 \int G^q )^{\frac{\mu}{2}}$ give the same $\beta-$dependence which can be computed by using the reparameterised Green's function $G_f(\tau_1,\tau_2)$ and expanding in orders of $\epsilon$.
    Performing that expansion, we recover the expansion in $\beta$
	\begin{align}
		\left(J^2\int_0^\beta \mathrm{d}\tau_1\mathrm{d}\tau_2 G^{q}_f(\tau_1,\tau_2)\right)^{\frac{\mu}{2}} \sim (\#_1)\beta^{\frac{2\mu}{\mu + 2}} + (\#_2)\beta^{0} + \underset{\sim \beta^{\frac{2-\mu}{2+\mu}}\int \{f,\tau\}\mathrm{d}\tau}{\underbrace{(\#_3)\beta^{-\frac{2\mu}{\mu + 2}}}} + \cdots\,, 
	\end{align}
	where we have indicated the Schwarzian contribution, which agrees with the Pfaffian term up to $1/q$ corrections. 
    Thus, the thermodynamic expansion of the free energy, in contrast with Gaussian SYK, is given by
	\begin{align}
		\frac{-\beta F}{N} \sim (\#_1)\beta^{\frac{2\mu}{\mu + 2}} + (\#_2) + (\#_3)\beta^{-\frac{2\mu}{\mu + 2}} + \cdots + \mathcal{O}(1/q)\,.
        \label{eq:free-en-cnflm}
	\end{align}
	This scaling of the free energy (and resulting scaling of the thermodynamic energy, entropy etc.) will be investigated in further detail in Sec.~\ref{sec:thermo}.

	\section{Chaos exponents at large-$q$}
	\label{sec:chaos-exp}

\begin{figure}%[ht]
		\centering
		\includegraphics[width=0.49\linewidth]{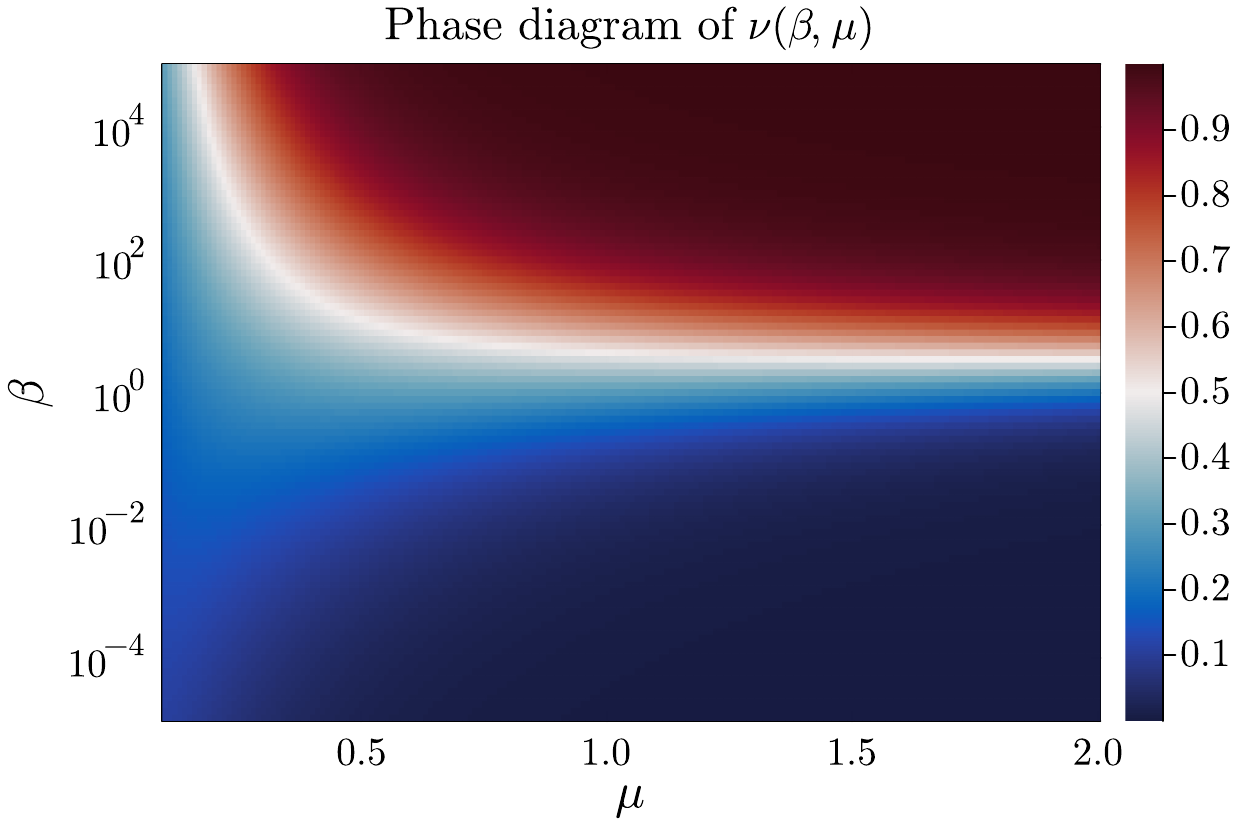}
		\includegraphics[width=0.49\linewidth]{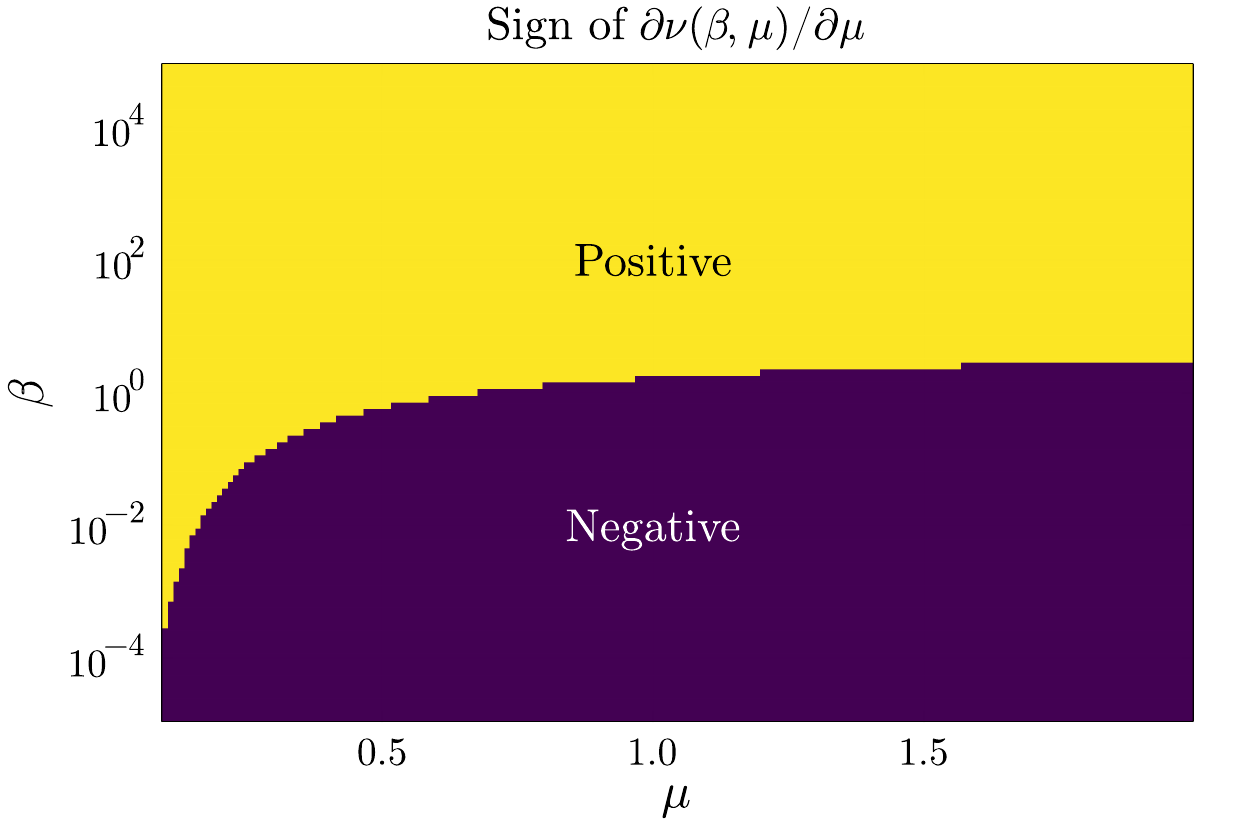}
		\caption{
			Dependence of $\nu(\beta,\mu)$ on parameters $\beta$ and $\mu$ .
			Left, (a): magnitude of $\nu$ as a function of \(\mu\) and $\beta$.
			Right, (b): the sign of the rate $\partial\nu/\partial\mu$ for fixed $\beta)$ fixed \(\beta\) and \(\mu\)
            Yellow/bright and violed/dark colors indicate regions with increase and decrease of \(\partial\nu/\partial\mu\) respectively.
			Recall that $\nu\in[0,1]$, corresponding to $\beta \in [0,\infty)$ for any $\mu$.
		}
		\label{fig:phase-diag-nu}
	\end{figure}

	The $2-$ point Greens' function $G_{R}(t)$ encodes information about the growth of operators, Majorana fermions $\psi_i$ in our case.
	This is extracted from the Krylov representation of operator growth~\cite{parker2018a}.
	For our SYK model with Green's function of the form~\ref{eq:greenf-largeq-exp}, the Lanczos coefficients are given by
	\begin{align}
		b_n = 
		\begin{cases}
			\frac{\pi\nu}{\beta}\sqrt{2/q} + O(1/q) \;\;\;\;\;\;\;\;\;\;\, n = 1\\
			\frac{\pi\nu}{\beta}\sqrt{n(n-1)} + O(1/q) \;\;\; n > 1
		\end{cases}\,.
	\end{align}
	Asymptotic linear growth of these coefficients $b_n \sim \alpha n$ corresponds to chaotic dynamics~\cite{parker2018a}.
	For such Lanczos coefficients, the quantity $\alpha =\pi\nu/\beta$ controls the rate at which $b_n$ grows.
	Furthermore, the Krylov complexity of the model endowed with such Lanczos coefficients grows exponentially with the exponent $2\alpha$, i.e. $C_K(t) \sim e^{2\alpha t}$.
	The analysis of quantum chaos in this model then reduces to the analysis of $\pi\nu(\nu,\beta)/\beta$.
	From Eqn.~\eqref{eq:beta-scaling} and Fig.~\ref{fig:nu0-vs-beta-and-betamu}, we note that for small $\beta$ 
	\begin{align}
		\nu(\mu,\beta) \sim \beta^{\frac{\mu}{2}}\;\;\;,\; \beta \ll 1
	\end{align}
	Therefore $\lim_{\beta\rightarrow0}\nu/\beta \rightarrow \infty$ for any $\mu < 2$.
	For finite $\beta$, we can solve Eqn~\eqref{eq:eff-int-mu-s} to determine $\nu(\mu,\beta)$.
	The phase diagram that emerges is presented in Fig.~\ref{fig:phase-diag-nu} for $q=4$. 
	For generic SYK models, $q = 4$ is a reasonably good approximation to the large$-q$ result.
	We use the large$-q$ consistency equations to solve for $\nu$, where $q$ does not qualitatively modify the properties of $\nu(\mu,\beta)$.
	For any fixed $\beta > 0$, the limit $\mu \rightarrow 0$ is not chaotic since $\nu \rightarrow 0$.
	Thus $\alpha\rightarrow 0$, as expected from Eqn.~\eqref{eq:sd-eqn-2} where $\mu\rightarrow 0$ corresponds to a free theory.
	This tells us that for the Green's function (and equivalently for the growth exponent $\alpha$), the limit $\mu\rightarrow 0$ and $\beta\rightarrow 0$ do not commute
	\begin{align}
		\lim_{\mu\rightarrow 0}\lim_{\beta\rightarrow 0} G_{R}(t) \neq \lim_{\beta\rightarrow 0} \lim_{\mu\rightarrow 0} G_{R}(t)\,. 
	\end{align}
	The same is, in fact, true for the action Eqn.~\eqref{eq:eff-action-largeN}.
	The difference between the two limits in the action is of $O(1)$. 

\begin{figure}[t]
		\centering
		\includegraphics[width=0.48\linewidth]{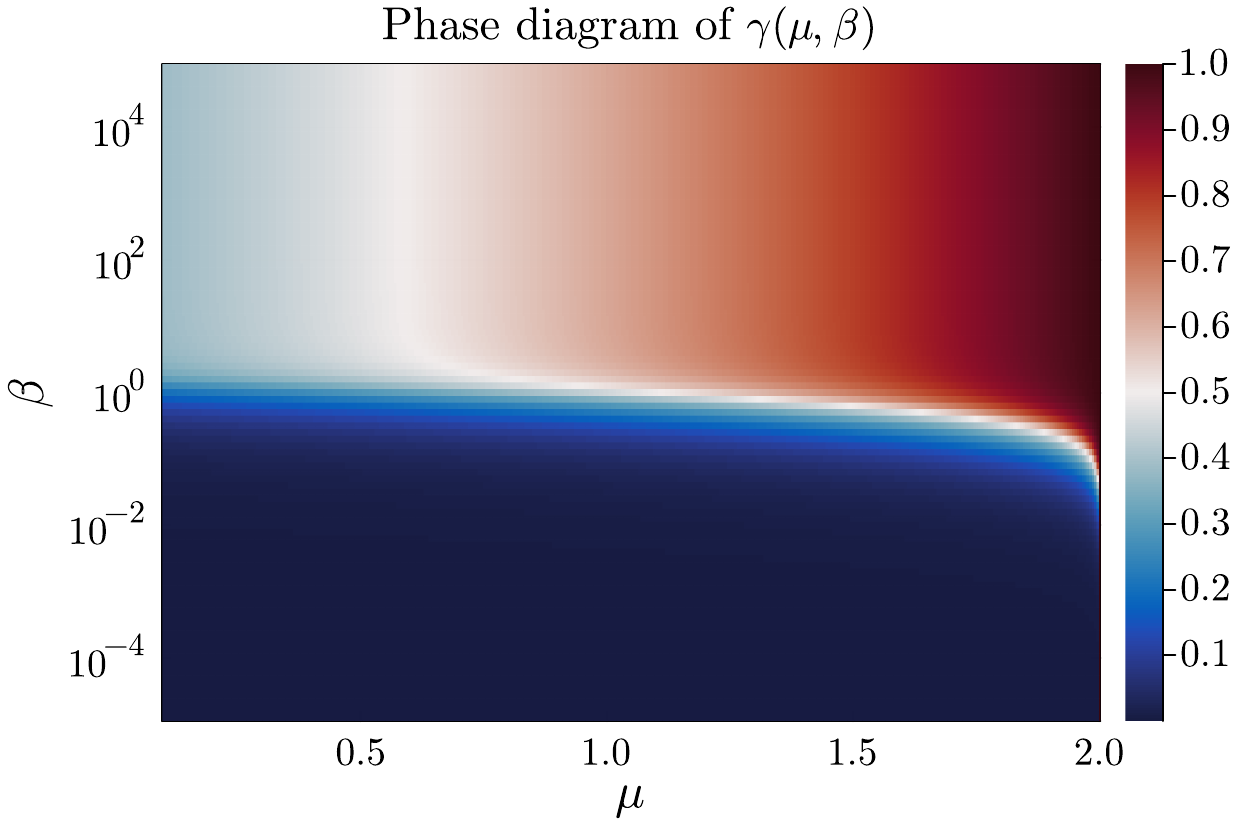}
		\includegraphics[width=0.48\linewidth]{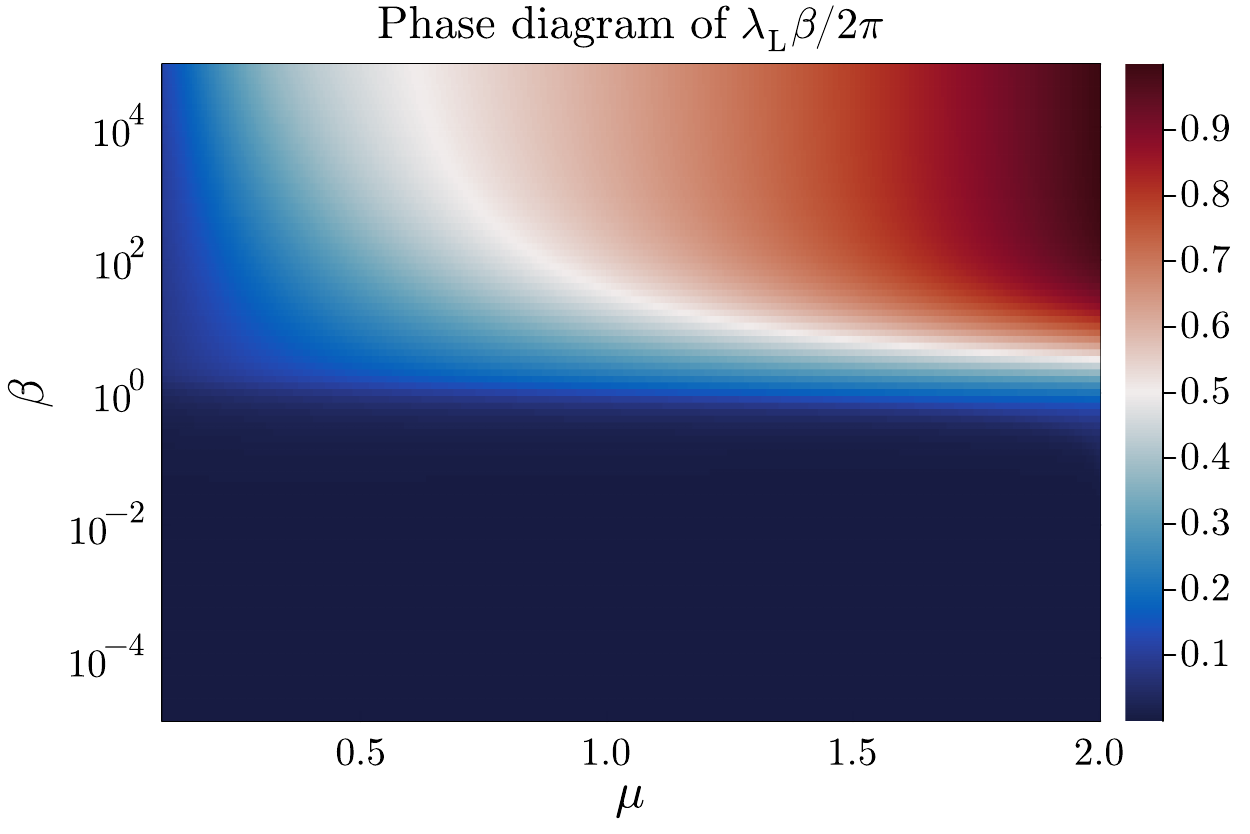}
		\caption{
			Dependence of $\gamma(\beta,\mu)$ and $\lambda_L\beta/2\pi$ on parameters $\beta$ and $\mu$.
			Left (a): the ratio \(\lambda_L/2\alpha\)  -- the non-saturation of the $\mathcal{Q}-$complexity bound.
			Right (b): the OTOC exponent $\lambda_L$ in units of $2\pi/\beta$.
            The Gaussian SYK result corresponds to the vertical line at $\mu = 2.0$ (dark red).
            %AA: a more detailed comparison with GSYK is needed here, e.g. which colorcodes correspond to GSYK?
		}
		\label{fig:phase-diag-lambda}
	\end{figure}

	The operator growth is also captured by the $4-$ point function of the Majorana fermions.
	It is known~\cite{maldacena2016remarks} that this is given by
	\begin{align}
		F(\tau_1,\tau_2, \tau_3,\tau_4) = \frac{1}{N^2}\sum_{i,j = 1}^{N}\langle T \psi_i (\tau_1)\psi_i(\tau_2)\psi_j(\tau_3)\psi_j(\tau_4)\rangle\,. 
	\end{align}
	where $T$ is the time-ordering operator.
	This expression can be written in the following form
	\begin{align}
		F(\tau_1,\tau_2, \tau_3,\tau_4) = \langle G(\tau_1,\tau_2)G(\tau_3,\tau_4)\rangle_{G,\Sigma} = G_{*}(\tau_1,\tau_2)G_{*}(\tau_3,\tau_4) + O(1/N)
	\end{align}
	For the LSYK, the kernel for the ladder construction of the $4-$point function is given by $K = (G_{*}\circ G_{*})(\delta^2 I_{\mathrm{eff}}/\delta G^2)\vert_{G = G_{*}}$, which is formally identical to that of the Gaussian SYK.
	The main difference from the Gaussian SYK occurs for the variational derivative of $I_\mathrm{eff}$ which gives a $\mu-$ dependent result.
	The kernel is written as~\footnote{Here we define $\tau_{i j} \equiv \tau_i - \tau_j$.}
	\begin{align}
		K(\tau_1,\tau_2;\tau_3,\tau_4) = G(\tau_{1 2})G(\tau_{3 4})\mathcal{G}(\tau_{13})\,,
	\end{align}
	where the function $\mathcal{G}(\tau_{13})$ is
	\begin{align}
		\mathcal{G}(\tau_{1 3}) = &\frac{(q-1)\mu}{2}J^{2}G^{q-2}(\tau_{1 3})\left(J^2 \int\mathrm{d}\tau\mathrm{d}\tau' G^{q}(\tau,\tau')\right)^{\frac{\mu}{2}-1}\notag\\
		&\times\left(1 + \frac{q (\mu-2)}{2(q-1)}\frac{G^q(\tau_{13})}{\int\mathrm{d}\tau\mathrm{d}\tau'G^{q}(\tau,\tau')}\right)\,.
	\end{align}
	This expression reduces to the Gaussian SYK case for $\mu = 2$ and vanishes for $\mu = 0$.
	The $1/N$ correction is constructed via the ladder diagram approach and sums to $(1 - K)^{-1}$.
	It is possible to investigate the chaos exponent in the large$-q$ limit without having to explicitly diagonalize the kernel.
	Since the ladder diagram construction is unchanged from the Gaussian SYK, we directly use the result~\cite{maldacena2016remarks} that the large$-q$ exponent can be extracted from the real-time retarded kernel $K_R = G_R G_R \mathcal{G}_{l r}$ by finding its' eigenfunction(s) $F(t_1,t_2)$ and using the ansatz $F(t_1,t_2) \sim e^{\lambda_L (t_1 + t_2)/2}f(t_{12})$.
	To leading order, we have $G_R (t) \sim \theta(t)$ and
	\begin{align}
		\mathcal{G}_{lr}(t) = &\frac{(q-1)\mu J^2}{2^{q-1}}\frac{\cos^2(\pi\nu/2)}{\cosh^2(\pi\nu t/\beta)}\left(\frac{J^2 \beta^2 \sin(\pi\nu)}{2^{q-2}\pi\nu}\right)^{\frac{\mu}{2}-1}\notag\\
		&\times\left(1 + \frac{\mu-2}{2}\frac{\pi\nu \cot(\pi\nu/2)}{2\beta^2 \cosh^2(\pi\nu t/\beta)}\right)\,.
		\label{eq:glr-levy}
	\end{align}
	We have to solve the eigenvalue equation $F(t_1,t_2) = \int\mathrm{d}t_3\mathrm{d}t_4 K_{R}(t_1,\cdots,t_4)F(t_1,t_2)$ with the exponential ansatz.
	This leads to the following differential equation satisfied by $f(t_1-t_2)\equiv f(x)$.
	\begin{align}
		\partial^2_x f + \frac{s_1 }{\cosh^2(x)}\left(1 + \frac{s_2}{\cosh^2(x)}\right)f(x) = \left(\frac{\lambda_L \beta}{2\pi\nu}\right)^2f(x)\,,
		\label{eq:diff-eq-fx}
	\end{align}
	where the constants $s_{1,2}$ are given by
	\begin{align}
		s_1 &= \frac{\beta^2(q-1)\mu J^2 \cos^2 (\pi\nu/2)}{2^{q-1}\pi^2\nu^2}\left(\frac{J^2 \beta^2 \sin(\pi\nu)}{2^{q-2}\pi\nu}\right)^{\frac{\mu}{2}-1} = 2^{\mu-1}\,, \\
		s_2 &= \frac{2-\mu}{4\beta^2}\frac{\pi\nu}{\tan(\pi\nu/2)}\,.
		\label{eq:s1-s2}
	\end{align}
	In the last equality for $s_1$, we use Eqn.~\eqref{eq:eff-int-mu-s}.
	Note that for $\mu = 2$, $s_2 = 0$ and $s_1 = 2$, which makes the above differential equation the P\"oschl-Teller type Schr\"odinger equation.
	This has a single bound state with eigenvalue $1$, which fixes $\lambda_L = 2\pi\nu/\beta$.
    In the presence of the $s_2$, the differential equation is more involved.
	The problem then reduces to the analysis of eigenvalues of a Schr\"odinger operator.
	This is discussed further in Appendix~\ref{app:sch-prob}, where we semi-analytically approximate the ground state eigenvalue $E_{\min}$ and compare with numerical diagonalisation.
	Given the ground state eigenvalue $E_{\min} \equiv -\gamma^2(\mu,\beta)$ of the LHS of Eqn.~\eqref{eq:diff-eq-fx}, then the chaos exponent is given by
	\begin{align}
		\lambda_L \approx \frac{2\pi\nu}{\beta}\gamma(\mu,\beta) = \frac{2\pi\nu}{\beta}\sqrt{-E_{\min}}\,.
	\end{align}
	For $\mu = 2.0$, it is known that $\gamma(\beta,2) = 1$.
	The expression for $E_{\min}$ is given by
	\begin{align}
		E_{\min} \approx
		\begin{cases}
			&-\epsilon(s_1) ( 1 - s_2)\;\;\;\;\;s_2 < 1/2\\
			&-\epsilon(s_1)/4s_2\;\;\;\;\;\;\;\;\;\;\,s_2 \geq 1/2
		\end{cases}\,.
		\label{eq:e-min}
	\end{align}
	where $\epsilon(x) = \frac{(1 - \sqrt{1 + 4 x})^2}{4}$.
	We emphasize that this is not an exact solution, but  merely an approximation, that, however agrees with the numerical result $E_{\min}$ to a high degree of accuracy.  
	At this stage, we identify the key difference between operator growth in Gaussian and L\'evy SYK.
	For Gaussian SYK, one finds $\lambda_L = 2\alpha$.
	This is one of the hints of maximal chaos, since the $\mathcal{Q}-$complexity bound~\cite{parker2018a} implies that $\lambda_L\leq 2\alpha$.
	For L\'evy SYK, we rather find that ($E_{\min}(\mu,\beta)$ is given by Eqn.~\eqref{eq:e-min})
	\begin{align}
		\lambda_L = \frac{2\pi\nu}{\beta}\sqrt{-E_{\min}} \leq \frac{2\pi\nu}{\beta} = 2\alpha\,.
	\end{align}
	The equality is saturated only when $\sqrt{-E_{\min}} = 1$, which happens only for $\mu = 2$.
    The phase diagrams of $\lambda_L(\beta/2\pi)$ and $\gamma(\beta,\mu)$ are shown in Fig.~\ref{fig:phase-diag-lambda}.
	
    In conclusion, we note that the Krylov exponent behaves as \(2\alpha = 2\pi\nu/\beta\), where the effect of the L\'evy factor is implicit in the dependence of $\nu$ on $\mu$ and $\beta$.
	On the other hand, the Lyapunov exponent $\lambda_L$ is further affected by the presence of the additional factor $\gamma = \sqrt{-E_{\min}} \leq 1$ which decreases as $\mu \rightarrow 0$.
	Thus, we can summarise all the results for finite $\beta$  as
	\begin{align}
		&\mu = 0\;\; \;\;\;\;\;\;\;\;\;\lambda_L = 2\alpha = 0\;\;\;\;\;\;\;\;\;\, \text{Non-chaotic}\notag\\    
		&\mu \in (0,2) \;\;\;\;\;\,\lambda_L < 2\alpha\;\;\;\;\;\;\;\;\;\;\;\;\;\;\;\; \text{Non-maximal chaos}\notag\\
        &\mu = 2 \;\;\;\;\;\;\;\;\;\;\;\lambda_L = 2\alpha  \;\;\;\;\;\;\;\;\;\;\;\;\;\;\;\; \text{Maximal chaos}\notag
	\end{align}
	At large$-\beta$, $\nu \rightarrow 1$, giving $2\alpha \rightarrow \frac{2\pi}{\beta}$ and $\lambda_L \rightarrow \frac{2\pi\sqrt{-E_{\text{min}}}}{\beta}$.
    The large$-\beta$ expansion of $\nu$ generates corrections to the chaos exponent $\lambda_L$, that follow from expanding Eqn.~\eqref{eq:eff-int-mu-J}.
    To the first few orders, this expansion behaves as
	\begin{align}
		\nu \sim 1 - \frac{(\#)}{\beta^{\frac{2\mu}{2 + \mu}}} + \frac{(\#)}{\beta^{\frac{4\mu}{2 + \mu}}} + \cdots\,.
	\end{align}
	Together with the previously studied scaling at small $\beta$ in Eqn.~\eqref{eq:beta-scaling}, we can summarize the small and large$-\beta$ behavior of $\nu$ as follows
	\begin{align}
		\nu \sim \begin{cases}
			&(\#)\beta^{\mu/2} \;\;\;\;\;\;\;,\;\;\beta \ll 1 \\
			&1 - \frac{(\#)}{\beta^{\frac{2\mu}{\mu+2}}}\;\;\;\;\;\,,\;\;\beta \gg 1
		\end{cases}
        \label{eq:nubetasc}
	\end{align}
	Similarly, using the above large-\(\beta\) behavior of \(\nu\) we can determine the large$-\beta$ scaling of the Lyapunov exponent $\lambda_L$
	\begin{align}
		\lambda_L \sim \frac{2\pi}{\beta}\left(1 - \frac{(\#)}{\beta^{\frac{2\mu}{\mu + 2}}} - \cdots\right)\,.
	\end{align} 
	In the next section, we investigate thermodynamic quantities governed by the saddle-point equation of LSYK.
    The scaling of $\nu$ with $\beta$~\eqref{eq:nubetasc} is controlling the low temperature equilibrium thermodynamics of LSYK.

	\section{Thermodynamics}
	\label{sec:thermo}

    The effective large-$N$ action~\eqref{eq:eff-action-largeN} is the starting point in order to compute thermodynamic quantities such as the thermal entropy or the free energy.
	The free energy $F$ is defined as $I_{\mathrm{eff}} = \frac{\beta F}{N}$.
	Let us recall that the full $I_\mathrm{eff}$ can be written as
	\begin{align}
		I_{\mathrm{eff}} = -\frac{1}{2}\log[\det(\partial_\tau - \Sigma)] - \frac{1}{2 q}\left(J^2 \int_0^\beta \mathrm{d}\tau\mathrm{d}\tau' G^q (\tau,\tau')\right)^{\frac{\mu}{2}} + \frac{1}{2}\int \mathrm{d}\tau\mathrm{d}\tau' \Sigma G\,.
		\label{eq:eff-action-largeN-rep}
	\end{align}
	Following the usual approach~\cite{maldacena2016remarks}, we take a derivative of $I_\mathrm{eff}$ with respect to $J_{\mu} \equiv J^{\frac{\mu}{2}}$.
	Since apriori $\Sigma, G$ are independent of $J$, the only explicit dependence comes from the term proportional to $J_\mu$, i.e. the second term in Eqn.~\eqref{eq:eff-action-largeN-rep}.
	This gives us the equation
	\begin{align}
		J_\mu \partial_{J_\mu}\left(-\frac{\beta F}{N}\right) = J_\mu \partial_{J_\mu}\left(\frac{J^2_\mu}{2 q}\left(\int_0^\beta \mathrm{d}\tau\mathrm{d}\tau' G^q (\tau,\tau')\right)^{\frac{\mu}{2}}\right)\,.
	\end{align}
	Using time translation invariance we can perform one of the integrals over $G^q$. 
	Using the solution of the SD equations $G_{*}$ the RHS can then be evaluated. 
	This gives us a differential equation for $F$, which is then solved. 
	To rewrite the LHS in a useful manner, the following substitution proves instrumental
	\begin{align}
		J_\mu \partial_{J_\mu} = J_{\mu} \frac{\partial (\beta K)^{\mu/2}}{\partial J_\mu} \frac{\partial \nu}{\partial (\beta K)^{\mu/2}}\partial_\nu\,.
	\end{align}
	These equations can be solved exactly in the large-$q$ limit, as we demonstrate in the following subsections.
    We also compare them with direct numerical evaluation of thermodynamic quantities for finite-$q$.
	%\begin{align}
	%    J_\mu \partial_{J_\mu}\left(-\frac{\beta F}{N}\right) = \frac{J^2_\mu \beta^{\frac{\mu}{2}}}{2 q}\left(\int_0^\beta \mathrm{d}\tau G_{*}^q (\tau)\right)^{\frac{\mu}{2}} = \frac{\beta}{q}\lim_{\tau\rightarrow 0_{+}}\partial_\tau G_{*}(\tau)\,.
	%    \label{eq:entropy-1}    
	%\end{align}
	%The last equality follows from Eqn.~\eqref{eq:sd-eqn-3} and~\eqref{eq:sd-eqn-4}.
	%The RHS of Eqn.~\eqref{eq:entropy-1} can be evaluated using the large$-q$ solution $G_{*}(\tau)$.
	
	%Using $\beta K = \frac{\pi\nu}{\cos(\pi\nu/2)}$ and Eqn.~\eqref{eq:k-nu-relation}, we obtain the following equation 

    \begin{figure}[ht]
		\centering
		\includegraphics[width=0.49\columnwidth]{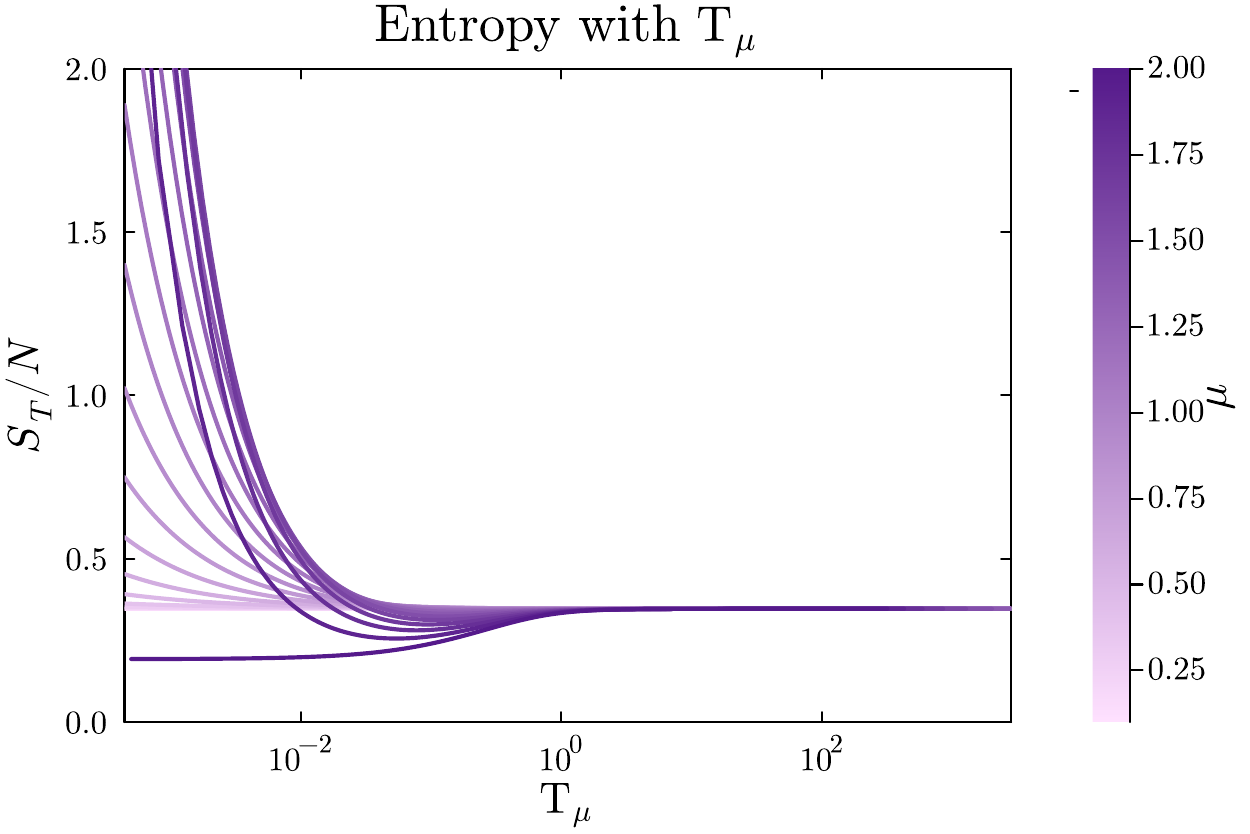}
		\includegraphics[width=0.49\columnwidth]{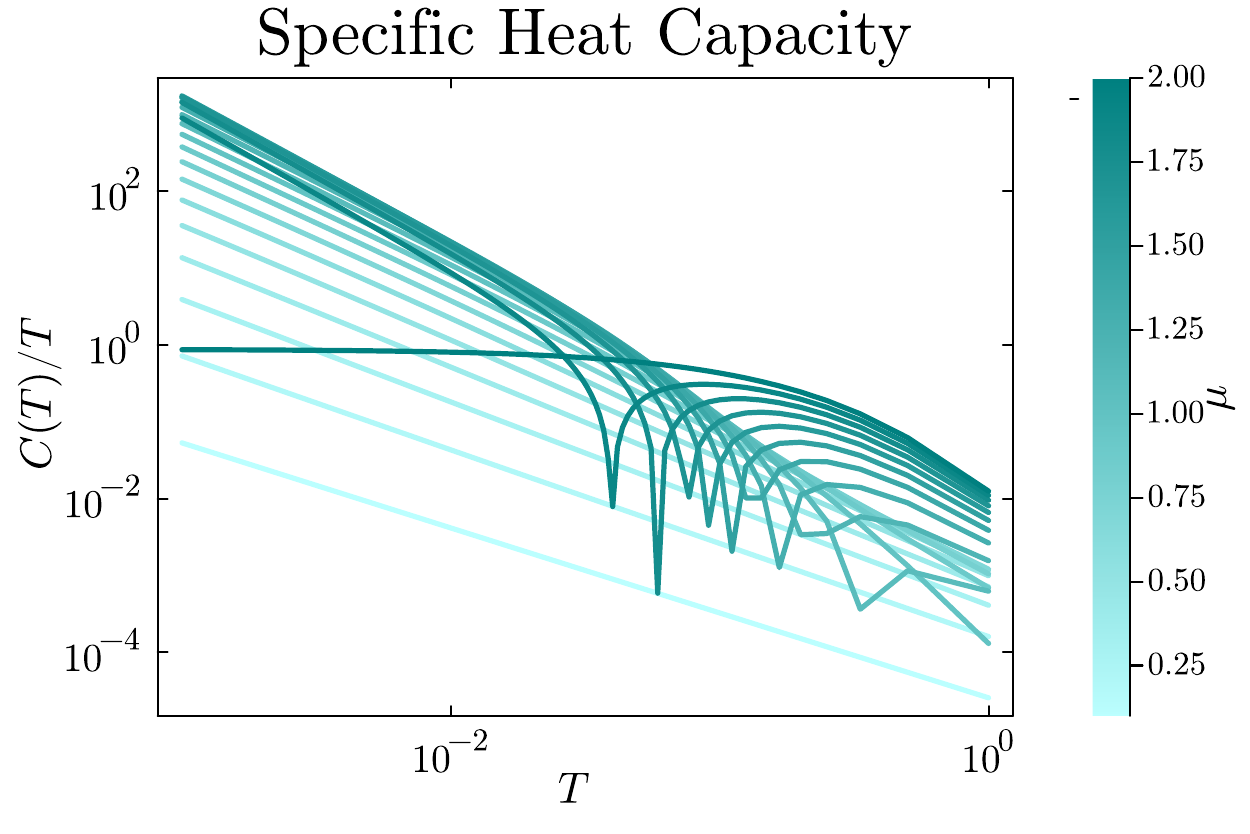}
		\caption{
			(Left) Thermodynamic entropy per fermion as a function of $T_\mu = (\beta \mathcal{J}_\mu)^{-1}$.
            (Right) Specific heat capacity per fermion per unit temperature $C(T)/T$, where $C(T) = C_T/N$, as a function of $T = \beta^{-1}$ on the \(\log\)-\(\log\) scale, indicating the scaling of $C(T)/T$ with $T$ for small $T$.
        }
		\label{fig:sthermo-mu}
	\end{figure}

    \subsection{Large-$q$}

    Using the large-$q$ solution of the SD equations, we obtain the relation
	\begin{align}
		\frac{\nu}{1 + \frac{\pi\nu}{2}\tan\left(\frac{\pi\nu}{2}\right)} \partial_\nu \left(-\frac{\beta F}{N}\right) = \frac{\beta \mu}{4 q}\lim_{\tau\rightarrow 0_+}\partial_\tau G_{*}(\tau) = \frac{\pi\mu\nu}{2 q^2}\tan\left(\frac{\pi\nu}{2}\right)\,.
	\end{align}
	This differs from the Gaussian SYK result~\cite{maldacena2016remarks} only by the overall factor of $\mu$.
	The solution of this differential equation is given by
	\begin{align}
		-\frac{\beta F}{N} = c_1 + \frac{\mu \pi\nu}{2 q^2}\left(\tan\left(\frac{\pi\nu}{2}\right) - \frac{\pi\nu}{4}\right)\,.
	\end{align}
	For the free theory \(\mu=0\), we expect that $\beta F/ N = \ln(2)/2$, and that fixes the constant $c_1 = \ln(2)/2$.
	Therefore the expression for the large-\(q\) free energy is
	\begin{align}
		-\frac{\beta F}{N} = \frac{1}{2}\ln(2) + \frac{\mu\pi\nu}{2 q^2}\left(\tan\left(\frac{\pi\nu}{2}\right) - \frac{\pi\nu}{4}\right)\,.
		\label{eq:free-energy}
	\end{align}
	As expected, at $\mu = 0$, we recover the free theory and the Gaussian SYK result at $\mu = 2$.

    The thermodynamic entropy per fermion is defined as $\frac{S_T}{N} = (1-\beta \partial_\beta)(-\beta F/N)$.
	%Evaluating this expression involves derivatives of $\nu$ with $\beta$.
	%To perform this evaluation, we note that
	%\begin{align}
	%    \beta\partial_\beta = \beta\frac{\partial\nu}{\partial \beta}\partial_\nu 
	%\end{align}
	%To evaluate the derivative $\partial\nu/\partial\beta$, we can differentiate both sides of Eqn.~\eqref{eq:eff-int-mu-s} by $\beta$.
	%This allows us to write
	%\begin{align}
	%    &\frac{\mu}{\beta} \left(\frac{\pi\nu}{\cos(\pi\nu/2)}\right)^2\left(\frac{\pi\nu}{\sin(\pi\nu)}\right)^{\frac{\mu}{2}-1} = \partial_\nu\left\{\left(\frac{\pi\nu}{\cos(\pi\nu/2)}\right)^2\left(\frac{\pi\nu}{\sin(\pi\nu)}\right)^{\frac{\mu}{2}-1}\right\}\frac{\partial\nu}{\partial\beta}\,,\notag\\
	%    \implies& \frac{2 \nu \mu }{(2-\mu )\pi   \nu  \cot (\pi  \nu )+\mu +2 \pi  \nu  \tan \left(\frac{\pi  \nu }{2}\right)+2} = \beta\frac{\partial\nu}{\partial\beta}\,.
	%\end{align}
	%Thus we obtain the following expression
	%\begin{align}
	%    \beta\partial_\beta \left(-\frac{\beta F}{N}\right) = \frac{2 \nu \mu }{(2-\mu )\pi   \nu  \cot (\pi  \nu )+\mu +2 \pi  \nu  \tan \left(\frac{\pi  \nu }{2}\right)+2}\partial_\nu \left(-\frac{\beta F}{N}\right)\,.
	%\end{align}
	This can be similarly evaluated to obtain the expression for the thermodynamic entropy
	\begin{align}
		\frac{S_T}{N} = \frac{1}{2}\ln(2) - \frac{\mu}{2}\left(\frac{\pi\nu}{2 q}\right)^2 - \frac{\pi  (2-\mu ) \mu  \nu  (\pi  \nu +\sin (\pi  \nu )) \tan \left(\frac{\pi  \nu }{2}\right)}{2 q^2 (\pi  \nu  (\mu  \cos (\pi  \nu )-2)-(\mu +2) \sin (\pi  \nu ))}\,.
		\label{eq:sthermo-mu}
	\end{align}
	In the $\mu = 2$ case, the last term vanishes leaving the first two, which is again the Gaussian SYK result.
	Conversely in the $\mu = 0$ case all terms except the constant vanish, giving us the expected free theory result, $\frac{1}{2}\ln(2)$.
	The Eqn.~\eqref{eq:sthermo-mu}, evaluated as a function of $T_\mu = (\beta \mathcal{J}_\mu)^{-1}$, is presented in Fig.~\ref{fig:sthermo-mu}.
	It is clear from the figure that the entropy diverges as $T_\mu \rightarrow 0$ for all $\mu < 2$.
	However, this divergence is apparent at lower and lower \(T\) for decreasing $\mu$.

    Similarly, we can calculate the specific heat capacity 
	\begin{align}
		\frac{C_T}{N} = \frac{T}{N}\frac{\mathrm{d}S_T}{\mathrm{d}T} = \frac{\pi  \mu ^2 \nu  \sin ^2\left(\frac{\pi  \nu }{2}\right) \csc ^4(\pi  \nu )}{2 q^2 (-\pi  \mu  \nu  \cot (\pi  \nu )+\mu +2 \pi  \nu  \csc (\pi  \nu )+2)^3}\mathcal{C}\,.
		\label{eq:capacity-large-q}
	\end{align}
	Here 
	\begin{align}
		\mathcal{C} &= \left(3 \left(\mu ^2-4\right)-2 \pi ^2 ((\mu -18) \mu +8) \nu ^2\right) \sin (\pi  \nu )+8 \pi  \left(\mu ^2+\mu -2\right) \nu +4 \pi ^3 \mu  \nu ^3\notag\\
		&-2 \pi ^2 (\mu -2) (5 \mu +4) \nu ^2 \sin (2 \pi  \nu )-(\mu +2) \left(2 \pi ^2 \mu  \nu ^2+\mu -2\right) \sin (3 \pi  \nu )\notag\\
		&+\pi  \nu  \left((\mu +2)^2-\pi ^2 (\mu  (\mu +8)-16) \nu ^2\right) \cos (\pi  \nu )\notag\\
		&+4 \pi  \nu  \left(\pi ^2 (\mu -3) \mu  \nu ^2-2 \mu  (\mu +1)+4\right) \cos (2 \pi  \nu )\notag\\
		&+\pi  \nu  (\pi  \mu  \nu -\mu -2) (\pi  \mu  \nu +\mu +2) \cos (3 \pi  \nu )\,,
	\end{align}
	which is a smooth function of $\nu,\mu$.
    At $\mu = 0$, we find $C_T/N$ = 0, as expected from a free theory.
    \(C(T)/T = C_T/T/N\) is presented as a function of $\beta^{-1} = T$ on the $\log-\log$ scale in Fig.~\ref{fig:sthermo-mu} and demonstrates the low temperature scaling behaviour.

	\subsection{Finite$-q$}

	Aside from the large$-q$ result, the analysis of the thermodynamic quantities as a function of $\beta$ (or temperature) for finite$-q$ is also discussed.
	For this, we begin by recalling the expression for the free energy, evaluated at the SD equation solutions $G_{*},\Sigma_{*}$. 
	\begin{align}
		\frac{\log Z}{N} &= -\frac{1}{2}\log[\det(\partial_\tau - \Sigma)] - \frac{1}{2 q}\left(J^2 \int_0^\beta \mathrm{d}\tau\mathrm{d}\tau' G_{*}^q (\tau,\tau')\right)^{\frac{\mu}{2}} + \frac{1}{2}\int \mathrm{d}\tau\mathrm{d}\tau' \Sigma_{*} G_{*}\notag\\
		&= - \frac{1}{2}\sum_{n}\log(-i\omega_n - \tilde{\Sigma}_{*}(\omega_n)) + \frac{\mu q-2}{4q}\left(J^2 \int_0^\beta \mathrm{d}\tau\mathrm{d}\tau' G_{*}^q (\tau,\tau')\right)^{\frac{\mu}{2}}\,,
	\end{align}
	where the sum in the first term is over the Matsubara modes $\omega_n$.
	This can be evaluated numerically from the solution of the SD equations.
	For the Gaussian SYK, the conformal solution and the Schwarzian action lead to the following low-$\beta$ expansion of $\log Z/N$ 
	\begin{align}
		\frac{\log Z}{N}\Big\vert_{\mu = 2} \sim -\beta e_0 + s_0 + \frac{2\pi\alpha^2_S}{\beta} + \cdots\,,
		\label{eq:logz-exp}
	\end{align}
	Furthermore, a similar expansion can be performed for the average energy per fermion~\cite{cotler2016black}
	\begin{align}
		\frac{\langle H \rangle}{N}\Big\vert_{\mu = 2} \sim e_0 + \frac{a}{N}T + \frac{c}{2}T^2 + c_2 T^3 + \cdots\,,\label{eq:en-exp}
	\end{align}
	where $a/N$ is a vanishingly small coefficient (as \(N\to\infty\)) and $c = {4\pi\alpha_S^2}$.
	%The expansions in Eqn.~\eqref{eq:logz-exp} and Eqn.~\eqref{eq:en-exp} are results from Gaussian SYK. 
	
	\begin{figure}[ht]
		\centering
		\includegraphics[width=0.48\linewidth]{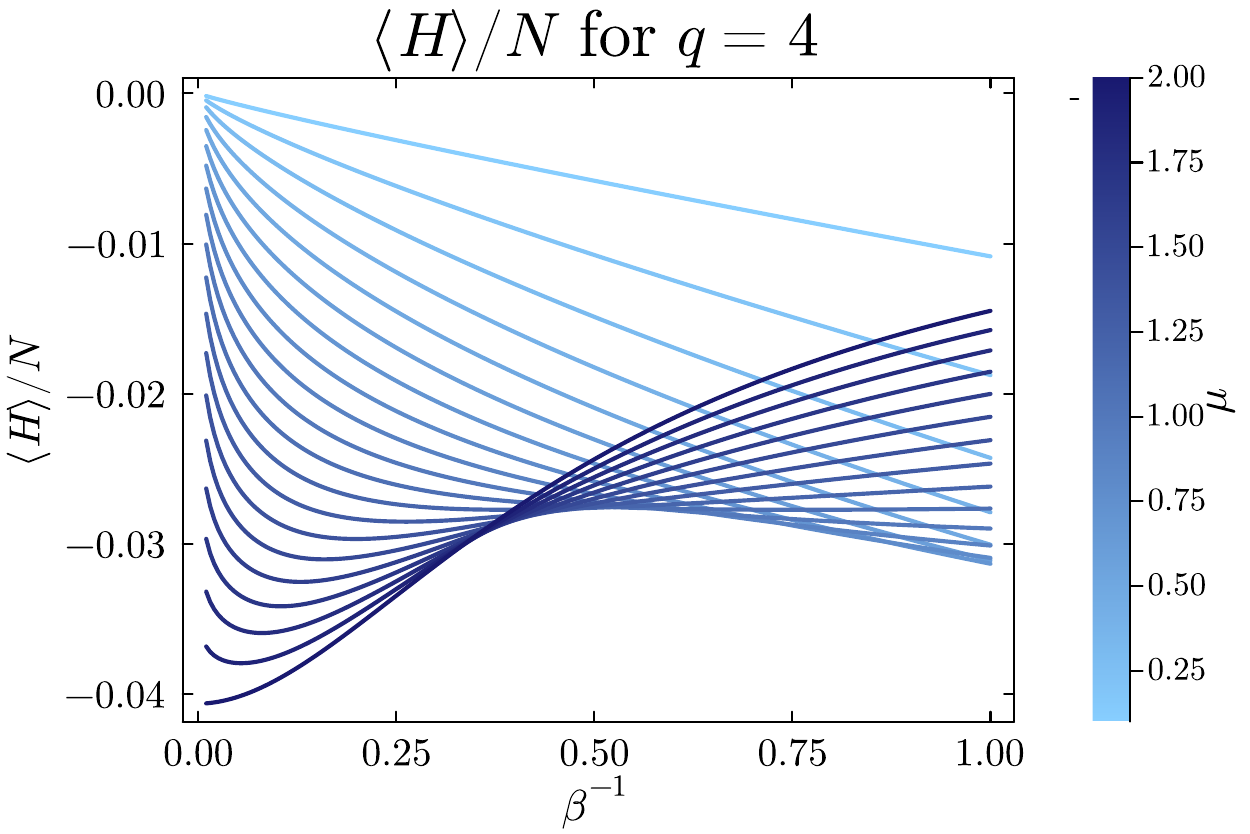}
		\includegraphics[width=0.48\linewidth]{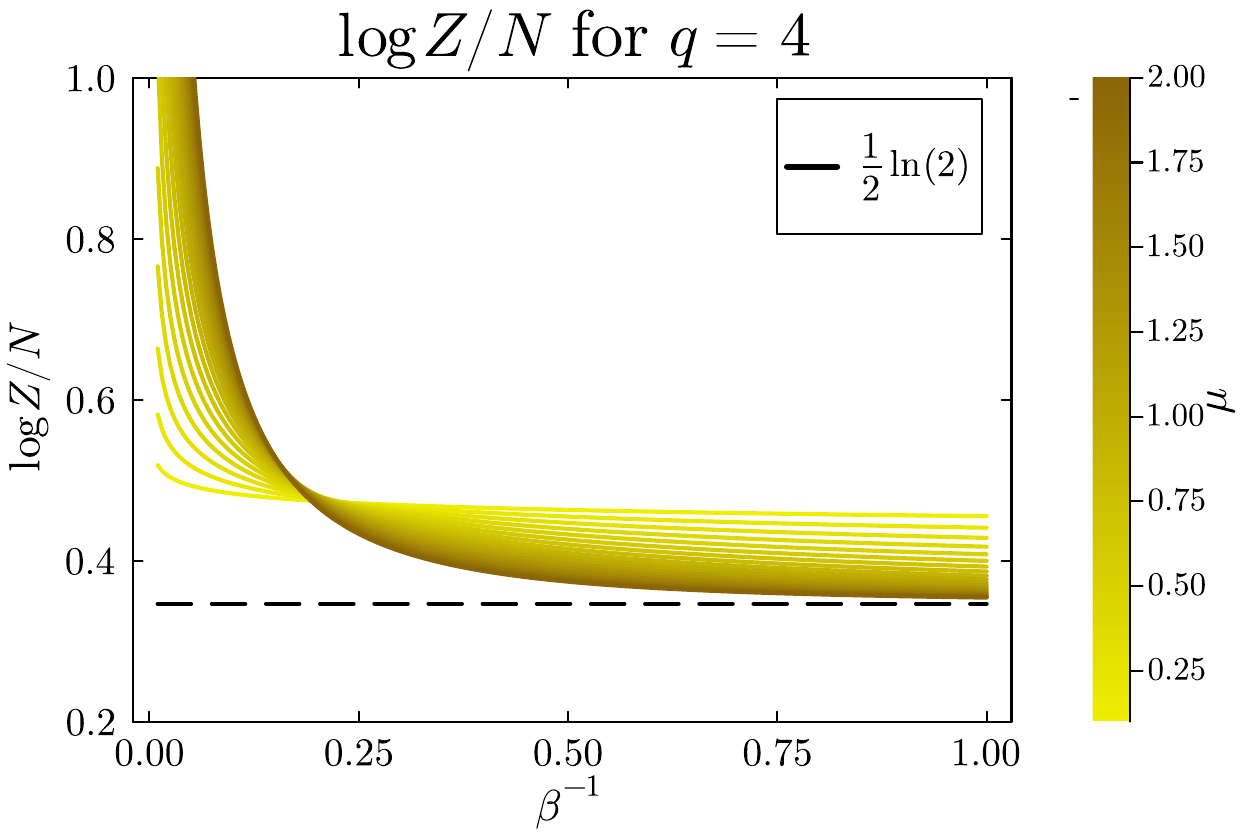}
		\includegraphics[width=0.48\linewidth]{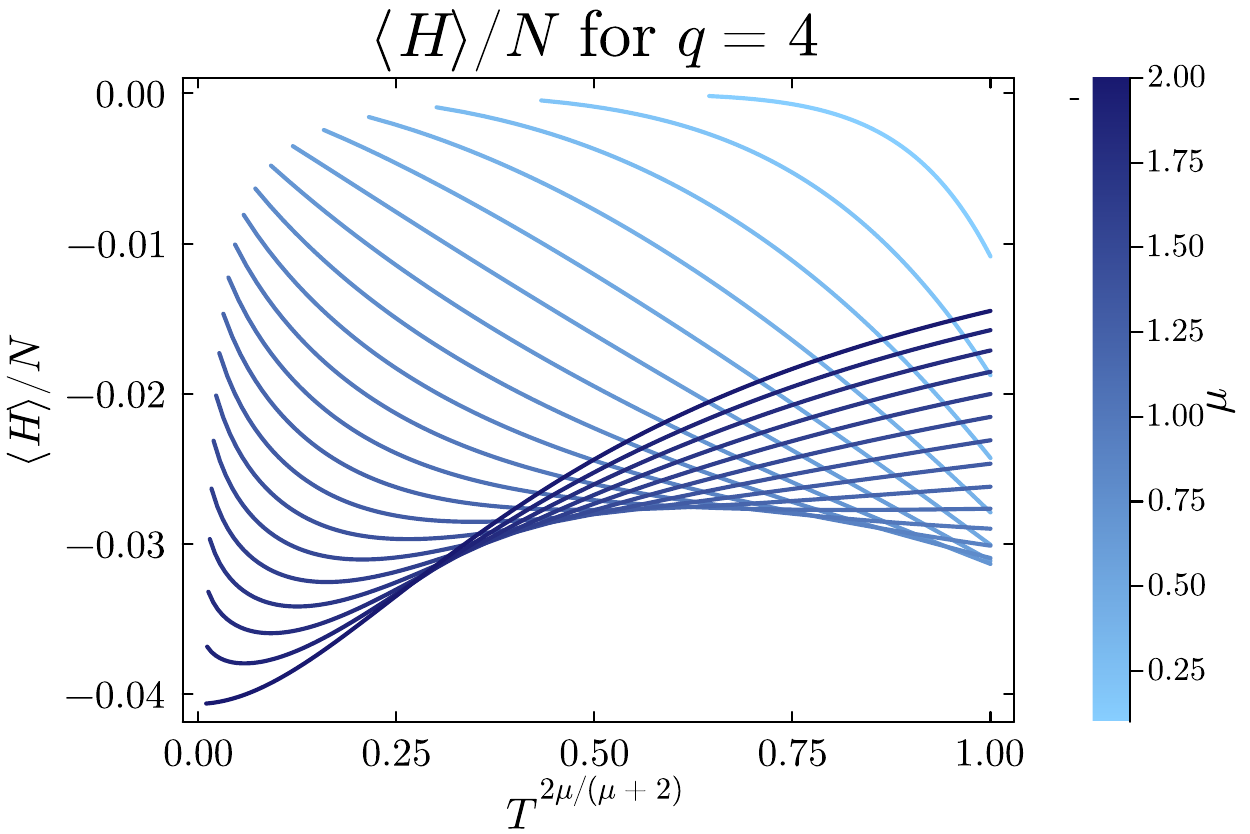}
		\includegraphics[width=0.48\linewidth]{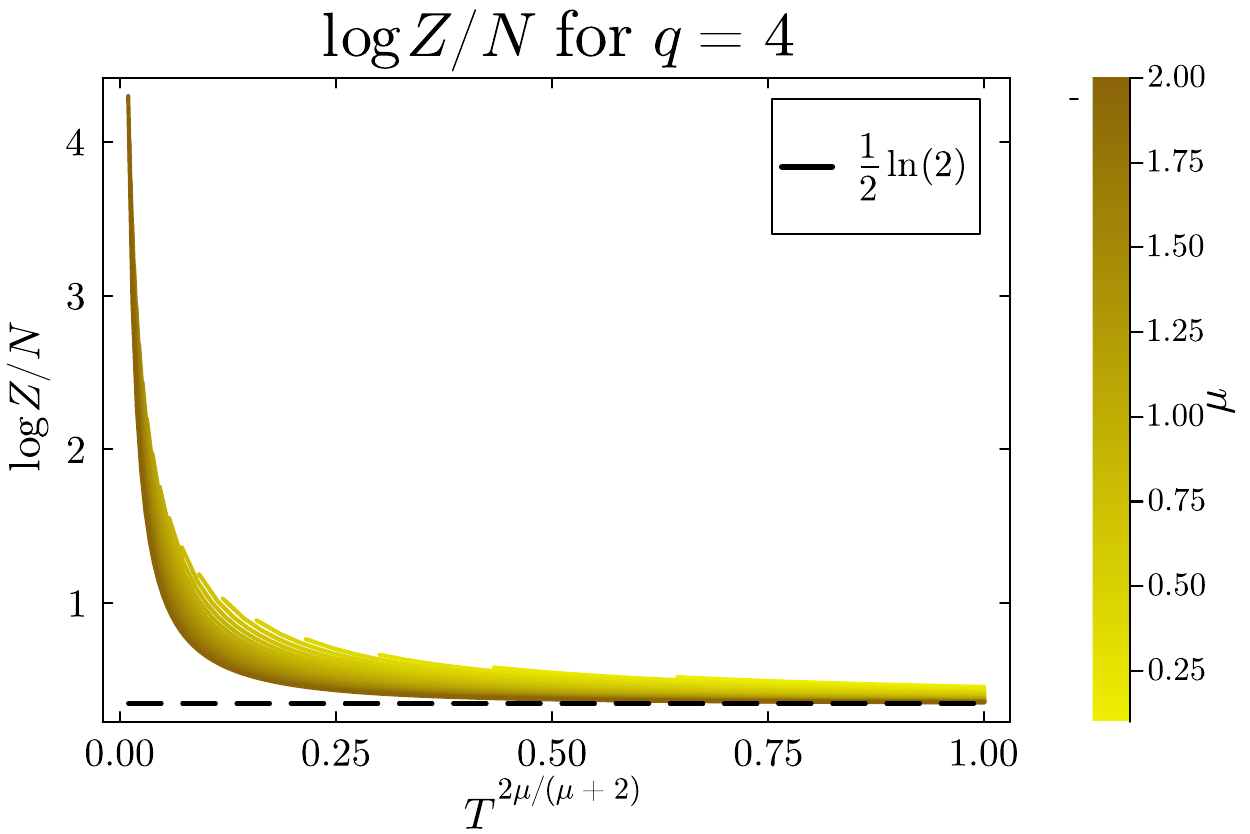}
		\caption{
            The average energy per fermion $\langle H \rangle/N$ (left) and log of partition function $\log Z/N$ (right) for different values of $\mu$ vs $\beta^{-1}$ (top) and on $T^{\frac{2\mu}{\mu + 2}} = \beta^{-\frac{2\mu}{\mu + 2}}$ (bottom). We use $q = 4$ for all the plots.
		}
		\label{fig:energy-logz}
	\end{figure}
	
	However, for the L\'evy SYK, we expect significant modifications of the above $\beta$ dependencies in the expansions.
	The primary modification is straightforward to observe from the large-$q$ expression(s) for $-\beta F/N$ and the expansion of Eqn.~\eqref{eq:eff-int-mu-J} for large$-\beta$.
	This leads to the following expansion
	\begin{align}
		\nu \sim 1 - \frac{\#}{\beta^{\frac{2\mu}{\mu+2}}} + \cdots\,,
	\end{align}
	which in turn produces the following expansion for the partition function $\log Z/N = -\beta F/M$
	\begin{align}
		\frac{\log Z}{N} \sim -\beta^{\frac{2\mu}{\mu+2}} \tilde{e}_0 + \tilde{s}_0 + \frac{\tilde{c}}{\beta^{\frac{2\mu}{\mu+2}}} + \frac{\tilde{c}_2}{\beta^{\frac{4\mu}{\mu+2}}}\cdots
        \label{eq:logz-mu-new}
	\end{align}
	This agrees with the scaling obtained from the reparameterization action in Eqn~\eqref{eq:free-en-cnflm}.
	Here we added a tilde to the coefficients by $\tilde{e}_0,\tilde{s}_0, \tilde{c}\dots$ to distinguish them from the coefficients in Eqn.~\eqref{eq:logz-exp}.
	These coefficients also depend on $\mu, q$.

\begin{figure}[ht]
		\centering
		\includegraphics[width=0.48\linewidth]{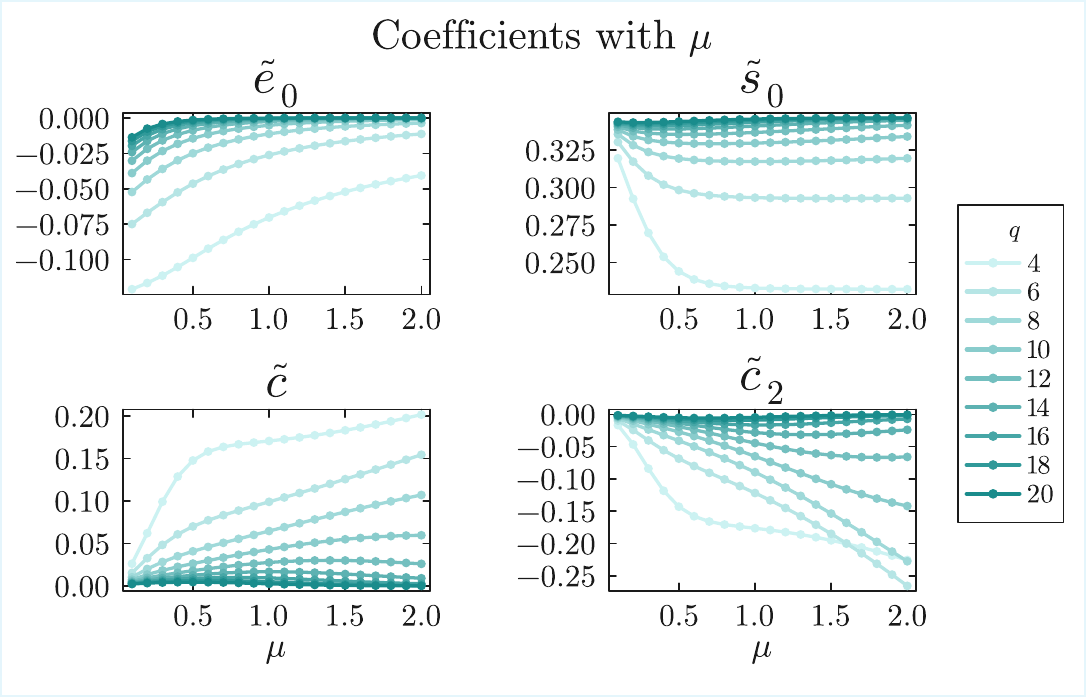}
		\includegraphics[width=0.48\linewidth]{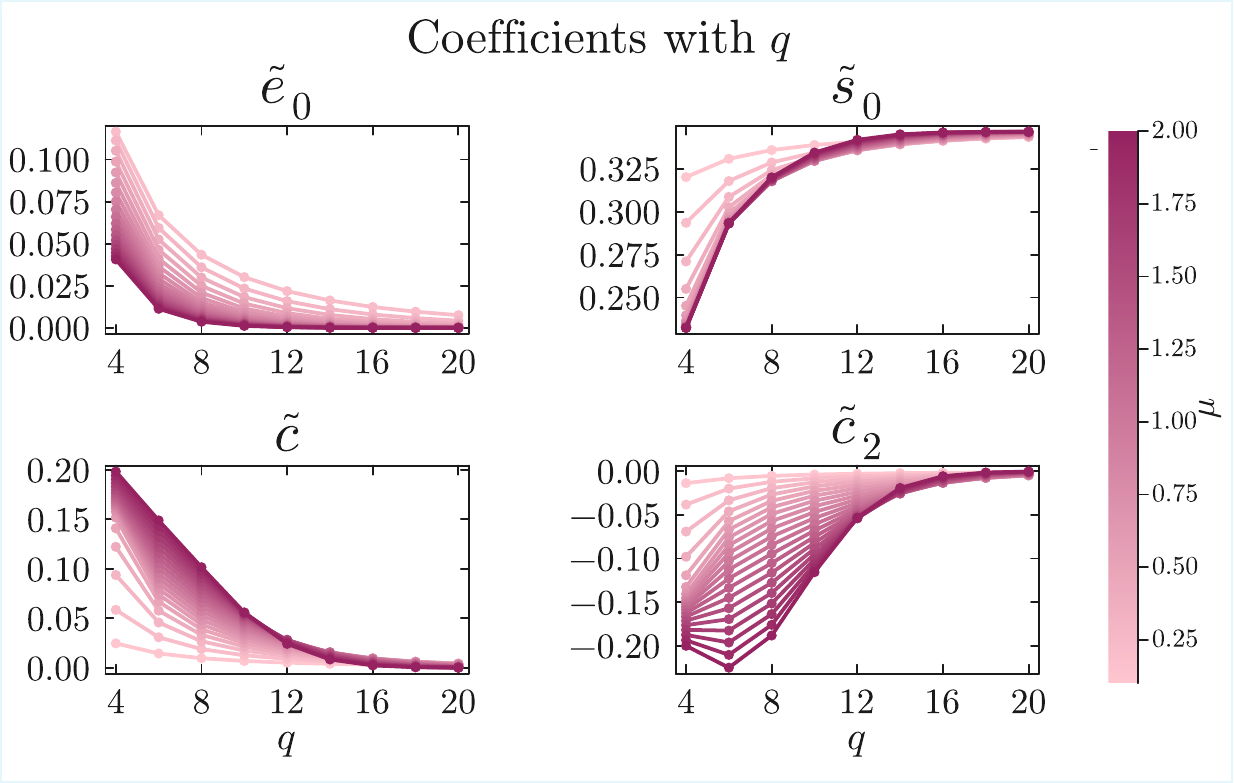}
		\caption{
			Thermodynamic coefficients $\tilde{e}_0,\tilde{s}_0,\tilde{c}$ and $\tilde{c}_2$ from the log of partition function $\log Z/N$~\eqref{eq:logz-mu-new} for different values of $\mu$.
            Left: dependence on $\mu$ for fixed values of $q$.
            Right: dependence on $q$ for fixed values of $\mu$.
			%In (a), we highlight the difference between $\lambda_L$ and $2\alpha$, in other words the non-saturation of the $\mathcal{Q}-$complexity bound.
			%In (b) we compute the OTOC exponent $\lambda_L$ in units of $2\pi/\beta$. 
		}
		\label{fig:logz-coeffs-new}
	\end{figure}

    We can compute the integral $(\beta\int \mathrm{d}\tau\, G^q)^{\mu/2}$, which corresponds to the energy per fermion
	\begin{align}
		\frac{\langle H \rangle}{N} \sim \tilde{e}_0 \beta^{\frac{\mu - 2}{\mu + 2}} + \tilde{a} \beta^{-1} + \tilde{c}\beta^{\frac{-3\mu - 2}{\mu + 2}} + \tilde{c}_2 \beta^{\frac{-5\mu - 2}{\mu + 2}} + \cdots \,.\label{eq:e-per-fermion}
	\end{align}
	Numerical results for $\langle H \rangle/N$ and $\log Z/N$ are presented in Fig.~\ref{fig:energy-logz} as a function of $\beta^{-1}$ and $\beta^{-2\mu/(\mu + 2)}$ for different values of $\mu$.
	The behaviour of the coefficients $\tilde{e}_0,\tilde{s}_0,\tilde{c},\tilde{c}_2$ is extracted from Eqn.~\eqref{eq:logz-mu-new} and presented in Fig.~\ref{fig:logz-coeffs-new} demonstrating their dependence on $\mu, q$. 
	
	\begin{figure}[ht]
		\centering
		\includegraphics[width=0.48\linewidth]{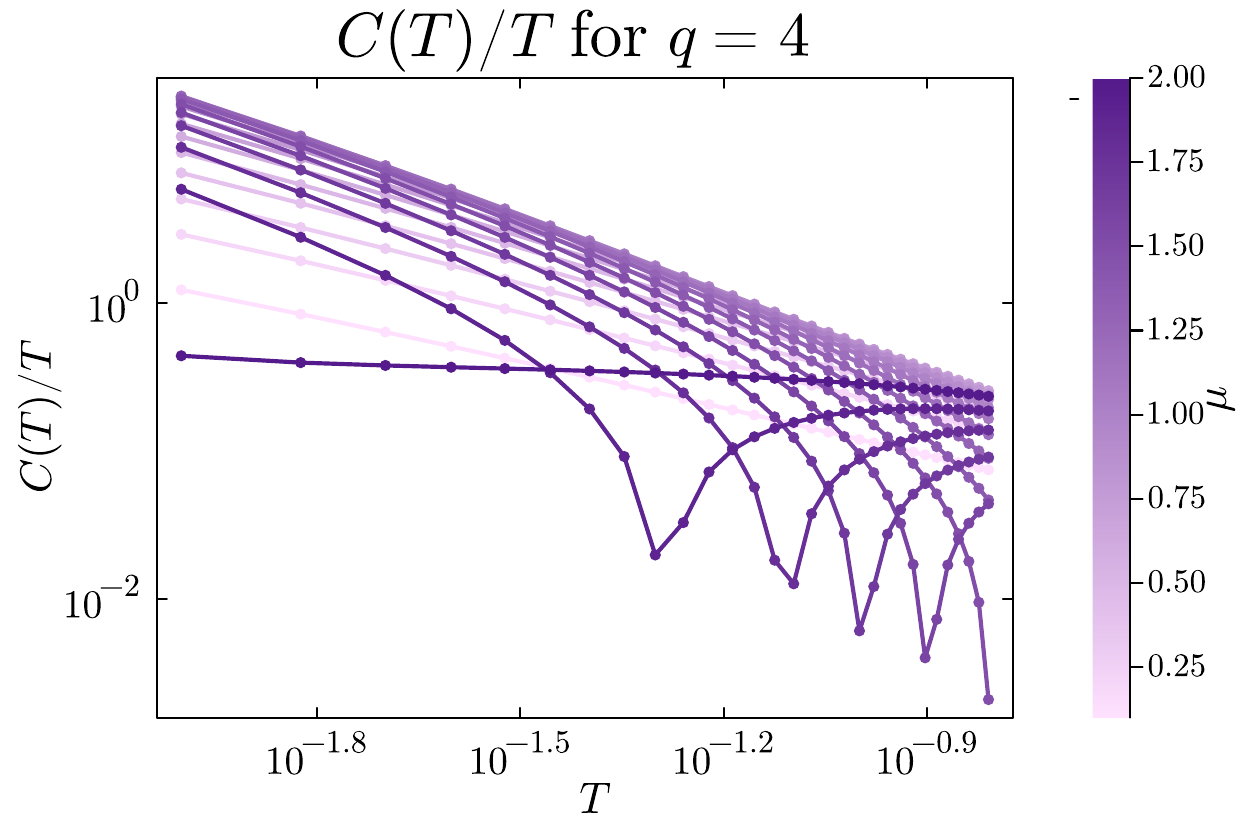}
		\includegraphics[width=0.48\linewidth]{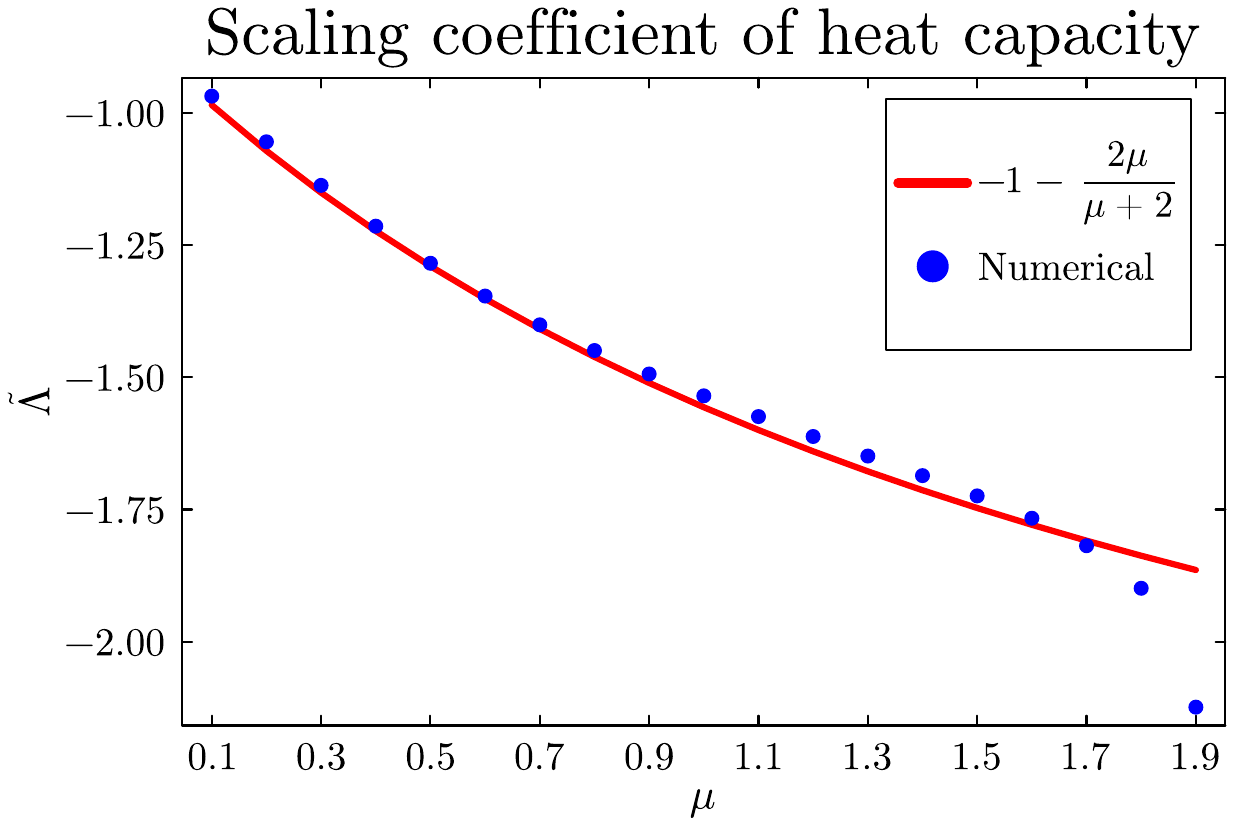}
		\caption{
			Heat capacity per unit temperature \(C(T)/T\) .
            Left: scaling with temperature $T$. 
			Right: The numerical scaling coefficient for small $T$, $\tilde{\Lambda}$, extracted and compared with the analytical expectation of $-1 - \frac{2\mu}{\mu + 2} = \frac{-3\mu - 2}{\mu + 2}$~\eqref{eq:ct-finite-q}.
			%In (a), we highlight the difference between $\lambda_L$ and $2\alpha$, in other words the non-saturation of the $\mathcal{Q}-$complexity bound.
			%In (b) we compute the OTOC exponent $\lambda_L$ in units of $2\pi/\beta$. 
		}
		\label{fig:specific-heat}
	\end{figure}

	Finally, we compute the specific heat $C(T) = T\frac{d S}{d T} = \frac{1}{N}\frac{d \langle H \rangle}{d T}$ and observe its' low temperature behaviour.
	From the $\beta \rightarrow 0$ expansion of $\langle H \rangle$~\eqref{eq:e-per-fermion}, we find the following scaling of specific heat
	\begin{align}
		C(T) \sim \tilde{c}\frac{4\mu}{\mu+2}T^{\frac{2\mu}{\mu+2}} + \tilde{e}_0\left(\frac{2-\mu}{2+\mu}\right)T^{-\frac{2\mu}{\mu+2}} + \cdots.
        \label{eq:ct-finite-q}
	\end{align}
	This is observed for $q=4$ in Fig.~\ref{fig:specific-heat}.
	In the limit $T\rightarrow 0$, the L\'evy contribution $T^{-\frac{2\mu}{\mu+2}}$ dominates over the regular contribution $T^{\frac{2\mu}{\mu+2}}$ for any $\mu < 2$.
	However, the L\'evy contribution vanishes exactly for $\mu=2$ and the regular contribution vanishes for $\mu = 0$ because of the prefactor.
	The regular contribution gives the usual linear scaling at $\mu = 2$.
	
	From the point of view of non-Fermi liquid theory~\cite{lee2018recent,luo2019quantum,chowdhury2022sachdev,davis2023nfl}, the ground state is exponentially degenerate i.e. $S_0 \neq 0$ as follows from Eq.~\eqref{eq:logz-mu-new}. 
	At a finite but low temperature, the system explores an anomalously large portion of its state space, leading to sub-linear entropy growth ($S\sim T^\nu,\, \nu = \frac{2\mu}{\mu +1}< 1$). 
	The effective low-energy description via the Schwarzian mode becomes increasingly rigid (indicated by the divergence of the coefficient of the Schwarzian for $T \to 0$). 
	The excitations cannot be described as quasiparticles since the specific heat diverges relative to the Fermi liquid expectation ($C/T \to \infty$), the spectral function is anomalous, and the system remains chaotic but with a suppressed scrambling amplitude. 
	The parameter $\mu$ continuously tunes between the marginal Fermi liquid ($\mu = 2$) and a frozen phase ($\mu \to 0$) where the ground state degeneracy dominates all thermodynamics.

	\subsection{Comments on the Bulk Dual}

    We comment on the holographic dual for L\'evy SYK under the AdS/CFT dictionary.
    To suggest a possible the gravity dual, we first recall the following facts
	\begin{itemize}
		\item The Lyapnuov exponent satisfies the usual~\cite{maldacena2016a} bound on chaos.
        The sub-leading correction terms (at large-$\beta$) are different from the Gaussian SYK and are given by
		\begin{align}
			\lambda_L \sim \frac{2\pi}{\beta}\left(1 - \frac{(\#)}{\beta^{\frac{2\mu}{\mu + 2}}} - \cdots\right)\notag 
		\end{align}
		\item The conformal theory in the deep IR regime has the same conformal dimension $\Delta = 1/q$ as the Gaussian SYK.
        However, the central charge depends on the temperature $\beta$ and scales as $b^q \sim \beta^{2\left(\frac{2-\mu}{2+\mu}\right)}$. 
		\item The Schwarzian action has the same form as that of the Gaussian SYK $S_f \sim C_\beta\int\mathrm{d}\tau \{f,\tau\}$, but the proportionality constant \(C_\beta\) scales as $\sim \beta^{\frac{2-\mu}{2+\mu}}$.
		\item The leading order contribution to the free energy $\log Z/N$ is also modified, compared to the Gaussian SYK case:
		\begin{align}
			\frac{\log Z}{N} \sim -\beta^{\frac{2\mu}{\mu+2}} \tilde{e}_0 + \tilde{s}_0 + \frac{\tilde{c}}{\beta^{\frac{2\mu}{\mu+2}}} + \frac{\tilde{c}_2}{\beta^{\frac{4\mu}{\mu+2}}}\cdots\notag 
		\end{align}
	\end{itemize}
	The appearance of the Schwarzian in the near-IR expansion indicates that the theory is potentially holographic in nature. 
	A general potential 2D dilaton theory has the following action~\cite{mertens2023solvable}, where the dilaton is represented by the field $\Phi$:
	\begin{align}
		I = \int_{\mathcal{M}} \sqrt{g}(\Phi R + V(\Phi)) - \int_{\partial \mathcal{M}}\sqrt{h} K\,.
	\end{align}
	The saddle equations for this action lead to the geometry $R + V'(\Phi) = 0$ and the dynamical equations $\nabla_\mu\nabla_\nu \Phi - g_{\mu \nu}\nabla^2 \Phi + g_{\mu \nu}V(\Phi) = 0$. 
	\begin{figure}[ht] 
		\centering
		\includegraphics[width=0.85\linewidth]{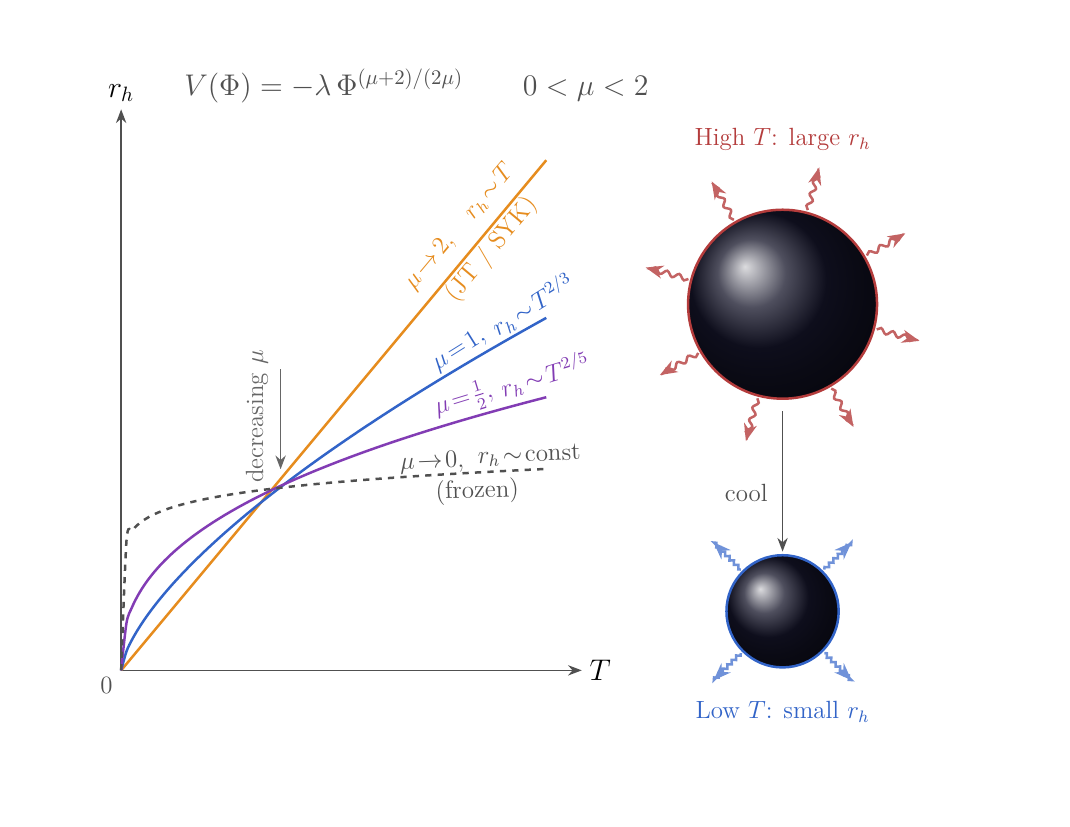}
		\caption{
			Schematic representation of the shrinking dual black hole picture of the Levy SYK model. 
		}
		\label{fig:bh-schem}
	\end{figure}
	The solution for this general set of equations is given by the metric $\mathrm{d}s^2 = -f(\Phi)\mathrm{d}t^2 + f^{-1}(\Phi)\mathrm{d}\Phi^2$, where the dilaton is identified as an effective radial coordinate $r = \frac{a}{2}\Phi$ and the metric component is $f(\Phi) = \frac{4}{a^2}\int_{\Phi_h}^{\Phi}V(\tilde{\Phi})\mathrm{d}\tilde{\Phi}$. 
	The lower bound $\Phi_h$ corresponds to the radius of the black hole horizon in the dilaton gauge. 
	Expanding near the horizon leads to the following thermodynamic relations
	\begin{align}
		T = \frac{V(\Phi_h)}{2 \pi a}\;\;\;,\;\;\; S = \frac{\Phi_h}{4 G_N} \;\;\;,\;\;\; C = \frac{1}{4 G_N} \frac{V(\Phi_h)}{V'(\Phi_h)}\,.
	\end{align}
	The Hawking temperature $T$, thermodynamic entropy $S$ and specific heat $C$ correspond to the thermodynamics of the black hole. 
	From the Schwarzian theory, we recall that the coefficient of the Schwarzian~\cite{maldacena2016conformal} has the scaling $\mathcal{C} = \frac{C}{T} \sim T^{\frac{\mu-2}{\mu+2}}$, which allows us to identify the following potential at the horizon $\Phi_h$
	\begin{align}
		V(\Phi_h) \sim \Phi_h^{\frac{\mu+2}{2\mu}}\,.
        \label{eq:dilaton-potential}
	\end{align}
	The implication of this scaling for the black hole geometry is that the radius of the black hole $\Phi_h$ depends upon the temperature of the boundary state (corresponding to the Hawking temperature in the bulk) with a modified exponent compared to the Gaussian SYK
	\begin{align}
		\Phi_h \sim T^{\frac{2\mu}{\mu + 2}}\,.
	\end{align}
	This describes a black hole whose horizon depends weakly on the boundary theory temperature, see Fig.~\ref{fig:bh-schem} for a schematic representation.
	The implication is that (at low temperatures), higher temperatures are required for smaller $\mu$ to excite the black hole to a larger horizon.
    
    Developing a higher dimensional theory for this effective dilaton potential remains an open problem. 
	Some potential theories could involve coupling with an operator $O$ that scales with the radius as $r^{(2-\mu)/(2\mu)}O$. 
	This would have to be included in such a way that the IR correlator is not modified.
	Power law scaling of dilaton potentials is also common in hyperscaling-violating Lifschitz gravity theories~\cite{Dong2012aspects,huijse2012hidden} arising from charged dilatonic black holes. 
	Further evidence is required to establish the bulk theory rigorously, such as the full computation of the spectrum of LSYK and its' entanglement profile (with comparison with to the holographic entanglement~\cite{ryu2006holographic}).
	A careful treatment of these aspects is beyond the scope of this work. 
	%However, in those theories the IR correlator also picks up a modified conformal dimension, which is not case for LSYK and therefore it is unlikely that the higher dimensional theory is Lifschitz-type. 

	\section{Conclusions}
	\label{sec:conclusions}

	We study the thermodynamic properties of the Sachdev-Ye-Kitaev model with L\'evy Stable disorder in limit of large number of fermions (large $-N$ limit). 
    Strong fluctuations in the interaction due to the L\'evy stable disorder provide a framework for analytically understanding effectively sparse SYK models, as we discussed in Ref.~\cite{bhattacharjee2025levy}.
	The model is controlled by a parameter $\mu \in [0,2]$ where $\mu = 0$ corresponds to a free theory (no interactions) and $\mu = 2$ corresponds to the usual, fully connected Gaussian SYK model.
	Since the method of solution of the Gaussian SYK does not work for the L\'evy SYK, we develop a new method to extract effective dynamical correlators from formally divergent functions that are characteristic of L\'evy Stable distributions.
	We derive the Schwinger-Dyson equations for this model and solve them (a) numerically, (b) in the limit of large-$q$ and (c) in the limit of strong coupling.
	We elucidate the phase diagram of the model.
	The model is found to be maximally chaotic in the $\mu = 2$ point, non-maximally chaotic for $0 < \mu < 2$, and free in the $\mu = 0$ point.
	The chaoticity of the model in the intermediate parameter regime is quantified, by computing the chaos exponents for $2-$ and $4-$ point functions.
	Using the solution of the SD equations, we compute the equilibrium thermodynamic observables -- entropy, average energy, free energy -- for the model, both analytically and numerically.
	We highlight the similarities and difference with Gaussian SYK, and comment on the a possible holographic dual. 

	The model studied in this manuscript represents a non-trivial extension of the Sachdev-Ye-Kitaev model that gives rise to a chaotic-to-integrable crossover in the dynamics of the model as well as temperature-modified thermodynamics compared to the Gaussian SYK. 
	It serves, in an appropriate sense~\cite{bhattacharjee2025levy}, as a solvable version of the sparse Sachdev-Ye-Kitaev model~\cite{garcia2021sparse}. 
	We emphasize that the model is purely integrable only at the point $\mu = 0$, and therefore does not have a \emph{transition} from chaos to integrability. 
	A natural future direction is to develop such a model which remains solvable and demonstrates a transition between chaos and integrability.
	Another open issue is whether the crossover in this model holds to higher orders in $1/N$. 
	The thermodynamics observed in this model warrants further careful analysis, along the lines of holographic duality or non-Fermi liquid behaviour. 
	These can be explored further by computing other holographic quantities such as entanglement, and comparing with bulk RT-surface calculation.
	Finally, several modifications of this model (coupled, dissipative, double-scaled etc.) promise interesting physics and warrant consideration.

	\section*{Acknowledgements}

    B.B. acknowledges financial support from the  Luxembourg National Research Fund under the Grant No. C24/MS/18940482/STAOpen.
    W.E.S is supported by the CQT PhD scholarship.
	A.A. acknowledges the financial support from the Institute for Basic Science (IBS) in the Republic of Korea through the Project No. IBS-R024-D1.
	D.~R. acknowledges FAPESP for the ICTP-SAIFR grant 2021/14335-0 and the Young Investigator grant 2023/11832-9.
	D.R. also acknowledges the Simons Foundation for the Targeted Grant to ICTP-SAIFR.

	% TODO: include author contributions
	\paragraph{Author contributions}
	A.A. and D.R. conceived the initial idea of mimicking the sparse SYK model with a L\'{e}vy-like disorder.
	All the technical developments and results have been obtained, with equal contributions, by B.B. and W.E.S.
	All the authors contributed to the writeup of the manuscript.
	
	% TODO: include funding information
	% \paragraph{Funding information}
	% Authors are required to provide funding information, including relevant agencies and grant numbers with linked author's initials. Correctly-provided data will be linked to funders listed in the %\href{https://www.crossref.org/services/funder-registry/}{\sf Fundref registry}.

	\begin{appendix}

		\section{Alternative Description}
		\label{app:alternate-desc}

		In this section, we present an alternative description of the L\'evy SYK model.
		The eventual partition function is identical; however, the model can now be interpreted in terms of a regular Gaussian SYK. 
		To achieve this, we consider the \emph{Stochastic Representation of L\'evy Random Variables}.
		
\begin{figure} 
			\centering
			\includegraphics[width=\linewidth]{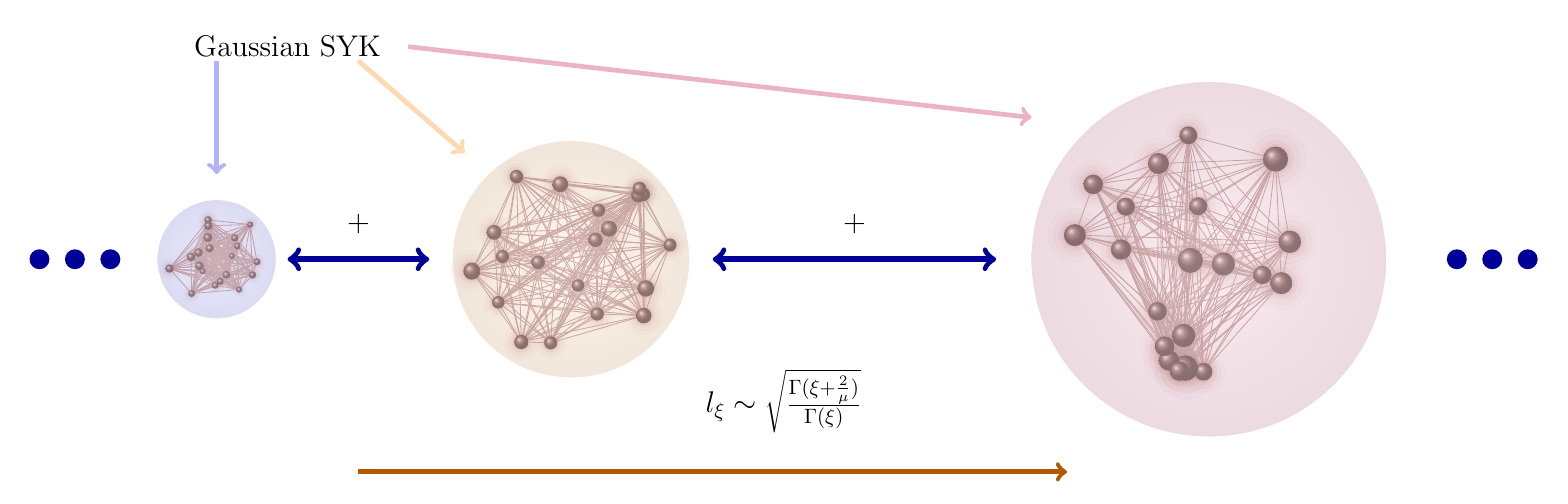}
			\caption{
				Stochastic representation of L\'evy SYK model in terms a sum of infinitely many correlated Gaussian SYK.
			}
			\label{fig:stoch-sykham}
		\end{figure}

		\begin{theorem}[Stochastic Representation of L\'evy Distribution]
			\label{stoch-rep} 
		\end{theorem}
		Consider a random variable $X \in \mathbf{L}_\mu (\eta,\gamma,\delta)$ where $\mu$ is the stability index, $\gamma \in \mathbb{R}^{+}$ is the scale parameter, $\eta \in [-1,1]$ is the ``skewness'' parameter and $\delta \in \mathbb{R}$ is a constant shift.
		Given this, let us construct the following
		\begin{itemize}
			\item[1.] For $\xi \in \mathbb{Z}^{+}$, generate the following set of i.i.d random numbers $\{e_1,e_2,e_3,\dots,e_{\xi}\}$ from the exponential distribution with unit rate (i.e. from the PDF $P(X) = \exp(-X)$).  
			\item[2.] Evaluate the sum of these random numbers $\Gamma_{\xi} = \sum_{l = 1}^{\xi}e_{l}$. 
			\item[3.] For $\xi \in [1,\infty)$, generate i.i.d random numbers $W_{\xi}$ from any distribution has finite moments $\langle \vert W \vert^{\mu} \rangle < \infty$ (for $\mu \neq 1$) or $\langle \vert W \vert \log \vert W \vert \rangle < \infty$ (for $\mu = 1$) .
			\item[4] Evaluate the following $\mu-$dependent constant 
			\begin{align}
				k_{\xi}(\mu) = \begin{cases}
					&0 \;\;\;\;\;\;\;\;\;\;\;\;\;\;\;\;\;\;\;\;\;\;\;\;\;\;\;\;\;\;\;\;\;\;\;\;\;\;\;\;\;\;\;\;\;\;\;\;\;\;\;\;\;\;\, \mu < 1  \\
					&\left\langle W_1 \int_{\vert W_1 \vert/\xi}^{\vert W_1 \vert/(\xi-1)}x^{-2}\sin(x) \mathrm{d}x \right\rangle \;\;\;\;\;\;\;\;\; \mu = 1 \\
					&-\frac{\mu}{\mu - 1}\langle W_1 \rangle \;\;\;\;\;\;\;\;\;\;\;\;\;\;\;\;\;\;\;\;\;\;\;\;\;\;\;\;\;\;\;\;\;\;\;\; 1  < \mu \leq 2
				\end{cases}\notag
			\end{align}
		\end{itemize}
		The Stochastic Representation theorem then implies the following result
		\begin{align}
			\sum_{\xi = 1}^{\infty}\left(\Gamma^{-1/\mu}_{\xi}W_{\xi} - k_{\xi}(\mu)\right) \xrightarrow[r.v.]{a.s.}
			\begin{cases}
				&\mathbf{L}_{\mu}(\eta,\gamma,0) \;\;\;\;\;\; \forall\; \mu \neq 1 \\
				&\mathbf{L}_{\mu}(\eta,\gamma,\delta) \;\;\;\;\;\;\;\;\; \mu = 1
			\end{cases}
		\end{align}
		where $\eta = \frac{\langle \sign(W)\vert W \vert^{\mu} \rangle}{\langle \vert W \vert^{\mu} \rangle}$ and $\gamma^{\mu} = \frac{\langle \vert W\vert^\mu \rangle}{\mathcal{N}_{\mu}}$ with $\mathcal{N}_{\mu} = \sin(\frac{\pi \mu}{2})\frac{\Gamma(\mu)}{\pi}$ and finally $\delta =  - \langle W \log W \rangle$.
		In other words, the theorem gives a recipe for constructing a L\'evy random variable from distributions which have finite moments.
		This is a realization of the Generalized Central Limit Theorem.
		Using this, we can rewrite the LSYK Hamiltonian as
		\begin{align}
			H = \sum_{I}J_{I}\Psi_{I} \rightarrow \sum_{I,\xi}\Gamma^{-1/\mu}_{I,\xi}W_{I,\xi}\Psi_{I}\,,~\label{eq:stoch-sykham}
		\end{align}
		which can now be used to evaluate the partition function. A schematic of this representation is given in Fig.~\ref{fig:stoch-sykham}.
		The correlation length $l_\xi$ is determined by $l^{-2}_\xi \sim \langle \Gamma_{\xi}^{-1/\mu}\Gamma_{\xi'}^{-1/\mu}\rangle \approx \Gamma(\xi)^{-1}\Gamma(\xi - \frac{2}{\mu})\cdots$.
		This let us write (after a rescaling) $l_\xi \sim \sqrt{\Gamma(\xi + \frac{2}{\mu})/\Gamma(\xi)}$.

		The key point is to choose $W_\xi$ in such a way that the scale parameter $\gamma$ coincides with the scale parameter of $J_I$ (which we denote by $\sigma$).
		It is most convenient to choose $W$ to be Gaussian distributed with variance $\sigma_W$, which gives us the following relation
		\begin{align}
			\sigma^\mu = \frac{2^{\frac{\mu}{2}}\sigma_W^\mu \Gamma(\frac{\mu + 1}{2})\sqrt{\pi}}{\sin(\frac{\mu\pi}{2})\Gamma(\mu)}
			\implies \sigma_W = \left(\frac{\sin(\frac{\pi \mu}{2})\Gamma(\mu)}{\sqrt{\pi}2^\frac{\mu}{2}\Gamma(\frac{\mu+1}{2})}\right)^{\frac{1}{\mu}}\sigma\,.
		\end{align}
		Using this, we can now write the partition function as 
		\begin{align}
			Z(\beta) &= \int \mathcal{D}\psi \exp\left\{\int_0^\beta \mathrm{d}\tau \left(-\frac{1}{2}\sum_{i}\psi_i\partial_\tau\psi_i\right)\right\}\exp\left\{-i^{\lfloor q/2 \rfloor}\sum_I J_I\Psi_I\right\}\notag\\
			&= \int \mathcal{D}\psi \exp\left\{\int_0^\beta \mathrm{d}\tau \left(-\frac{1}{2}\sum_{i}\psi_i\partial_\tau\psi_i\right)\right\} \exp\left\{-i^{\lfloor q/2 \rfloor}\sum_{I,\xi} \Gamma^{-1/\mu}_{I,\xi}W_{I,\xi}\Psi_I\right\}\,.
			\label{eq:partition-function-stochrep}
		\end{align}
		Here we note that the couplings are uncorrelated in the $I$ index.
		However while $W_{I,\xi}$ are also uncorrelated in the $\xi$ index, the $\Gamma_{I,\xi}$ are correlated with each other for different $\xi$ values (for fixed $I$).
		The model can thus be interpreted in terms of a series of correlated Gaussian SYK models.
		To average the partition function in~\eqref{eq:partition-function-stochrep}, we can first average over the i.i.d Gaussian random variables $W_{I,\xi}$, which gives us the following result
		\begin{align}
			\langle Z(\beta)\rangle_{W} = &= \int \mathcal{D}\psi \exp\left\{\int_0^\beta \mathrm{d}\tau \left(-\frac{1}{2}\sum_{i}\psi_i\partial_\tau\psi_i\right)\right\} \exp\left\{\frac{\sigma^2_W}{2}\sum_{I,\xi} \Gamma^{-2/\mu}_{I,\xi}V[G_I]\right\}\,,
		\end{align}
		where $V[G_I] = \left( i^{\lfloor q/2 \rfloor}\int_0 ^\beta \mathrm{d}\tau \Psi_I \right)^2$.
		The next step is to average over the random variable $T_I = \sum_\xi \Gamma^{-2/\mu}_{I,\xi}$.
		Let the PDF of $T_I$ be denoted by $f(T_I)$.
		Then we are required to perform the integral
		\begin{align}
			\left\langle \exp\{v T_I\} \right\rangle_{T_I} = \int \mathrm{d}T_I f(T_I)\exp\{v T_I\}\notag
		\end{align}
		where $v = \frac{\sigma^2_W}{2}V[G_I]$.
		Once again, we can use the stochastic representation (by setting $W_\xi = 1$), which tells us that $T_I$ is described by $\mathbf{L}_{\mu/2}\left(1,\mathcal{N}^{-2/\mu}_{\mu/2},0\right)$.
		As such, we are unable to perform the integral, since a closed-form expression is not known for the distribution of $T_I$. 
		The solution is to first perform a Wick rotation, setting $\beta = -i k$, which transforms $v \rightarrow -v$, allowing us to apply the results for the Laplace transform of L\'evy random variables:
		\begin{align}
			&\left\langle e^{-s X} \right\rangle_{X \sim \mathbf{L}_{\alpha}(1,\theta_\alpha,0)}\notag\\
			&= 
			\begin{cases}
				&\exp\{-\sec(\pi \alpha/2)(\theta_\alpha s)^{\alpha}\}\;\;\;\;\; \alpha \neq 1 \\
				&\exp\{-\frac{2 \theta_\alpha}{\pi}s \log s\}\;\;\;\;\;\;\;\;\;\;\;\;\;\;\;\;\, \alpha = 1
			\end{cases} 
		\end{align}
		This allows us to evaluate the average over \(T_I\)
		\begin{align}
			\left\langle \exp\{-v T_I\} \right\rangle_{T_I \sim \mathbf{L}_{\mu/2}(1,\mathcal{N}^{-2/\mu}_{\mu/2},0)}= \exp\left\{ -\frac{\sec(\pi \mu/4)}{\mathcal{N}_{\mu/2}}v^{\mu/2} \right\}\,,
		\end{align}
		from which we can restore the original integral by rotating back via replacing $k = i\beta$.
		Inserting the expression for $v$ and $\sigma_W$ in terms of $\sigma$ leads to all terms canceling out, simply giving us $\exp\{-(\sigma^2 V[G_I])^{\mu/2}\}$.
		Therefore, we obtain the final action as
		\begin{align}
			\langle Z(\beta)\rangle_{W,\Gamma} = \int \mathcal{D}\psi \exp\left\{\int_0^\beta \mathrm{d}\tau \left(-\frac{1}{2}\sum_{i}\psi_i\partial_\tau\psi_i\right)\right\} \exp\left\{-\sum_{I}\left(\sigma^2 V[G_I]\right)^{\frac{\mu}{2}}\right\}\,,
		\end{align}
		which is the same result that we had obtained previously via the characteristic function approach~\eqref{eq:partition-function-k-avg}.

        \section{Numerical results}
        \label{app:num-res}

        \begin{figure}[ht]
			\centering
			\includegraphics[width=0.49\linewidth]{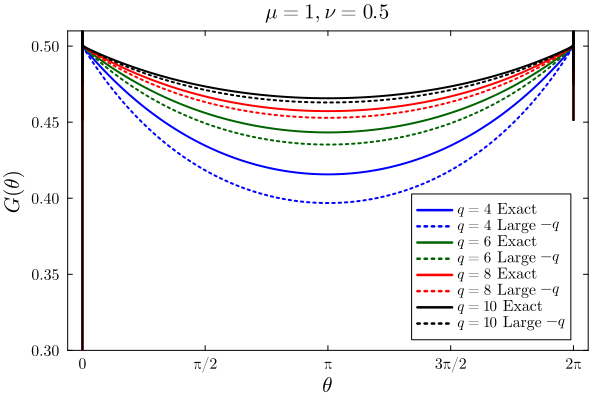}
			\includegraphics[width=0.49\linewidth]{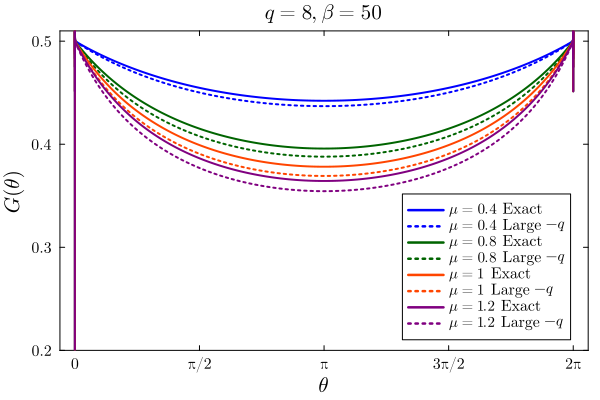}
			\caption{
                Comparison between the numerical and the analytical large-\(q\) solutions of the Schwinger-Dyson equations. 
				Left: comparison for fixed $\mu$ and $\nu$, but for varying $q$.
				Right: comparison for fixed $q$ and $\beta$, but varying $\mu$.
            }
			\label{fig:large-q-soln-comparision}
		\end{figure}
		\begin{figure}[ht]
			\centering
			\includegraphics[width=0.49\linewidth]{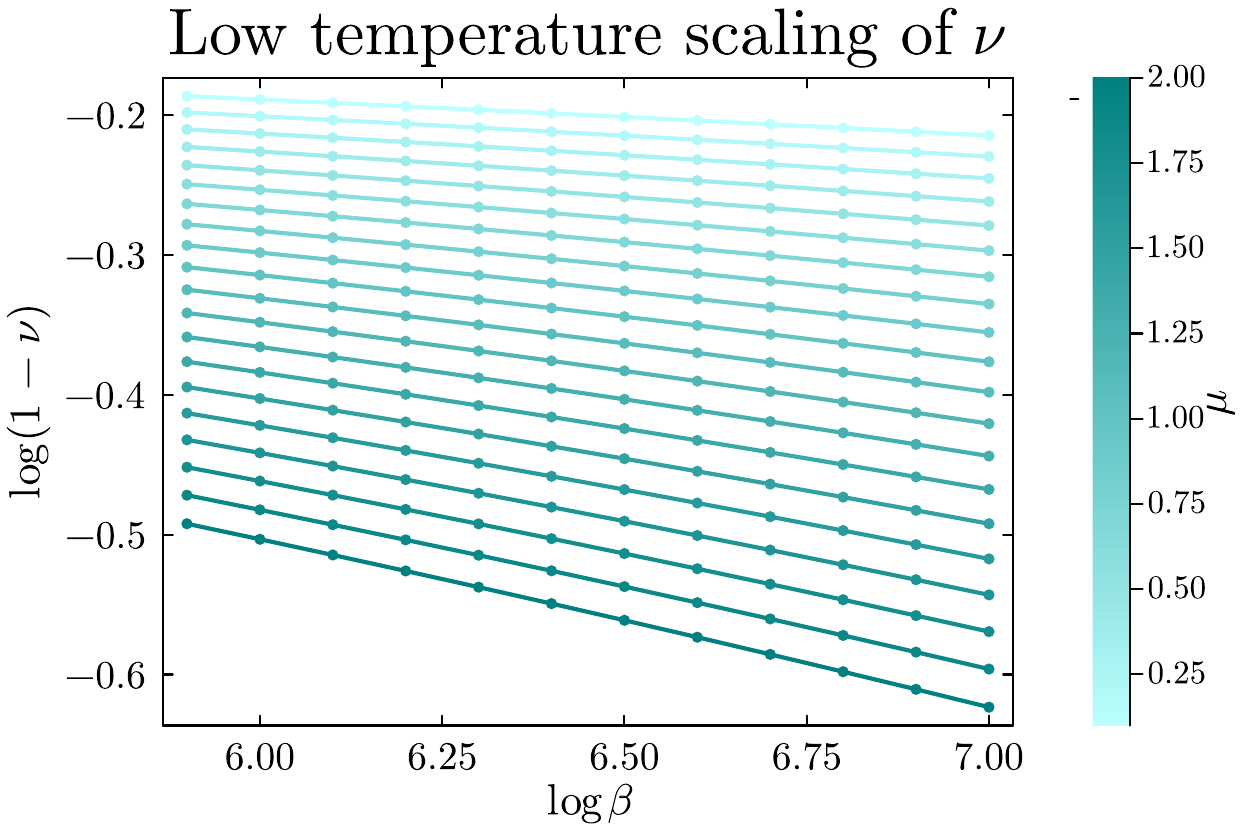}
			\includegraphics[width=0.49\linewidth]{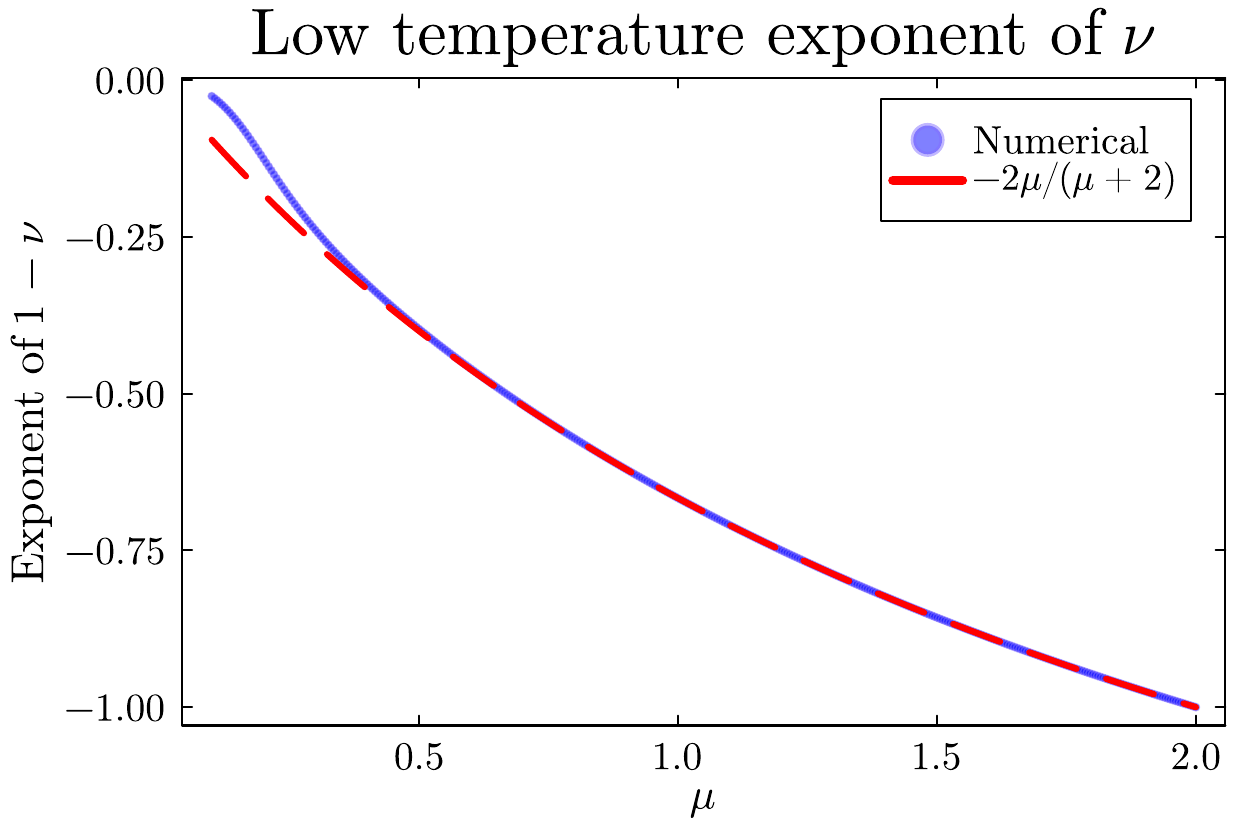}
			\caption{
				Numerical scaling of $1-\nu$ for large $\beta$ (low temperatures). 
				(Left) scaling of $\log(1-\nu)$ with respect to $\log \beta$ indicates power-law scaling of $1-\nu$ with a single dominant exponent.
				(Right) The slope of linear fit of $\log(1-\nu)$ with $\log\beta$ is compared with the analytical expectation $-\frac{2\mu}{\mu + 2}$.
			}
			\label{fig:lowT-nu}
		\end{figure}

		The large-$q$ solution agrees well with the exact numerical solution of the SD equations, as evidenced by the comparison shown in Fig.~\ref{fig:large-q-soln-comparision}.
		It is evident that the difference between the large$-q$ solution and the exact numerical result decreases as $q$ increases, which is expected.
		Surprisingly, the agreement also becomes better for small $\mu$. 
		The scaling of $\nu$ at low temperatures has also been verified numerically (by solving Eqn.~\eqref{eq:eff-int-mu-J}). The scaling exponent is extracted by fitting $1 - \nu$ to $\beta$, and the fit is shown in Fig.~\ref{fig:lowT-nu}.
		The discrepancy between the fit of the exponent and the analytical prediction arises from the poor convergence of iterative solvers used to solve Eqn.~\eqref{eq:eff-int-mu-J} for small $\mu$.

		\section{The Schr\"odinger problem}
		\label{app:sch-prob}
		%\label{app:sch-prob}
		
\begin{figure}[h]
			\centering
			\includegraphics[width=0.85\columnwidth]{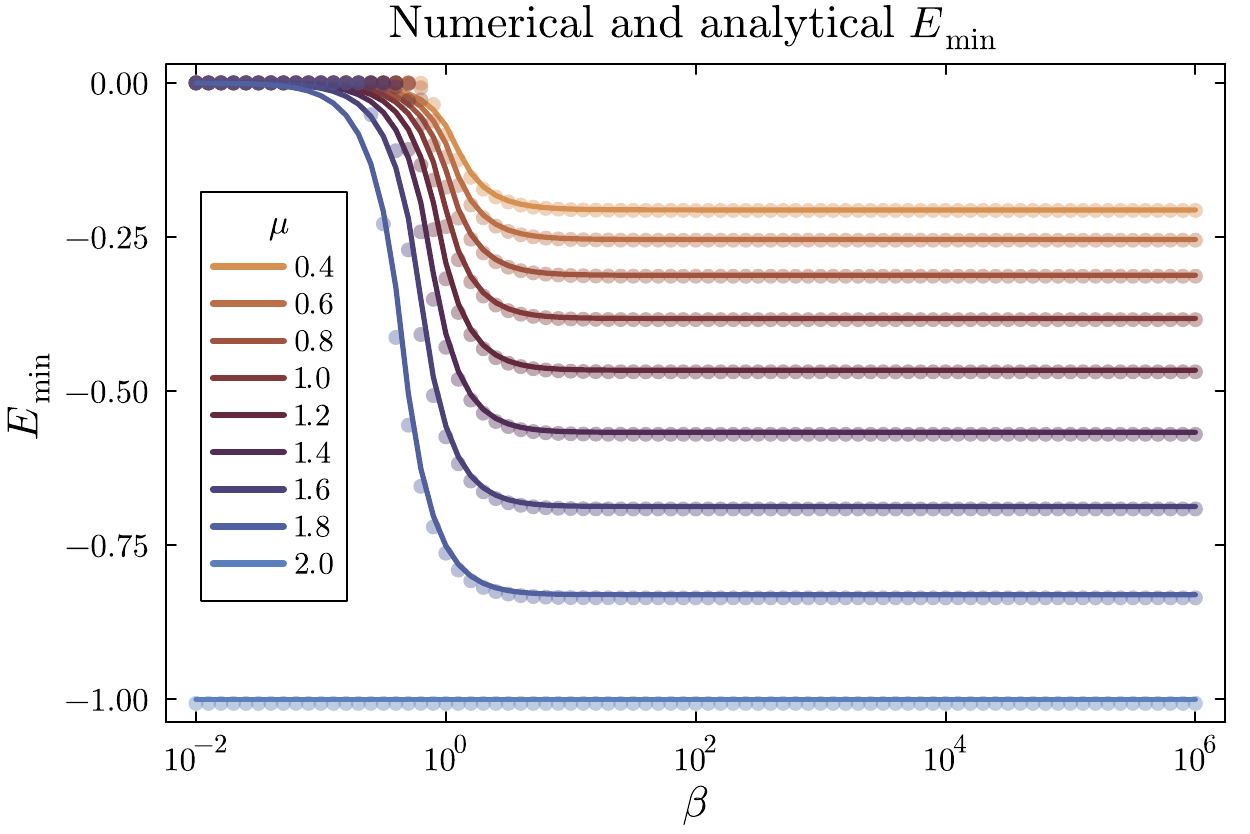}
			\caption{
				The lowest eigenvalue of the Schr\"odinger operator~\eqref{eq:sch-eqn-1}:
                 numerical solution (circles) and analytical approximation (solid curves) as a function of $\beta$ for fixed values of $\mu$.
			}
			\label{fig:emin-comp}
		\end{figure}

		Let us consider the following Schr\"ordinger problem
		\begin{align}
			D_x f \equiv -\partial^2_x f(x) + V(x) f(x)\,,
			\label{eq:sch-eqn-1}
		\end{align}
		where $V(x)$ is some potential.
		In the problem of interest, we have~\eqref{eq:diff-eq-fx}
		\begin{align}
			V(x) = - 2^{\mu-1}\sech^2(x)\left(1 - \frac{2-\mu}{4\beta^2}\frac{\pi\nu}{\tan(\pi\nu/2)}\sech^2(x) \right)\,.
		\end{align}
		We denote $s_1 = 2^{\mu-1}$ and $s_2 = (2-\mu)\pi\nu/4\beta^2\tan(\pi\nu/2)$ and $s_1, s_2 \geq 0$ for all values of $\mu, \beta$.
		We note that the difference
		\begin{align}
			\int \left(\vert V(x) \vert - V(x)\right)\mathrm{d}x = s_1 \int_{x\in \Lambda}\sech^2(x) (1 - s_2 \sech^2(x))\mathrm{d}x \geq 0\,
		\end{align}
		where $\Lambda = \{x\, \vert \,\sech^2(x) \leq s^{-1}_2\}$.
		Since the integrand is positive, the potential $V(x)$ has a finite interval where it is negative.
		Thus, there exists at least $1$ bound state.
		A $0-$th order estimate of the ground state energy is $V_{\min} = \min_{x}V(x)$.
		Differentiating $V(x)$ tells us that
		\begin{align}
			V'(x) = -4 s_1 s_2\tanh (x) \text{sech}^2(x) \left(\text{sech}^2(x)-\frac{1}{2 s_2}\right)\,,
		\end{align}
		which has the zeros $x_0 = 0, \infty, \pm\arcosh(\sqrt{2 s_2})$.
		The only non-trivial zero is the third one, which appears for $s_2 \geq \frac{1}{2}$.
		For $s_2 < \frac{1}{2}$, the relevant zero is $x_0 = 0$.
		Therefore, we have
		\begin{align}
			V_{\min} = 
			\begin{cases}
				&-s_1 ( 1 - s_2)\;\;\;\;\;s_2 < 1/2\\
				&-s_1/4s_2\;\;\;\;\;\;\;\;\;\;\,s_2 \geq 1/2
			\end{cases}\,.
		\end{align}
		To make the estimate agree with the $\mu = 2$ limiting case, as well as the case where $s_2 = 0$ but $2 > s_1 > 0$, we choose the approximation to the minimum energy as 
		\begin{align}
			E_{\min} = 
			\begin{cases}
				&-\epsilon(s_1) ( 1 - s_2)\;\;\;\;\;s_2 < 1/2\\
				&-\epsilon(s_1)/4s_2\;\;\;\;\;\;\;\;\;\;\,s_2 \geq 1/2
			\end{cases}\,.
			\label{app:emin-eqn}
		\end{align}
		where $\epsilon(x) = \frac{(1 - \sqrt{1 + 4 x})^2}{4}$, which is the eigenvalue of the P\"oschl-Teller problem (i.e. $s_2 = 0$) for arbitary $s_1 = x$.
		The numerical results obtained by diagonalizing the operator $D_x$ in Eqn.~\eqref{eq:sch-eqn-1} are compared with the approximation~\eqref{app:emin-eqn}, see Fig.~\ref{fig:emin-comp}.

		On the other hand using the transformation $u = \tanh(x)$, one can write the eigenvalue equation~\eqref{eq:sch-eqn-1} as
		\begin{align}
			\partial_u ((1-u^2)\partial_u f) + \left(s_1 - s_1 s_2 (1-u^2) - \frac{\alpha^2}{1-u^2}\right)f = 0\,,
		\end{align}
		where we denote the eigenvalues as $E = -\alpha^2$.
		This is the \emph{oblate spheroidal wave equation}~\cite{DLMF}. 
		In general, it is difficult to compute the eigenvalues for this equation analytically
		However, these can be computed numerically, which we evaluate and compare with the estimate coming from $E_{\min}$ in Fig.~\ref{fig:emin-comp}.
		It is evident that the agreement is good for any $\mu$ for $\beta \gtrsim 10^{-1}$.

    \end{appendix}

	% TODO:
	% Provide your bibliography here. You have two options:
	
	% FIRST OPTION - write your entries here directly, following the example below, including Author(s), Title, Journal Ref. with year in parentheses at the end, followed by the DOI number.
	%\begin{thebibliography}{99}
	%\bibitem{1931_Bethe_ZP_71} H. A. Bethe, {\it Zur Theorie der Metalle. i. Eigenwerte und Eigenfunktionen der linearen Atomkette}, Zeit. f{\"u}r Phys. {\bf 71}, 205 (1931), \doi{10.1007\%2FBF01341708}.
	%\bibitem{arXiv:1108.2700} P. Ginsparg, {\it It was twenty years ago today... }, \url{http://arxiv.org/abs/1108.2700}.
	%\end{thebibliography}
	
	% SECOND OPTION:
	% Use your bibtex library
	% \bibliographystyle{SciPost_bibstyle} % Include this style file here only if you are not using our template
	%\bibliographystyle{SciPost_bibstyle}
	\bibliography{local,mbl,ergodicity,glass,general,software}
	
	\nolinenumbers
\end{document}

%% file: Figures/rg-flow.tikz
\begin{tikzpicture}[scale=0.65,node style/.style={circle, fill=blue!50, inner sep=1.5pt}]
	%\node [style={red_ball}] (0) at (3.75, 4.25) {};
	%\node [style={blue_ball}] (1) at (-1.5, -1.5) {};
    %\clip (0,0) circle (1);
	\node (1) at (9.5, 0.0) {$\mu=2$};
	\node (2) at (-1.5, -0.0) {$\mu=0$};
	%\node (5) at (3.5, 5.25) {$\rm{SYK}_{\mu}\vert_{\mu=2}=\rm{SYK}_{2}$};
	\node (3) at (4.0, 0.8) {$0<\mu<2$};
	%\draw[black, thick, in=-165, out=15] (0,0) to (3,3);

    %\node (0) at (4.0,2.4) {Chaos Profile};
    %\node (00) at (4.0,1.7) {of L\'evy SYK};
    %\node (000) at (4.0,1.0) {as function of $\mu$};
    % Simulated blur: stack thick-to-thin lines with decreasing opacity
    \draw[RedViolet!20!NavyBlue, line width=10pt, opacity=0.1] (0,0) to (8,0);
    \draw[RedViolet!40!NavyBlue, line width=7pt,  opacity=0.2] (0,0) to (8,0);
    \draw[RedViolet!60!NavyBlue, line width=4pt,  opacity=0.3] (0,0) to (8,0);
    \draw[RedViolet!80!NavyBlue, line width=2pt,  opacity=0.4] (0,0) to (8,0);
    \draw[RedViolet!100!NavyBlue,line width=0.8pt, opacity=0.6] (0,0) to (8,0);
  
    \draw[fill=RedViolet] (0,0) circle (1em);
    \draw[fill=NavyBlue] (8,0) circle (1em);

    \node (4) at (-0.0,-1.0) {\color{RedViolet} Free};

    \node (5) at (8.0,-1.0) {\color{NavyBlue} Maximal};
    \node (8) at (8.0,-1.5) {\color{NavyBlue} Chaos};

    \node (6) at (4.0,-0.7) {\color{Mulberry} Non-maximal};
    \node (7) at (4.0,-1.3) {\color{Mulberry} Chaos};

    % Object 1: Fully Connected Graph (K_20)
    \tikzset{
        fullyConnected/.pic={
            \foreach \i in {1,...,20} {
                \node[node style] (n\i) at (\i*360/20:1.5) {};
            }
            \foreach \i in {1,...,20} {
                \foreach \j in {\i,...,20} {
                    \draw[thin, opacity=0.3] (n\i) -- (n\j);
                }
            }
        }
    }
    
    % Object 2: Sparsely Connected Graph (7 Manual/Random edges)
    \tikzset{
        sparseGraph/.pic={
            \foreach \i in {1,...,20} {
                \node[node style] (s\i) at (\i*360/20:1.5) {};
            }
            % Define 7 specific edges to simulate randomness
            \draw (s1) -- (s5);
            \draw (s3) -- (s12);
            \draw (s18) -- (s2);
            \draw (s7) -- (s8);
            \draw (s20) -- (s11);
            \draw (s15) -- (s4);
            \draw (s9) -- (s14);
        }
    }
    
    % Object 3: Single Edge Graph
    \tikzset{
        singleEdge/.pic={
            \foreach \i in {1,...,20} {
                \node[node style] (e\i) at (\i*360/20:1.5) {};
            }
            \draw[thick] (e1) -- (e10);
        }
    }

    \pic[scale=0.5] at (0,-3.5)   {singleEdge};
    \pic[scale=0.5] at (4,-3.5)   {sparseGraph};
    \pic[scale=0.5] at (8,-3.5)  {fullyConnected};
    
    % Labels for clarity
    %\node at (0,-2)  {Fully Connected};
    %\node at (5,-2)  {Sparsely Connected};
    %\node at (10,-2) {Single Edge}; 

\end{tikzpicture}

%% file: SciPost-version.bbl
\begin{thebibliography}{10}
\providecommand{\url}[1]{\texttt{#1}}
\providecommand{\urlprefix}{URL }
\expandafter\ifx\csname urlstyle\endcsname\relax
  \providecommand{\doi}[1]{doi:\discretionary{}{}{}#1}\else
  \providecommand{\doi}{doi:\discretionary{}{}{}\begingroup
  \urlstyle{rm}\Url}\fi
\providecommand{\eprint}[2][]{\url{#2}}

\bibitem{bhattacharjee2025levy}
B.~Bhattacharjee, W.~E. Salazar, D.~Rosa and A.~Andreanov,
\newblock \emph{L\'evy sachdev-ye-kitaev model} (2025), \eprint{2506.04343}.

\bibitem{bohigas1984characterization}
O.~Bohigas, M.~J. Giannoni and C.~Schmit,
\newblock \emph{Characterization of chaotic quantum spectra and universality of
  level fluctuation laws},
\newblock Phys. Rev. Lett. \textbf{52}, 1 (1984),
\newblock \doi{10.1103/PhysRevLett.52.1}.

\bibitem{haake1991quantum}
F.~Haake,
\newblock \emph{Quantum signatures of chaos},
\newblock In \emph{Quantum Coherence in Mesoscopic Systems}. Springer (1991).

\bibitem{mehta2004random}
M.~L. Mehta,
\newblock \emph{Random Matrices},
\newblock Pure and applied mathematics: v. 142. Elsevier/Academic Press,
  Amsterdam, 3rd ed. edn.,
\newblock ISBN 0120884097 (2004).

\bibitem{guhr1988random}
T.~Guhr, A.~Müller–Groeling and H.~A. Weidenmüller,
\newblock \emph{Random-matrix theories in quantum physics: common concepts},
\newblock Physics Reports \textbf{299}(4), 189 (1998),
\newblock \doi{https://doi.org/10.1016/S0370-1573(97)00088-4}.

\bibitem{srednicki1994chaos}
M.~Srednicki,
\newblock \emph{Chaos and quantum thermalization},
\newblock Phys. Rev. E \textbf{50}, 888 (1994),
\newblock \doi{10.1103/PhysRevE.50.888}.

\bibitem{deutsch2018eigenstate}
J.~M. Deutsch,
\newblock \emph{Eigenstate thermalization hypothesis},
\newblock Reports on Progress in Physics \textbf{81}(8), 082001 (2018),
\newblock \doi{10.1088/1361-6633/aac9f1}.

\bibitem{fava2025designs}
M.~Fava, J.~Kurchan and S.~Pappalardi,
\newblock \emph{Designs via free probability},
\newblock Phys. Rev. X \textbf{15}, 011031 (2025),
\newblock \doi{10.1103/PhysRevX.15.011031}.

\bibitem{wang2004entanglement}
X.~Wang, S.~Ghose, B.~C. Sanders and B.~Hu,
\newblock \emph{Entanglement as a signature of quantum chaos},
\newblock Phys. Rev. E \textbf{70}, 016217 (2004),
\newblock \doi{10.1103/PhysRevE.70.016217}.

\bibitem{xu2024scrambling}
S.~Xu and B.~Swingle,
\newblock \emph{Scrambling dynamics and out-of-time-ordered correlators in
  quantum many-body systems},
\newblock PRX Quantum \textbf{5}, 010201 (2024),
\newblock \doi{10.1103/PRXQuantum.5.010201}.

\bibitem{parker2018a}
D.~E. Parker, X.~Cao, A.~Avdoshkin, T.~Scaffidi and E.~Altman,
\newblock \emph{A universal operator growth hypothesis},
\newblock Phys. Rev. X \textbf{9}, 041017 (2019),
\newblock \doi{10.1103/PhysRevX.9.041017}.

\bibitem{sachdev1993spin}
S.~Sachdev and J.~Ye,
\newblock \emph{Gapless spin-fluid ground state in a random quantum heisenberg
  magnet},
\newblock Phys. Rev. Lett. \textbf{70}, 3339 (1993),
\newblock \doi{10.1103/PhysRevLett.70.3339}.

\bibitem{kitaevLectures}
A.~Kitaev,
\newblock \emph{{A simple model of quantum holography}}.

\bibitem{polchinski2016the}
J.~Polchinski and V.~Rosenhaus,
\newblock \emph{The spectrum in the sachdev-ye-kitaev model},
\newblock Journal of High Energy Physics \textbf{2016}(4), 1–25 (2016),
\newblock \doi{10.1007/jhep04(2016)001}.

\bibitem{maldacena2016remarks}
J.~Maldacena and D.~Stanford,
\newblock \emph{Remarks on the sachdev-ye-kitaev model},
\newblock Phys. Rev. D \textbf{94}, 106002 (2016),
\newblock \doi{10.1103/PhysRevD.94.106002}.

\bibitem{cotler2016black}
J.~S. Cotler, G.~Gur-Ari, M.~Hanada, J.~Polchinski, P.~Saad, S.~H. Shenker,
  D.~Stanford, A.~Streicher and M.~Tezuka,
\newblock \emph{{Black Holes and Random Matrices}},
\newblock JHEP \textbf{05}, 118 (2017),
\newblock \doi{10.1007/JHEP05(2017)118},
\newblock [Erratum: JHEP 09, 002 (2018)],
\newblock \eprint{1611.04650}.

\bibitem{garcia2016spectral}
A.~M. Garc\'{\i}a-Garc\'{\i}a and J.~J.~M. Verbaarschot,
\newblock \emph{Spectral and thermodynamic properties of the sachdev-ye-kitaev
  model},
\newblock Phys. Rev. D \textbf{94}, 126010 (2016),
\newblock \doi{10.1103/PhysRevD.94.126010}.

\bibitem{sekino2008fast}
Y.~Sekino and L.~Susskind,
\newblock \emph{Fast scramblers},
\newblock Journal of High Energy Physics \textbf{2008}(10), 065 (2008),
\newblock \doi{10.1088/1126-6708/2008/10/065}.

\bibitem{shenker2014black}
S.~H. Shenker and D.~Stanford,
\newblock \emph{Black holes and the butterfly effect},
\newblock Journal of High Energy Physics \textbf{2014}(3), 67 (2014),
\newblock \doi{10.1007/JHEP03(2014)067}.

\bibitem{mertens2023solvable}
T.~G. Mertens and G.~J. Turiaci,
\newblock \emph{Solvable models of quantum black holes: a review on
  jackiw--teitelboim gravity},
\newblock Living Reviews in Relativity \textbf{26}(1), 4 (2023),
\newblock \doi{10.1007/s41114-023-00046-1}.

\bibitem{sarosi2018ads2}
G.~Sarosi,
\newblock \emph{{AdS$_{2}$ holography and the SYK model}},
\newblock In \emph{Proceedings of XIII Modave Summer School in Mathematical
  Physics {\textemdash} PoS(Modave2017)}, vol. 323, p. 001,
\newblock \doi{10.22323/1.323.0001} (2018).

\bibitem{rosenhaus2019an}
V.~Rosenhaus,
\newblock \emph{An introduction to the syk model},
\newblock Journal of Physics A: Mathematical and Theoretical \textbf{52}(32),
  323001 (2019),
\newblock \doi{10.1088/1751-8121/ab2ce1}.

\bibitem{chowdhury2022sachdev}
D.~Chowdhury, A.~Georges, O.~Parcollet and S.~Sachdev,
\newblock \emph{Sachdev-ye-kitaev models and beyond: Window into non-fermi
  liquids},
\newblock Rev. Mod. Phys. \textbf{94}, 035004 (2022),
\newblock \doi{10.1103/RevModPhys.94.035004}.

\bibitem{jha2025introduction}
R.~Jha,
\newblock \emph{{Introduction to Sachdev-Ye-Kitaev Model: A Strongly Correlated
  System Perspective}}  (2025),
\newblock \eprint{2507.07195}.

\bibitem{tezuka2023binary}
M.~Tezuka, O.~Oktay, E.~Rinaldi, M.~Hanada and F.~Nori,
\newblock \emph{Binary-coupling sparse sachdev-ye-kitaev model: An improved
  model of quantum chaos and holography},
\newblock Phys. Rev. B \textbf{107}, L081103 (2023),
\newblock \doi{10.1103/PhysRevB.107.L081103}.

\bibitem{garcia-garcia2018chaotic}
A.~M. Garc\'{\i}a-Garc\'{\i}a, B.~Loureiro, A.~Romero-Berm\'udez and M.~Tezuka,
\newblock \emph{Chaotic-integrable transition in the sachdev-ye-kitaev model},
\newblock Phys. Rev. Lett. \textbf{120}, 241603 (2018),
\newblock \doi{10.1103/PhysRevLett.120.241603}.

\bibitem{garcia2021sparse}
A.~M. Garc\'{\i}a-Garc\'{\i}a, Y.~Jia, D.~Rosa and J.~J.~M. Verbaarschot,
\newblock \emph{Sparse sachdev-ye-kitaev model, quantum chaos, and gravity
  duals},
\newblock Phys. Rev. D \textbf{103}, 106002 (2021),
\newblock \doi{10.1103/PhysRevD.103.106002}.

\bibitem{andreanov2025from}
A.~Andreanov, M.~Carrega, J.~Murugan, J.~Olle, D.~Rosa and R.~Shir,
\newblock \emph{From dyson models to many-body quantum chaos},
\newblock Phys. Rev. B \textbf{111}, 035147 (2025),
\newblock \doi{10.1103/PhysRevB.111.035147}.

\bibitem{monthus2016localization}
C.~Monthus,
\newblock \emph{Localization transition in random lévy matrices:
  multifractality of eigenvectors in the localized phase and at criticality},
\newblock Journal of Statistical Mechanics: Theory and Experiment
  \textbf{2016}(9), 093304 (2016),
\newblock \doi{10.1088/1742-5468/2016/09/093304}.

\bibitem{kutlin2024anatomy}
A.~Kutlin and I.~M. Khaymovich,
\newblock \emph{{Anatomy of the eigenstates distribution: A quest for a genuine
  multifractality}},
\newblock SciPost Phys. \textbf{16}, 008 (2024),
\newblock \doi{10.21468/SciPostPhys.16.1.008},
\newblock \eprint{2309.06468}.

\bibitem{janzen2008replica}
K.~Janzen, A.~K. Hartmann and A.~Engel,
\newblock \emph{Replica theory for levy spin glasses*},
\newblock Journal of Statistical Mechanics: Theory and Experiment
  \textbf{2008}(04), P04006 (2008),
\newblock \doi{10.1088/1742-5468/2008/04/P04006}.

\bibitem{janzen2010the}
K.~Janzen, A.~Engel and M.~Mézard,
\newblock \emph{The lévy spin glass transition},
\newblock Europhysics Letters \textbf{89}(6), 67002 (2010),
\newblock \doi{10.1209/0295-5075/89/67002}.

\bibitem{janzen2010thermodynamics}
K.~Janzen, A.~Engel and M.~M\'ezard,
\newblock \emph{Thermodynamics of the l\'evy spin glass},
\newblock Phys. Rev. E \textbf{82}, 021127 (2010),
\newblock \doi{10.1103/PhysRevE.82.021127}.

\bibitem{maldacena2016a}
J.~Maldacena, S.~H. Shenker and D.~Stanford,
\newblock \emph{A bound on chaos},
\newblock Journal of High Energy Physics \textbf{2016}(8), 106 (2016),
\newblock \doi{10.1007/JHEP08(2016)106}.

\bibitem{nolan2020univariate}
J.~P. Nolan,
\newblock \emph{Univariate Stable Distributions},
\newblock Springer Cham,
\newblock \doi{https://doi.org/10.1007/978-3-030-52915-4} (2020).

\bibitem{neri2010the}
I.~Neri, F.~L. Metz and D.~Bollé,
\newblock \emph{The phase diagram of lévy spin glasses},
\newblock Journal of Statistical Mechanics: Theory and Experiment
  \textbf{2010}(01), P01010 (2010),
\newblock \doi{10.1088/1742-5468/2010/01/P01010}.

\bibitem{lee2018recent}
S.-S. Lee,
\newblock \emph{Recent developments in non-fermi liquid theory},
\newblock Annu. Rev. Condens. Matter Phys. \textbf{9}(1), 227 (2018).

\bibitem{luo2019quantum}
Z.~Luo, Y.-Z. You, J.~Li, C.-M. Jian, D.~Lu, C.~Xu, B.~Zeng and R.~Laflamme,
\newblock \emph{Quantum simulation of the non-fermi-liquid state of
  sachdev-ye-kitaev model},
\newblock npj Quantum Information \textbf{5}(1), 53 (2019),
\newblock \doi{10.1038/s41534-019-0166-7}.

\bibitem{davis2023nfl}
A.~E. Davis,
\newblock \emph{{Non-Fermi Liquid Physics and Chaos in Yukawa-SYK Models}},
\newblock Ph.D. thesis, Florida U. (2023).

\bibitem{maldacena2016conformal}
J.~Maldacena, D.~Stanford and Z.~Yang,
\newblock \emph{Conformal symmetry and its breaking in two-dimensional nearly
  anti-de sitter space},
\newblock Progress of Theoretical and Experimental Physics \textbf{2016}(12),
  12C104 (2016),
\newblock \doi{10.1093/ptep/ptw124}.

\bibitem{Dong2012aspects}
X.~Dong, S.~Harrison, S.~Kachru, G.~Torroba and H.~Wang,
\newblock \emph{Aspects of holography for theories with hyperscaling
  violation},
\newblock Journal of High Energy Physics \textbf{2012}(6), 41 (2012),
\newblock \doi{10.1007/JHEP06(2012)041}.

\bibitem{huijse2012hidden}
L.~Huijse, S.~Sachdev and B.~Swingle,
\newblock \emph{Hidden fermi surfaces in compressible states of gauge-gravity
  duality},
\newblock Phys. Rev. B \textbf{85}, 035121 (2012),
\newblock \doi{10.1103/PhysRevB.85.035121}.

\bibitem{ryu2006holographic}
S.~Ryu and T.~Takayanagi,
\newblock \emph{Holographic derivation of entanglement entropy from the
  anti--de sitter space/conformal field theory correspondence},
\newblock Phys. Rev. Lett. \textbf{96}, 181602 (2006),
\newblock \doi{10.1103/PhysRevLett.96.181602}.

\bibitem{DLMF}
\emph{Wave equation in oblate spheroidal coordinates},
\newblock \doi{DLMF-S30.14}.

\end{thebibliography}
